\documentclass[a4paper,fleqn,usenatbib]{mnras}

\usepackage{newtxtext,newtxmath}

\usepackage[T1]{fontenc}
\usepackage{ae,aecompl}

\usepackage{graphicx}	
\usepackage{amsmath}	
\usepackage{amssymb}	
\usepackage{multirow}

\usepackage[utf8]{inputenc}
\usepackage[export]{adjustbox}
\usepackage{wrapfig}

\graphicspath{ {images/} }

\newcommand{\lya}{\mbox{$\rmn{Ly}\alpha$}}
\newcommand{\lunits}{\mbox{$[\rm erg\ s^{-1}]$}}
\newcommand{\munits}{\mbox{$[M_{\odot}/h]$}}
\newcommand{\llya}{\mbox{$L_{{\rm Ly}\alpha}$}}
\newcommand{\galform}{\texttt{GALFORM}}
\newcommand{\pmill}{\texttt{P-Millennium}}
\newcommand{\NoRT}{AM-noRT}
\newcommand{\fesc}{\mbox{$f_{\rmn esc}^{\rm Ly \alpha}$}}
\newcommand{\colorThin}{green}
\newcommand{\colorWind}{blue}
\newcommand{\colorBicone}{red}

\title [\lya\ emitters in a cosmological volume I]{\lya\ emitters in a cosmological volume I: the impact of radiative transfer}

\author[S. Gurung-L\'opez. et al.]{
Siddhartha Gurung-L\'opez,$^{1}$\thanks{E-mail: sidgurung@cefca.es}
\'Alvaro A. Orsi$^{1}$,
Silvia Bonoli$^{1}$, 
Carlton M. Baugh$^{2}$,
\newauthor
and
Cedric G. Lacey$^{2}$.
\\
$^{1}$Centro de Estudios de F\'isica del Cosmos de Arag\'on, Plaza San Juan 1, piso 2, Teruel, 44001, Spain. \\
$^{2}$Institute for Computational Cosmology, Department of Physics, Durham University, Science Laboratories, South Road, Durham, DH1 3LE.   
}

\date{Accepted XXX. Received YYY; in original form ZZZ}

\pubyear{2018}

\begin{document}
\label{firstpage}
\pagerange{\pageref{firstpage}--\pageref{lastpage}}
\maketitle

\begin{abstract}

Lyman-$\alpha$ emitters (LAEs) are a promising target to probe the large scale structure of the Universe at high redshifts, $z\gtrsim 2$. 
However, their detection is sensitive to radiative transfer effects that depend on local astrophysical conditions. 
Thus, modeling the bulk properties of this galaxy population remains challenging for theoretical models.
Here we develop a physically-motivated scheme to predict LAEs in cosmological simulations. 
The escape of \lya\ photons is computed using a Monte Carlo radiative transfer code which outputs a \lya\ escape fraction. To speed-up the process of assigning escape fractions to individual galaxies, we employ fitting formulae that approximate the full Monte Carlo results within an accuracy of 10\% for a broad range of column densities,  gas metallicities and gas bulk velocities. 
We apply our methodology to the semi-analytical model {\texttt{GALFORM}} on a large $N$-body simulation. 
The \lya\  photons escape through an outflowing neutral gas medium, implemented assuming different geometries.
This results in different predictions for the typical column density and outflow velocities of the LAE population.
To understand the impact of radiative transfer on our predictions, we contrast our models against a simple abundance matching assignment. Our full models populate LAEs in less massive haloes than what is obtained with abundance matching. Overall, radiative transfer effects result in better agreement when confronting the properties of LAEs against observational measurements.
This suggest that incorporating the effects of \lya\ radiative transfer in the analysis of this galaxy population, including their clustering, can be important for obtaining an unbiased interpretation of future datasets.

\end{abstract}




\section{Introduction}


During the past two decades, surveys targeting the \lya\ emission in star-forming galaxies, the so-called \lya\ emitters (LAEs), have detected objects out to redshift $z \sim 7$ \citep[e.g.][]{steidel96, hu98,rhoads00, malhotra02, taniguchi05, kashikawa06, guaita10, Konno2016, Sobral2017}.The study of this galaxy population has allowed us to explore the kinematics of the interstellar medium (ISM) in high redshift galaxies \citep{shapley03, steidel10, steidel11, kulas11, Guaita2017, Chisholm2017}, the large scale structure \citep{gawiser07, orsi08, ouchi10, Bielby2016, Kusakabe2018, Ouchi2018a}, the epoch of reionization \citep{santos04, kashikawa06, dayal11,Inoue2018} and to test galaxy formation models \citep{ledelliou06, kobayashi07, nagamine10, orsi12}.


Despite the success in detecting progressively larger samples of LAEs,  their physical interpretation has proven to be a difficult challenge \citep[see][for a review]{dijkstra17}. \lya\ photons are easily scattered by neutral hydrogen, causing a large increase in the path that the photon needs to travel through neutral hydrogen clouds \citep[e.g.][]{harrington73, neufeld90}.  This results in an increased probability of interaction with dust grains, and thus, absorption. Hence, the \lya\ radiative transfer through a neutral medium reduces the \lya\ flux that escapes the galaxy and also modifies the line profile, since each scattering event changes the frequency of the photons. These physical processes also take place in the surrounding intergalactic medium (IGM) of galaxies and can also modify the observed \lya\ flux and line profile \citep{santos04, dijkstra11}.

Analytical approximations for \lya\ radiative transfer have been derived for over-simplistic neutral gas configurations \citep[e.g.][]{harrington73,neufeld90, dijkstra06}. More realistic configurations can be explored with a Monte Carlo algorithm. Individual \lya\ photons are generated inside a neutral hydrogen cloud with a given geometry, kinematics and temperature. The path of \lya\ photons is tracked including their interactions, which produce scattering events, until the photons escape or are absorbed by dust. This approach has been studied in several scenarios \citep{ahn00, zheng02, ahn03, verhamme06, Gronke_2016}. Most notably, Monte Carlo radiative transfer has shown to reproduce the diversity of observed \lya\ line profiles by allowing photons to escape through an outflowing medium \citep[e.g.][]{schaerer08, orsi12}.

Theoretical models of galaxy formation have introduced the effect of radiative transfer in different approximate ways to predict the properties of the LAE population. The first model of LAEs in a hierarchical galaxy formation framework implemented a constant escape fraction of \lya\ photons to reproduce their observed abundance and clustering \citep{ledelliou05, ledelliou06, orsi08}. Further attempts introduced radiative transfer effects over simple geometries in semi-analytical models \citep{orsi12, garel12}. Cosmological hydrodynamical simulations also incorporated \lya\ radiative transfer in post-processing. One approach has been to track \lya\ rays to simulate different lines of sight \citep[e.g.][]{laursen07, laursen09a, Laursen2011} over small volumes. With a Monte Carlo radiative transfer code, \citet{zheng10} showed that the proper treatment of  \lya\ photons radiative transfer has dramatic effects on the clustering of LAEs. However, recently, \cite{Behrens2017} found no significant change in the clustering of LAEs after implementing \lya\ radiative transfer  in the Illustris simulation \citep{Nelson2015}, and attribute the claims of \citet{zheng10}  about the clustering of LAEs to resolution effects.

In the next years many ground-based large surveys such as HETDEX \citep{Hill2008}, J-PAS \citep{J-PAS} and space missions like ATLAS-Probe \citep{wang18}, will aim to detect LAEs over large areas to trace the large scale structure (LSS) at high redshifts. Such measurements could potentially deliver cosmological constraints in redshift ranges well above those currently targeted by Multi-Object Spectroscopic surveys. With progressively larger and more accurate datasets, it becomes crucial to improve our theoretical understanding of galaxies as tracers of the underlying matter distribution \citep{orsi18}. One of our aims in this work is to understand the impact of radiative transfer effects on clustering measurements. 

The model for the \lya\ luminosity of star-forming galaxies presented here is based on a fast implementation of a Monte Carlo radiative transfer. To avoid the prohibitively long time that it would take to run a Monte Carlo code over millions of galaxies, we develop fitting formulae that reproduce the full Monte Carlo results accurately. To illustrate the potential of our model, we apply this methodology to the semi-analytic model \galform\ run over an $N$-body simulation. 
This is a first paper in a series that explores the properties of galaxies selected by their \lya\ luminosity. Here we focus on the impact of the \lya\ RT in defining the properties of the LAE galaxy population. In a forthcoming paper we implement the impact of the intergalactic medium (IGM) and the effects of reionization on the LAE population.

The outline of this paper is as follows: in \S 2 we develop fitting formulae to predict the escape fraction of Lyman alpha photons through outflows. In \S 3, we describe our model for LAEs that combines galaxy formation physics and \lya\ radiative transfer in addition to the implementation of the \lya\ RT in a galaxy formation model is presented. We analyze the LAE population predicted by our model in \S4. We discuss our results in \S5. Finally, conclusions and future work are summarized in \S6.

\section{Model ingredients}

\begin{figure*} 
    \centering
    \includegraphics[width=7.2in]{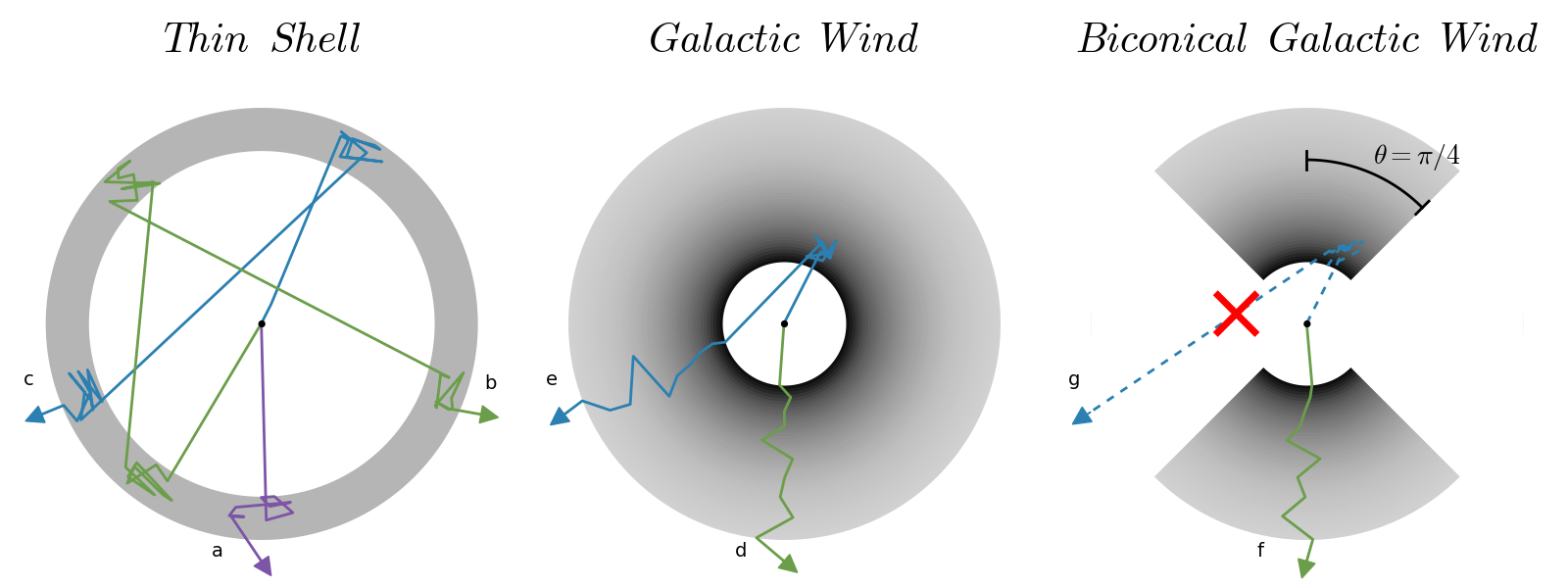}
    \caption{A schematic illustration of the different outflow geometries implemented in this work:  {\it Thin Shell} (left), {\it Wind} (middle) and {\it Biconical Wind} (right). The gas density is represented by the gray colour scale. Different possible trajectories of photons are labeled from $a$ to $g$. The red cross over photon $g$ illustrates the point where this photon is absorbed by the medium.}
    \label{fig:geometries}
\end{figure*}

\indent In this section we describe our model ingredients and the methodology we follow to predict the properties of LAEs in a cosmological simulation. 
    
\subsection{Ly$\alpha$ radiative transfer}\label{ssec:LyaRT}
    
We track the scattering, absorption and escape of \lya\ photons making use of the Monte Carlo radiative transfer code described in \citet{orsi12}, which has been made publicly available \footnote{\url{https://github.com/aaorsi/LyaRT}}. This code is similar to others in the literature \citep[e.g.][and references therein]{zheng02,ahn03,ahn04, dijkstra06,verhamme06,laursen07, barnes10}. A detailed review of \lya\ radiative transfer can be found in \citet{dijkstra17}. Below we summaries the main features of the \citet{orsi12} code that are most relevant to this work.
    
The code receives as input a configuration of a 3D neutral gas geometry, temperature, expansion velocity $\rm V_{exp}$, neutral hydrogen column density $\rm N_{H}$ and optical depth of dust $\tau_{a}$. For a given gas distribution, the code generates a \lya\ photon with a random direction and follows its interactions with hydrogen and dust until it is either absorbed by dust or escapes from the neutral gas medium. Every interaction with a hydrogen atom results in a scattering event that changes the direction and frequency of the photon. Interactions with dust, on the other hand, can change the direction of the photon or result in absorption depending on the assumed albedo of the dust grains. The process is repeated for $N_p$ photons, recording in the end the frequency of every photon that escaped and those that were absorbed by dust grains. This allows us to compute the escape fraction \fesc\ and wavelength distribution (i.e. the \lya\ line profile) for every outflow geometry over which both the neutral gas and the dust are distributed. In this work we implemented three different outflow geometries, which are illustrated in Fig.~\ref{fig:geometries}. 
    
    \begin{enumerate}
    
    \item {\it Thin Shell}.    
    This geometry consists of an expanding isothermal homogeneous spherical shell. This spherical shell is thin and can be described by an inner and an outer radius, $ R_{\rm in}$ and $ R_{\rm out}$ respectively, which satisfy $\rm R_{in}/R_{out} = 0.9$. The shell is expanding outwards, thus it has a radial macroscopic velocity $\rm V_{exp}>0$. The neutral hydrogen column density is given by:

    \begin{equation}\label{eq:Thi_shells_column_density}
    N_H = \frac{M_H}{4 \pi m_H R_{\rm out}^2}, 
    \end{equation} 
where $M_{H}$ is the total neutral hydrogen mass and $ m_{H}$ is the mass of a hydrogen atom.

    The empty cavity in the center of the shell produces photon {\it backscatterings}, i.e. photons can bounce back into the empty cavity multiple times, as illustrated by photons $b$ and $c$ in Fig.~\ref{fig:geometries}.

    \item {\it Galactic Wind}. 
    This geometry consists of an expanding spherical gas distribution with a central empty cavity of radius $ R_{\rm Wind}$. The gas is isothermal and is expanding radially at a constant velocity $ V_{\rm exp}$. Unlike the {\it Thin Shell}, the gas is distributed with a radial density profile given by: 

            \begin{equation}\label{eq:Wind-density}
            \rm 
            \rho_H(r) = \left\{
            \begin{array}{c l}
            0 & r < R_{\rm Wind}\\
            \frac{\dot M_H}{4 \pi m_H r^2  V_{\rm exp}} & r \geqslant R_{\rm Wind},
            \end{array}
            \right.
            \end{equation} 
where $\dot M_H$ is the ejection neutral hydrogen mass rate. Thus, the column density in the {\it Wind} geometry is 

        \begin{equation}\label{eq:Wind-column-density} 
        N_H =  \frac{\dot M_H}{4 \pi m_H  R_{\rm Wind} V_{\rm exp}}.
        \end{equation} 

    This geometry is illustrated in the middle panel of Fig.~\ref{fig:geometries}. 

    We define a large outer radius $R_{\rm out} = 20 R_{\rm Wind}$ where the computation is forced to end and any photon that have reached this radius is considered to have escaped. We have checked that for greater values of $R_{\rm out}$ the code provides the same line profile and  escape fraction. Thus, we conclude that our results converge for our choice of $ R_{\rm out}$.

     \begin{figure*} 
            \centering
            \includegraphics[width=8cm]{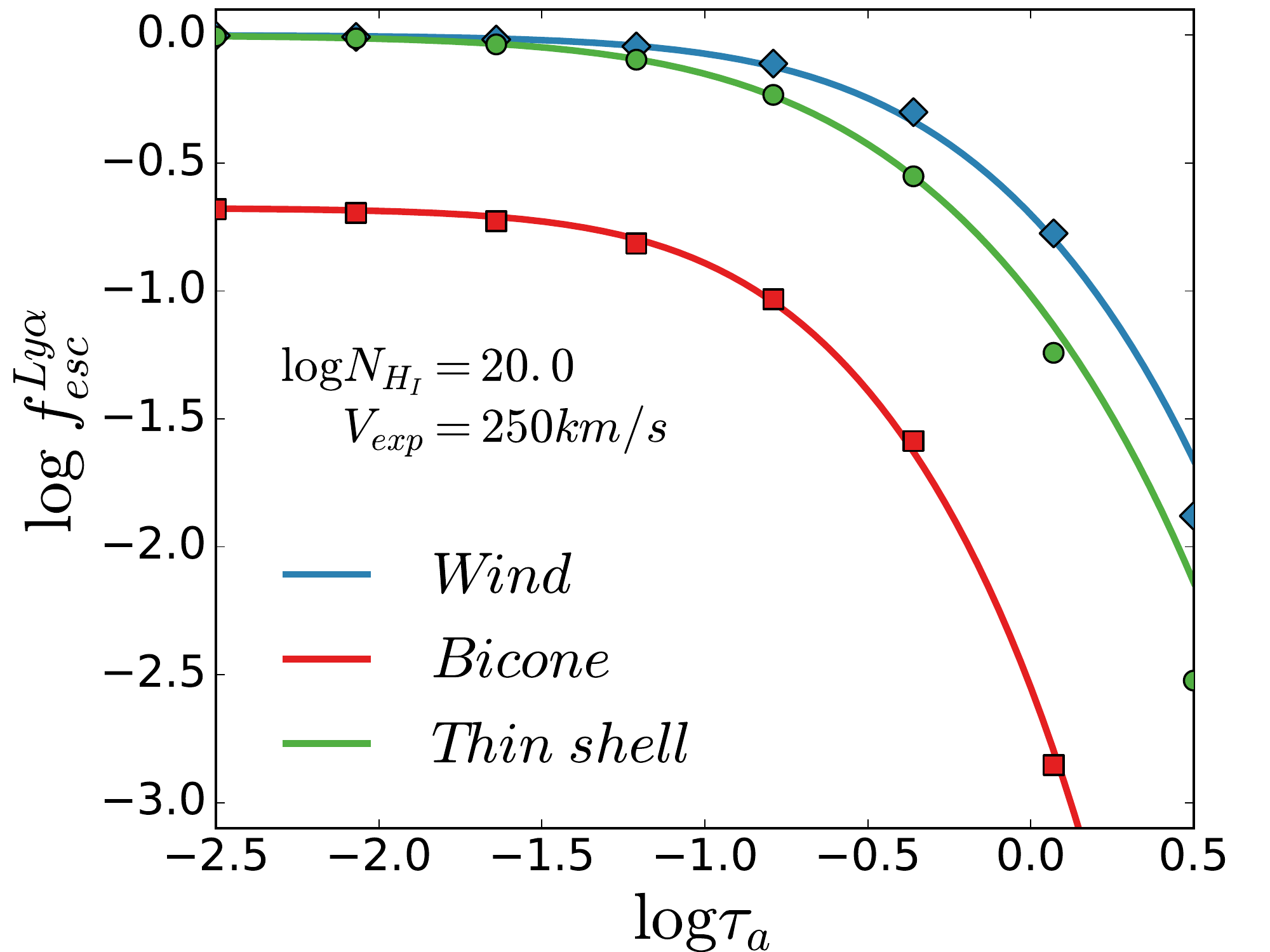}
            \includegraphics[width=8cm]{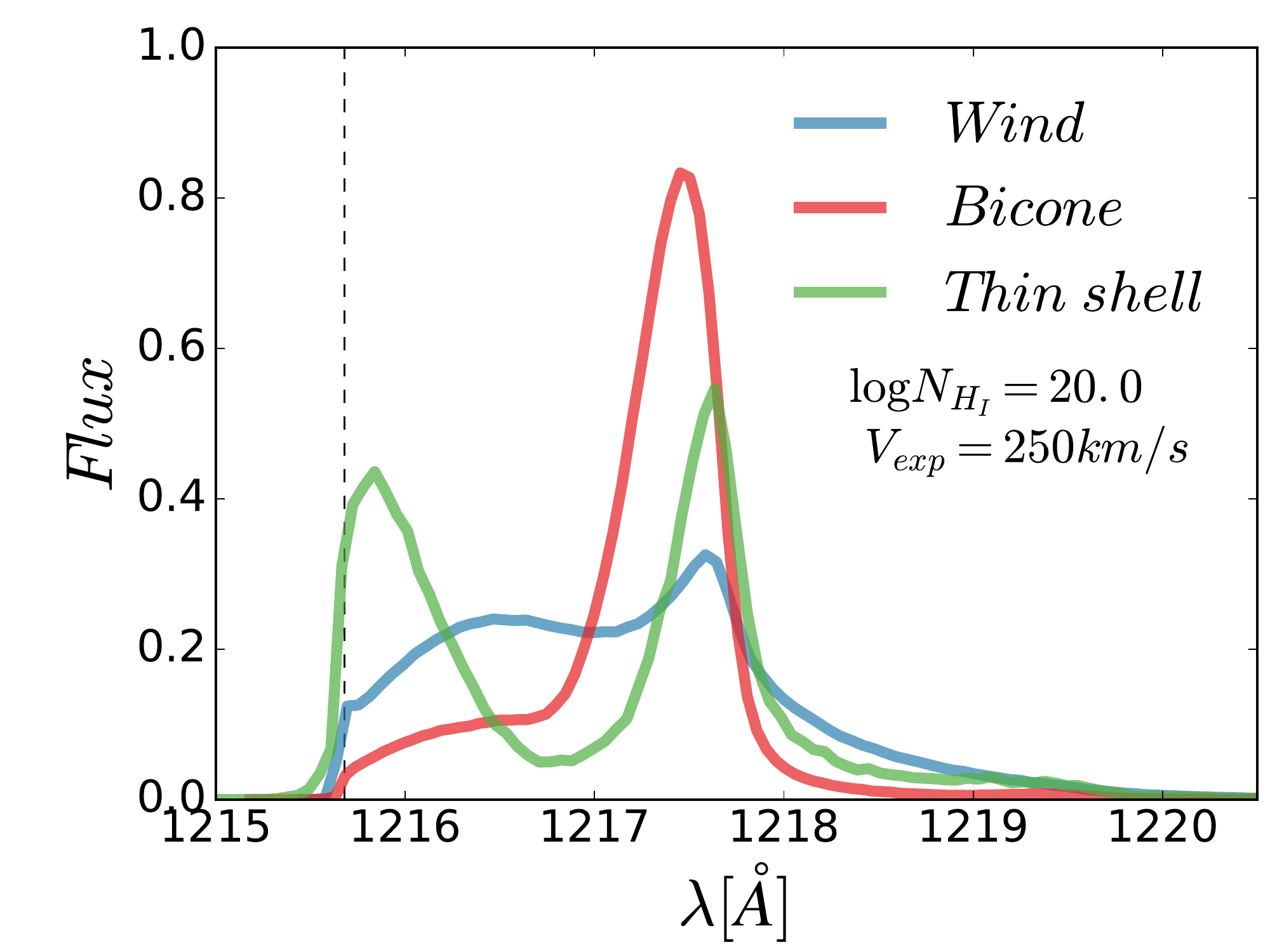}
            \caption{(Left)  $f^{Ly\alpha}_{esc}$ versus the dust optical depth $\tau_a$ for different geometries in outflows with the same physical properties ($V_{\rm exp}$ and $N_{\rm H}$), as indicated in the figures. The output of the radiative transfer code is represented by \colorThin\ circles, \colorWind\ diamonds and \colorBicone\ squares for the {\it Thin Shell}, galactic wind and biconical geometries respectively. Additionally, our analytical fit is represented by solid lines with the same color code as the code's output. (Right) \lya\ line profile for different geometries with the same physical properties. In colored lines the radiative transfer code output is plotted for the {\it Thin Shell} geometry (\colorThin), the galactic wind (\colorWind) and the biconical galactic wind (\colorBicone).  }
            \label{fig:fesc_comparison}
        \end{figure*}

  \item {\it Biconical Wind}. This geometry shares the same properties of the{\it Wind}, but additionally it features an aperture angle, $\theta_{\rm cone}$, which defines the volume of gas and dust. In particular we arbitrarily set $\theta_{\rm cone} = \pi /4 $. The resulting polar asymmetry is thus the main difference between the two previous geometries and this one. This is shown in the right panel of Fig.~\ref{fig:geometries}. \\

Furthermore, in this geometry we force photons to be emitted from the center of the geometry (as in the other geometries) and within the aperture of the bicone, i.e. no photons are emitted outside the bicone. Additionally, due to the empty regions in this geometry, photons that scatter off the internal cavity and escape off the the bicone are considered absorbed by the external medium (e.g. photon $g$ in Fig.~\ref{fig:geometries}). This is equivalent to assuming that there is a dusty optically thick medium surrounding the bicone.

    \end{enumerate}

Fig.~\ref{fig:fesc_comparison} illustrates the difference between the \lya\ escape fraction (left panel) and line profile (right panel) predicted by each geometry, for a particular choice of column density and expansion velocity. As expected, the escape fraction decreases towards higher values of $\tau_a$ in all geometries, as greater amounts of dust absorb more photons. However, the impact on the geometry of the medium is evident: even if the three configurations have the same $ N_{\rm H}$ and $ V_{\rm exp}$, photons have the highest escape fractions from the {\it Wind} geometry, and the lowest from the {\it Bicone}. This is due to the complicated \lya\ RT. For example, as in the {\it Bicone} configuration photons that leak through the empty cavity are considered absorbed, the escape fraction does not reach 1 even if there is no dust in the outflow, making a great difference with respect to the other two geometries. Additionally, even if the {\it Wind} and {\it Thin Shell} configurations share spherical symmetry (unlike the {\it Bicone}) the dependence of \fesc\ on $ N_{\rm H}$, $ V_{\rm exp}$ and $\tau_{a}$ is different due to the distinctive hydrogen density radial profiles of the two configurations. This dependence on the geometry does not only affect the \fesc\ but also the line profile of the \lya\ emission. The predicted line shape changes dramatically from a geometry to another:  in the case of the {\it Wind} it is a broad line, for the {\it Bicone} it is a narrow line and  for the {\it Thin Shell} it assumes a double-peak profile. We use these three different outflow geometries to estimate the variance in the LAEs population depending on the geometry.

\subsection{Fitting formulae for \lya\ radiative transfer} \label{ssec:escape_fraction}

        \begin{table*}
        \centering
        \caption{Constant parameter values used to derive the escape fraction of the different geometries}
        \label{tab:equations}
        \begin{tabular}{ll}
        \hline
        Thin Shell & \\
        \hline
        & \\
        $k_{1} = k_{11} V_{\rm exp} ^ {k_{12}} $ &  $k_{2} = k_{21} V_{\rm exp} ^ {k_{22}} $ \\
        $k_{11} = k_{111} (\log N_{H_{18}})^2+ k_{112} \log N_{H_{18}}+ k_{113} $ & $k_{21}=10^{ -0.0368 }$ \\
        $k_{12} = k_{121} ( \log N_{H_{18}} ) ^ 3 + k_{122} ( \log N_{H_{18}} ) ^ 2 + 
        k_{123}\log N_{H_{18}}+ k_{124} $ & $k_{22}=10^{ -1.556}$ \\
        $k_{111} =10^{2.109}$ &    \\
        $k_{112} = -10^{2.745}$ &    \\
        $k_{113} =10^{2.954}$ &    \\
        $k_{121} =10^{ -1.785}$ &    \\
        $k_{122} = -10^{ -0.730}$ &    \\
        $k_{123} =10^{ -0.155}$ &    \\
        $k_{124} = -10^{0.151}$ &  \\
        \hline
    	Galactic Wind & \\
        \hline
        $k_{1} = k_{11} V_{\rm exp}^{k_{12}}$  & $k_{2} = k_{21} V_{\rm exp}^{k_{22}} $ \\
    
        $k_{11}= k_{111} N_{H_{18}} ^ {k_{112}}$  & $k_{21} =10^{0.0137 }$ \\
        $k_{12}= k_{121} ( \log N_{H_{18}} )^2 + k_{122}\log N_{H_{18}} $  & $k_{22} =10^{ -1.62}$       \\
                        
        $k_{111} =10^{0.471}$ &         \\
        $k_{112} =10^{ -0.244 }$    &  \\
        $k_{121} =10^{ -1.82}$ &        \\
        $k_{122} = -10^{ -0.667}$  & \\
        \hline
        Biconical Wind & \\
        \hline     
        $k_{1} = k_{11}V_{\rm exp}^ {k_{12}} + k_{13} $ &  $k_{2} = k_{21}V_{\rm exp} ^{ k_{22} }$ \\
        $k_{11 } = 10 ^ {3.229}$ &  $k_{21 } =10 ^ {0.0470 }$  \\
        $k_{12 } = -10 ^ {-0.0752}$ & $k_{22 } =10 ^ {-1.490 }$ \\
        $k_{13} = k_{131} \; N_{H_{18}} ^ { k_{132} } +k_{133} $ &  \\
        $k_{131} =10 ^ {-0.580 }$  &  \\
        $k_{132} =10 ^ {-0.238 }$  &  \\
        $k_{133} =10 ^ {0.700}$  &  \\
        $k_{3} = k_{31} V_{\rm exp} ^ { k_{32} } + k_{33} $  &  \\
        $k_{31} = 10 ^{ k_{311} ( \log N_{H_{18}} )^{2} + k_{312} } $ &  \\
        $k_{32} = k_{321} ( \log N_{H_{18}} )^{2} + k_{322} \log N_{H_{18}} + k_{323}$ &  \\
        $k_{33 } =10 ^ {-0.0779}$  &  \\
        $k_{311} =10 ^ {-0.874 }$ & $k_{321} = -10 ^ {-1.226 }$ \\
        $k_{312} =10 ^ {0.571}$ & $k_{322} = 10 ^ {-0.477 }$ \\
        & $k_{323} = -10 ^ {0.292}$ \\
        \hline
        \end{tabular}
        \end{table*}

As discussed in \S\ref{ssec:LyaRT}, the Monte Carlo radiative transfer code can take a long time to run for a given configuration of parameters. For a single photon, the average number of scatterings, and thus, calculations, grows as a power-law function of the column density of the medium \citep{harrington73}. In the parameter space explored here, the completion time of the code can vary from a few seconds up to a few hours in the most extreme cases. Applying this directly in a catalog of millions of objects would result in prohibitively long execution times.

To overcome this, we develop empirical (measured from the radiative transfer Monte Carlo code) expressions that approximate the results of the Monte Carlo runs. We start by constructing a grid to scan the parameter space with $\sim 450$ configurations spanning the ranges $\rm 18 \leq \log(N_H [{\rm cm^{-2}}]) \leq 21$, $\rm 10 \leq V_{\rm exp}[km \ s^{-1}] \leq 1000$ and $-2.5 \leq \log \tau_{a} \leq 0.5$. We run the Monte Carlo code with $10^4$ photons and obtain the \lya\ escape fraction, \fesc\, as a function of $\tau_a$ , $N_H$ and, $V_{\rm exp}$. 

To construct an analytic expression for \fesc\ we start from a generalized form of the expression for the \fesc\ in a homogeneous, static slab derived in \citet{neufeld90}:

\begin{equation}\label{eq:fesc-func}
f_{\rm esc}^{\rm an} = k_3 \left[\cosh\sqrt{ k_1 \tau_{a}^{k_2}}\right]^{-1},
\end{equation}         
where $k_1$ and $k_2$ are functions of $N_H$ and $V_{\rm exp}$ for all geometries. Additionally, $k_3$ is set to $1$ in the {\it Thin Shell} and {\it Wind} geometries, but is a function, $k_3( {\rm  N_{\rm H} , V_{\rm exp}}) < 1$,  in the {\it Bicone}, since, in this geometry, the escape fraction is always less than 1 (see section \ref{ssec:LyaRT}). We perform a Monte Carlo Markov Chain (MCMC) with the \texttt{emcee}\footnote{\url{http://dfm.io/emcee/current/}} code \citep{emcee} to determine the functional form of $k_1$, $k_2$ and $k_3$, by minimizing the function

\begin{equation}\label{eq:MCMC_raditive_transfer_Xi2}
	\chi^2 = \sum\limits_{N_H, V_{\rm exp}, \tau_a} \left(
	\dfrac{f_{\rm esc}^{\rm MC} - f_{\rm esc}^{\rm an}}{\sigma_{MC}}\right)^2,
\end{equation} 
where $f_{\rm esc}^{MC}$ corresponds to the escape fraction of photons obtained with the MC code over each configuration in the grid, and $\sigma_{MC}$ is  the error in the calculation of the escape fraction, given by the dispersion in a binomial distribution with probability of success $f_{esc}^{MC}$ :
        
\begin{equation}\label{eq:fescError}
	\sigma_{MC} = z_{1-\alpha / 2} \sqrt{ \frac{f_{\rm esc}^{MC}(1-f_{\rm esc}^{MC})}{N}},
\end{equation} 
        
where $z_{1-\alpha / 2}$ is the $100(1- \alpha/2)$-th percentile of the standard normal distribution. In particular we use the quantile 95, i.e. $\alpha=0.1$. Additionally, $N$ is the number of generated photons in each configuration.

The functional form and parameter values of the fits for $k_1(N_H , V_{\rm exp}) $, $k_{2}( N_H , V_{\rm exp})$ and $k_3(N_H, V_{\rm exp})$ for each geometry are shown in Table \ref{tab:equations}.
 
Fig.~\ref{fig:fesc_comparison} compares the \fesc\ computed analytically with Eq.~(\ref{eq:fesc-func}) and with the free parameters obtained with the MCMC (lines), and that obtained with the full MC RT  code (symbols) for a given values of $\rm N_{H}$ and $\rm V_{exp}$ and the three different geometries. The analytical expression reproduce remarkably well the results of the full MC RT code. 

\indent The accuracy of our analytic expressions varies with  $\tau_{\rm a}$, $\rm V_{exp}$, $\rm N_{H}$ and the geometry. In particular, there is a strong dependence on $\tau_{\rm a}$:  for every geometry we find that the accuracy decreases with increasing $\tau_{\rm a}$. We find that, in general, the discrepancy with the full MC RT code in configurations with $\tau_{\rm a}>10^{-0.5}$ becomes greater than 10\%. Galaxies with such a large dust absorption, in general, will not be observed as a LAE so we are not concerned about the low accuracy at high $\tau_{\rm a}$. Additionally, we checked that, after calibration of our LAEs model (see \S\ref{sssec:Calibrating}), less than  2\% of the galaxies in every geometry have $\tau_{\rm a}>10^{-0.5}$, making  the contribution of these galaxies negligible. For galaxies with $10^{-1.5}<\tau_{\rm a}<10^{-0.5}$, the discrepancy is just a few percents for $\rm N_{H}$ between $10^{19}$ and $\rm 10^{22.5} cm^{-2}$ and $\rm V_{exp}$ between $80$ and $\rm 1000 \ km \; s^{-1}$. Moreover, for $\tau_{\rm a}<10^{-1.5}$ the discrepancy is typically below the 1\% in the same parameter range.

\indent A detailed assessment of the accuracy of the analytical expressions for \fesc\  is presented and discussed in  Appendix \ref{Ap:A} 

\begin{table*}
\centering
\caption{ Free parameters as defined in equations \ref{eq:recipe-thin-V} and \ref{eq:column_density} after the calibration with the observed luminosity function for different geometries and redshifts. }
\label{tab:parameters}

\begin{tabular}{l|c|cccc}

         redshift & Geometry   & $\log \kappa_{V,disk}$                      & $\log \kappa_{V,bulge}$                     & $\log \kappa_{N,disk}$                        & $\log \kappa_{N,bulge}$  \\ \hline
$z=2.2$  & Thin Shell & \multicolumn{1}{c|}{ 4.440 } & \multicolumn{1}{c|}{ 4.911 } & \multicolumn{1}{c|}{ -12.367 } &  -11.839  \\
         & Wind       & \multicolumn{1}{c|}{ 4.857 } & \multicolumn{1}{c|}{ 4.914 } & \multicolumn{1}{c|}{ -7.065 } &  -5.338   \\
         & Bicone     & \multicolumn{1}{c|}{ 4.982 } & \multicolumn{1}{c|}{ 4.258 } & \multicolumn{1}{c|}{ -8.140 } &  -7.249   \\ \hline
$z=3.0$  & Thin Shell & \multicolumn{1}{c|}{ 4.337 } & \multicolumn{1}{c|}{ 4.549 } & \multicolumn{1}{c|}{ -12.465 } &  -11.915  \\
         & Wind       & \multicolumn{1}{c|}{ 4.691 } & \multicolumn{1}{c|}{ 4.769 } & \multicolumn{1}{c|}{ -7.440 } &  -5.166   \\
         & Bicone     & \multicolumn{1}{c|}{ 4.896 } & \multicolumn{1}{c|}{ 4.338 } & \multicolumn{1}{c|}{ -8.436 } &  -6.404   \\ \hline
$z=5.7$  & Thin Shell & \multicolumn{1}{c|}{ 4.737 } & \multicolumn{1}{c|}{ 4.428 } & \multicolumn{1}{c|}{ -13.906 } &  -11.808  \\
         & Wind       & \multicolumn{1}{c|}{ 4.660 } & \multicolumn{1}{c|}{ 3.782 } & \multicolumn{1}{c|}{ -8.292 } &  -6.180   \\
         & Bicone     & \multicolumn{1}{c|}{ 4.612 } & \multicolumn{1}{c|}{ 3.590 } & \multicolumn{1}{c|}{ -8.078 } &  -7.614  \\ \hline
$z=6.7$  & Thin Shell & \multicolumn{1}{c|}{ 4.659 } & \multicolumn{1}{c|}{ 4.279 } & \multicolumn{1}{c|}{ -13.81 } &  -11.934  \\
         & Wind       & \multicolumn{1}{c|}{ 4.589 } & \multicolumn{1}{c|}{ 3.871 } & \multicolumn{1}{c|}{ -8.073 } &  -5.910   \\
         & Bicone     & \multicolumn{1}{c|}{ 4.455 } & \multicolumn{1}{c|}{ 3.561 } & \multicolumn{1}{c|}{ -7.848 } &  -7.647   \\ \hline

\end{tabular}

\end{table*}

    \subsection{ Simulation and semi-analytical model.}
    
We combine the radiative transfer code described above with the semi-analytical model of galaxy formation \galform\ \citep{lacey16} run on the \pmill\ $N$-body simulation (Baugh et al., in prep.). 
   
The \pmill\ is a state-of-the-art dark matter only N-body simulation using the Plank cosmology: $\rm H_{0}=67.77 \;km\;s^{-1}Mpc^{-1}$, $\Omega_{\Lambda}=0.693$, $\Omega_{M} = 0.307$ , $\sigma_{8}=0.8288$ \citep{Planck_2016}. The box size is ${\rm 542.16\;cMpc\;}h^{-1}$ and the particle mass ${\rm M_{p} = 1.061 \times 10^{8} \;M_{\odot}\;}h^{-1}$ ($5040^{3}$ dark matter particles). Between the initial redshift, $z=127$, and the present, $z=0$, there are 272 snapshots. In this work we use snapshots 77, 84, 120 and 136 corresponding to redshifts 6.7, 5.7, 3.0, 2.2, respectively.

\indent A full review on semi-analytical models of galaxy formation can be found in \citet{baugh06}. The variant of \galform\ used in this work is based on earlier versions described in \citet{cole00, baugh05} and \citet{bower06}. In brief, \galform\ computes the properties of the galaxy population following the hierarchical growth of dark matter halos. Halo merger trees are extracted from an $N$-body simulation (the \pmill\, in our case), so the model can also predict the spatial distribution and peculiar velocities of galaxies.

\indent In \galform, galaxies are formed and evolve as a result of the following processes: i) the radiative cooling and the shock-heating of gas inside halos; ii) the subsequent cooling of gas forming a disk at the bottom of the potential well; iii) quiescent star formation in the disk and starbursts in bulges resulting from disk instabilities and galaxy mergers; iv) feedback processes (supernovae, AGN and photoionization) regulating the star formation, and v) the chemical enrichment of stars and gas that results from star-formation and feedback episodes. Additionally, the variant of \galform\ used in this work assumes different initial mass functions (IMFs) for quiescent and starburst modes of star-formation \citep[see][for more details]{lacey16}.

\indent \galform\ generates a composite spectral energy distribution (SED) for each individual galaxy based on its star-forming history and computes the  rate of emission of hydrogen ionizing photons, $\dot Q_{H}$, by integrating the galaxy SED over  wavelengths bluer  than the Lyman break at $\lambda=912$\AA{}. All ionizing photons are assumed to be absorbed by the neutral medium. Then case B recombination \citep{osterbrock89} is used to compute the intrinsic line luminosity of \lya, where a fraction of $0.66$ of ionizing photons contribute to generating  \lya\ photons.

  \subsubsection{Radiative transfer parameters}
  
    \indent To combine the Monte Carlo radiative transfer code with \galform, we need to derive the parameters that define the neutral gas configuration from the galaxy output properties. In particular, the column density $N_{\rm H}$, expansion velocity $V_{\rm exp}$ and the optical depth of dust $\tau_{\rm a}$ are key to determine the escape fraction. The expansion velocity is computed for the three geometries as: 
   \begin{align} 
\label{eq:recipe-thin-V}
	\rm
    V_{exp , c }  & =  \kappa_{V , c } {{\rm SFR}_{c}} \frac{ r_{c}}{M_{*}}, 
    \end{align}
where the index $c$ denotes the galaxy component (disk or bulge), ${\rm SFR}_{c}$ and  $r_c$ are the SFR and half mass radius of each galaxy component, $M_{*}$ is the total stellar mass of the galaxy and $\rm \kappa_{V , c }$ are two (one per galaxy component) free parameters.\\

\indent The neutral hydrogen column density is computed in different ways depending on the geometry (see section \ref{ssec:LyaRT}) :

\begin{equation}\label{eq:column_density}
{\rm N_{H,c}} = \left\{
\begin{array}{c l}
\kappa_{N , c }  \frac{M_{\rm cold, c}}{r^2_c} & Thin \; Shell \\
\kappa_{N , c }  \frac{  M_{{\rm cold}, c}}{r_c V_{{\rm exp},c} } & Wind \; {\rm and} \; Bicone
\end{array}
\right.
\end{equation} 
where $M_{\rm cold , c}$ and $\kappa_{\rm N , c }$ are, respectively, the cold gas mass and a free parameter of the galaxy component $c$. 

All the free parameters linking \galform\ properties to $\rm V_{exp}$ and $\rm N_H$ are calibrated by fitting the observed LAE luminosity function at different redshifts. For further details see \S \ref{ssec:Calibrating}. 

Finally, the $\tau_a$ is computed for every geometry as: 
	\begin{align}
	\label{eq:recipe-ta}
    \tau_{a,c} & =  ( 1 - A_{{\rm Ly}\alpha} ) \frac{E_{\odot}}{Z_{\odot}}{N_{H,c}} Z_c,
    \end{align}

where $A_{Ly\alpha} = 0.39 $ is the albedo at the Ly$\alpha$ wavelength, $E_{\odot} = 1.77 \times 10^{-21} {\rm cm^{-2}}$ is the ratio $\tau_a/N_H$ for solar metallicity, $Z_{\odot} = 0.02$  \citep{granato00} and  $Z_c$ is the cold gas metallicity of the galaxy component $c$.

\begin{figure*} 
        \includegraphics[width=7.0in]{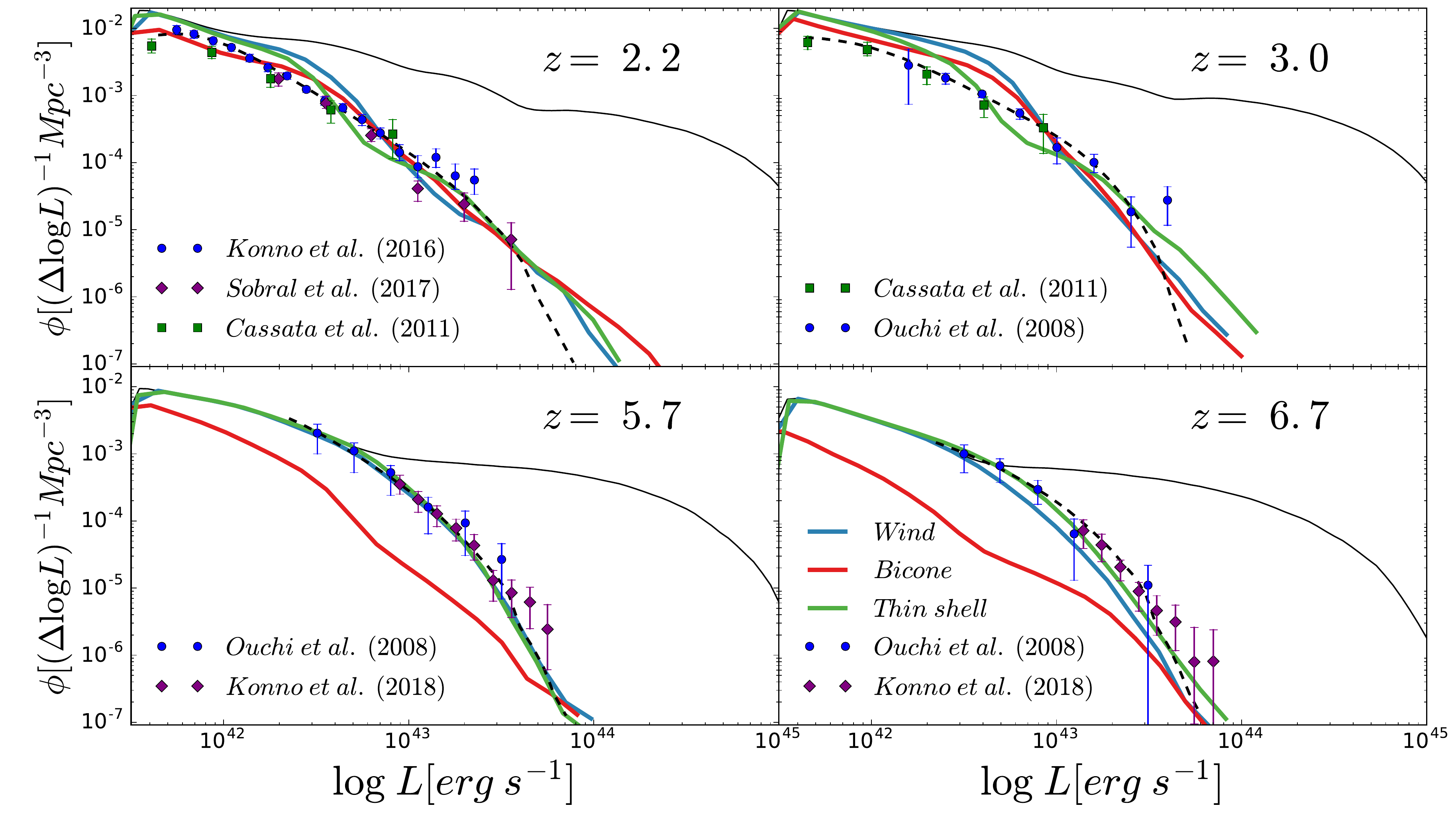} 
        \caption{  LAE LF at redshift 2.2 (top left), 3.0, (top right), 5.7 (bottom left) and 6.7 (bottom right). The LF computed for different geometries is plotted as colored continuum lines, in \colorWind\ for the {\it Wind} geometry, in \colorBicone\ for the {\it Bicone} geometry and in \colorThin\ the {\it Thin Shell} geometry. In continuum black we show the intrinsic \lya\ LF. The black dashed lines show the combined LF that is fitted that, at the same time, is the \NoRT\ LF (detailed in \S \ref{ssec:Results}) LF. At redshift 2.2 we also show the LF observed by Kono et al 2016 (blue dots), Sobral et al. 2016 (purple diamonds) and Cassata et al 2011 (green squares). At redshift 3.0 we show the LF observed by Cassata et al 2015 (green squares) and Ouchi et al. 2008 (blue dots). At redshift 5.7 and 6.7 we show the LF observed by Ouchi et al. 2008 (blue dots) and Konno et al. 2018 (purple diamonds).    }\label{fig:LF_across_redshift_fit}
    \end{figure*}

The intrinsic \lya\ LF predicted by \galform\ (see Figure \ref{fig:LF_across_redshift_fit}) results from two populations: normal star forming galaxies (populating the low luminosity range) and galaxies with an ongoing star formation burst (populating the high luminosity range). Consequently, the values of $\kappa_{N,disk}$ and $\kappa_{V,disk}$ control the shape of the faint-end LF, whereas $\kappa_{N,bulge}$ and $\kappa_{V,bulge}$ control the bright end of the LF. In both regimes, increasing (decreasing) $\kappa_{N,c}$ leads to an increase (decrease) of the $N_H$ distribution. This leads to a decreasing (increasing) in the resulting \fesc\ distribution and thus lowers (increases) the number of galaxies with higher luminosities. Also, increasing (decreasing) $\kappa_{V,c}$ leads to a increase (decrease) of the $\rm V_{exp}$ distribution, increasing (decreasing) \fesc\ and the number of galaxies with high luminosities.

\section{Implementing \lya\ radiative transfer in a semi-analytical model.}

In this section we describe how we incorporate the \lya\ radiative transfer processes inside the semi-analytical galaxies from \galform. We make use of the fitting formula  described above to predict the \lya\ escape fraction and line profiles. The strategy to fit the value of the free parameters of Eqs. \ref{eq:recipe-thin-V} and \ref{eq:column_density} is described below.



\subsection{Calibrating the model.}\label{ssec:Calibrating}

\indent In order to calibrate the model and compute the values of the free parameters for each geometry, we fit our model to the observed LAE luminosity function at redshifts $z=2.2, 3, 5.7$ and $6.7$. We run \texttt{emcee} \citep{Foreman_Mackey_2013} to perform an MCMC to find the values of $\kappa_{\rm N,c}, \kappa_{\rm V,c}$. The dynamical range of each free parameter is determined by limiting the expansion velocity and column densities of each component to lie within $80<V_{\rm exp} [{\rm  km~s^{-1}}] < 1000$ and $19.0 < \log(N_H [{\rm cm^{-2}}]) < 22.5$  for at least 90\% of the resulting galaxy population with \lya\ rest frame equivalent width $\rm EW_{0} > 20$ \AA{} and \lya\ luminosity $\rm L_{Ly\alpha}>10^{41.5}erg\ s^{-1}$. These limits are imposed by the range of validity of the fitting formulae to derive the escape fraction (see \S \ref{ssec:escape_fraction}).

This calibration is done independently for each outflow geometry and individual redshift bin. To combine multiple observed LFs at redshift 2.2 and 3.0 we compute a 5th-order polynomial fit (in logarithm of \lya\ luminosity - logarithm LF space) taking into account the uncertainties of each survey to obtain a single curve that represents the observational measurements. We choose to use a 5th-order polynomial at these redshifts as some recent works suggest that the typical Schechter function is not able to reproduce the observe LF \citep{Konno2016,Sobral2017}. Additionally, at redshift 5.7 and 6.7 we use the best fitting Schechter function to the observed LAE LF computed by \cite{Konno_2018}. The LF used to calibrate our model are shown in Fig.\ref{fig:LF_across_redshift_fit} in black dashed lines.
        
The model \lya\ luminosity of galaxies, for each geometry and choice of $[\kappa_{V,disk}$ , $\kappa_{V,bulge}$ , $\kappa_{N,disk}$ , $\kappa_{N,bulge}]$ is computed as follows: i) we compute the intrinsic \lya\ luminosity of each component, $L_{\rm Ly\alpha}^0$, of each galaxy, which is directly proportional to the ionizing photon production $\dot Q_{H}$ predicted by \galform; ii) we compute $V_{exp,disk}$ , $N_{H_I,disk}$ and $\tau_{a,disk}$ using Eqs.~(\ref{eq:column_density}) and (\ref{eq:recipe-ta}); iii) we obtain $\fesc$ for each galaxy component using Eq.~(\ref{eq:fesc-func}); iv) the observed \lya\ luminosity of each component is obtained by multiplying the intrinsic luminosities by their respective \fesc; and v) the total \lya\ luminosity for each galaxy is the sum of the observed luminosity of each component (disk $+$ bulge).

Fig.~\ref{fig:LF_across_redshift_fit} shows the observed LAE LF (points), the full \galform\ intrinsic \lya\ LF (thin black line), the predictions for each geometry (thick colored lines) using the free parameters that result from the MCMC (listed in table \ref{tab:parameters}) at the different redshifts implemented in this work.

The intrinsic \lya\ LF in divided into two populations: normal SFR galaxies in the low luminosity range and starburst galaxies in the high luminosity range. In general, in \galform\ the galaxy disk component in dominated by a quiescent SFR while in bulges the main mode of star formation is starburst, although quiescent star formation is also included. Additionally, in  \galform\ the quiescent SFR and the starburst have different IMFs, which produces the bumps in the LF. On one hand, at lower redshifts, the predicted intrinsic LF is above the observations at all luminosities, thus galaxies at these redshifts require a significant $f_{\rm esc}^{\rm Ly\alpha}<1$ in order to reduce the amplitude of the LF. On the other hand, at redshifts 5.7 and 6.7, the intrinsic LF at low $\rm L_{Ly\alpha}$ (disk-dominated region) matches observations, implying that galaxies in this range must have $f_{\rm esc}^{\rm Ly\alpha}\sim 1$. Additionally, the intrinsic high redshift LF at high luminosities (bulge-dominated regime) requires $f_{\rm esc}^{\rm Ly\alpha}<1$. 

In general, the MCMC approach finds good matching solutions for the models including the \lya\ radiative transfer. First, we find that the {\it Thin Shell} is consistent with the measured LF at at all redshifts. Secondly, the {\it Wind} geometry performs quite well at $z = 2.2, 3.0$ and 5.7 while at $z=6.7$ it underpredicts the number density of LAE. However, we have checked that by allowing $\rm V_{exp}$ to be slightly higher, the observed LF is matched at redshift 6.7 as well. In the third place, the {\it Bicone} geometry matches the observed LF at $z=2.2$ and 3.0 while at $z=5.7$ and 6.7 it fails. The low abundance of LAEs predicted with the {\it Bicone} geometry arises due to the low escape fractions predicted by this geometry. In fact, at high redshifts, faint \lya\ emitters require escape fractions close to 1 to match the observed LFs, and this is not possible in the {\it Bicone} geometry by construction, as shown in Fig.~\ref{fig:fesc_comparison}.


\subsection{A simplified model with no \lya\ radiative transfer} \label{sssec:SwoRT}

In order to highlight how radiative transfer changes the properties of LAEs, we compare the properties of our model with an abundance matching approach. We perform a simple SFR-\lya\ mapping where no \lya\ radiative transfer is taken into account. We refer to this model variant as '\NoRT'.

To construct the \NoRT\ model, we rank galaxies by their SFR. We assign a \lya\ luminosity to each galaxy based on their total SFR in a monotonic way. Objects with the highest SFR are assigned the brightest \lya\ luminosity. We compute the \lya\ equivalent width using the assigned \lya\ luminosity and continuum luminosity around the \lya\ frequency provided by \galform. \lya\ luminosities are assigned recursively towards lower luminosities such that the Ly$\alpha$ observed luminosity function (using the $EW_0$ cut of each survey) is recovered at each redshift. The resulting \lya\ luminosity distribution is shown in Fig.~\ref{fig:LF_across_redshift_fit} as dashed black line. We compute a \fesc, which corresponds to the ratio between the assigned $Ly\alpha$ luminosity and the intrinsic one. 

In contrast with our RT models, the \fesc\ in the SFR-only model does not depend on properties such as the cold gas mass or the galaxy metallicity. Due to the way that \lya\ luminosities are computed, the resulting \fesc\ can be higher than 1 in some cases.

\section{ Results. }\label{ssec:Results}

\indent In this section we describe the main predictions of our radiative transfer  model when applied to \galform\ with the different outflow geometries.

 \subsection{ The $\rm N_H$ and $\rm V_{ exp}$ distributions.} \label{sec:outprop}
 
\begin{figure*}
\includegraphics[width=7.0in]{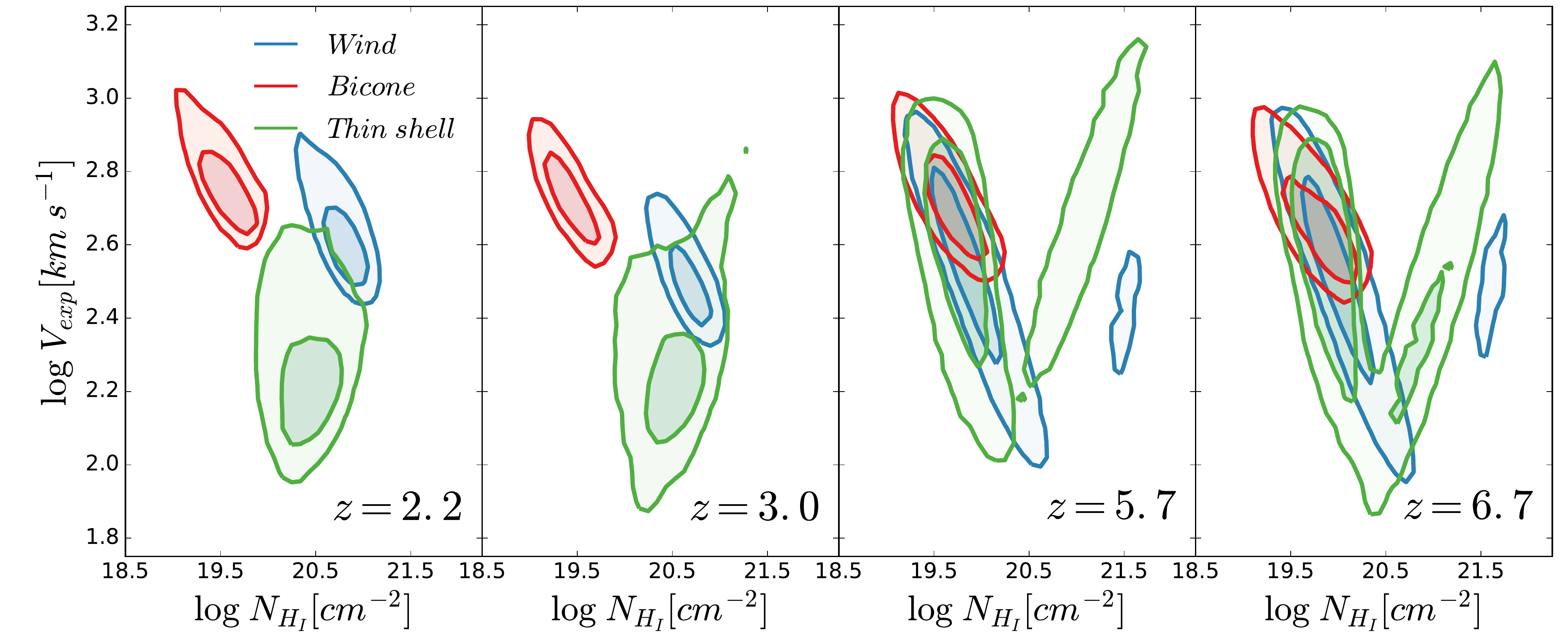} 
\caption{Ouflow expansion velocity and neutral hydrogen column density distributions for each redshift ($z=2.2 , 3.0 , 5.7$ and $6.7$ from left to right) and for each geometry color coded as stated in the legend. The dark and light shaded contours enclose the 40 and 80 percentiles of the galaxy population, respectively.  }
\label{fig:vexp_and_nh}
\end{figure*}

\indent Since the parameters in our model are calibrated to match the observed LFs for each geometry independently, the resulting distributions of $\rm N_{\rm H}$ and $\rm V_{\rm exp}$ are different for each configuration. Though this work unless it is different stated, we define LAE as a galaxy with a \lya\ restframe equivalent width $\rm EW_{0}>20$\AA{} as typically in the literature \citep[e.g.][]{Ouchi2018a}.  In this section, we use a subsample of the full LAE population obtained from each model by imposing a number density cut in \lya\ luminosity of $\rm 10^{-3}h^{3}cMpc^{-3}$. 

Fig.~\ref{fig:vexp_and_nh} shows the distribution of $V_{\rm exp}$ and $\rm N_{\rm H}$ for each geometry. Since each quantity is computed for the disk and bulge component of each galaxy separately, we weight each component by their  observed \lya\ luminosity to build the distributions shown in Fig.~\ref{fig:vexp_and_nh}.

Overall, the $\rm V_{exp} - N_H$ distribution is relatively compact at redshifts (z=2.2,3.0) and more extended at higher redshift (z=5.7,6.7). The {\it Thin Shell} tends to have lower $V_{exp}$ and $\rm N_H$ than the {\it Wind} geometry. Additionally, there is a strong difference between low and high redshift for these two distributions, while, in the case of the {\it Bicone},  remains generally unchanged across cosmic time. Additionally, most of the galaxies lie within the \fesc\ analytic expression optimal accuracy region defined in \S 3. Moreover, we have checked that the fraction of galaxies outside the this region is lower than a 7\% for every geometry and redshift.

Typical values the $V_{exp}$ are found to be around $\rm 150 km/s$ and $\rm 300 km/s$ for the {\it Thin Shell} and {\it Wind} geometries respectively at $z = 2.2, 3.0$. Meanwhile, $\rm N_H$ is found at $\rm \sim 10^{20.5} cm^{-2}$ for the {\it Thin Shell} and $\rm \sim 10^{20.8} cm^{-2}$ for the {\it Wind}. Notably, at higher redshifts, $z=5.7$ and 6.7, the distributions acquire a 'V' shape (especially visible for the {\it Thin Shell}) due to the division of each \galform\ galaxy into a  disk and bulge  and the significant difference in \fesc\  for starburst and normal SFR galaxies at these redshifts. Lower column densities are favored by disk-dominated galaxies, requiring a higher \fesc\ in order to fit the LF. the distribution of these galaxies peak around $\rm N_H \sim 10^{19.7} cm^{-2}$ and $\rm V_{exp} \sim 300 km/s$. Bulge-dominates starbursts require a lower \fesc\ to fit the LF, thus they favor high $N_H$ and low $V_{\rm exp}$ distributions centered around $\rm 10^{21.0} cm^{-2}$ and $\rm 200 km/s$ respectively.

The {\it Bicone} geometry displays noticeable differences with respect to the other two geometries. The {\it Bicone} $\rm V_{exp} - N_{H}$ distributions are very similar across the different redshifts used in this work and present the available highest $\rm V_{exp}$ and lowest $\rm N_{H}$ distributions (peaking around $\rm 600km/s$ and $\rm 10^{19.2}cm^{-2}$ respectively), maximizing as much as possible the escape of \lya\ photons. This is due to the fact that the typical \fesc\ is always lower in the {\it Bicone} compared to the other geometries, and it never reaches $1$. Thus, the \lya\ LF with this geometry is not able to fit the observed LF, as shown above.

\subsection{Breaking down the \lya\ LF} \label{ssec:lyaLF}
\begin{figure*}
        \centering
        \includegraphics[width=.48\linewidth]{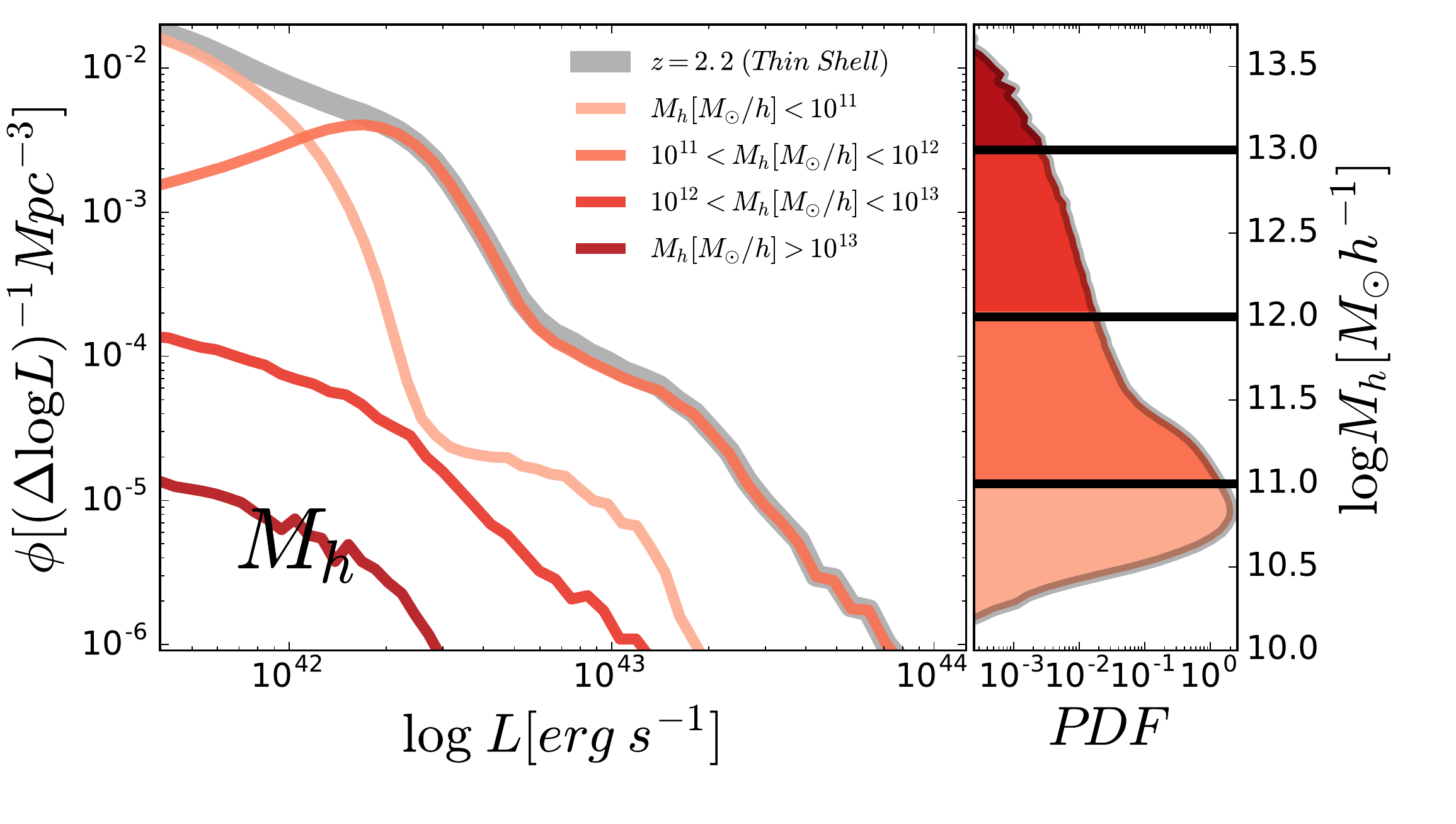}
        \includegraphics[width=.48\linewidth]{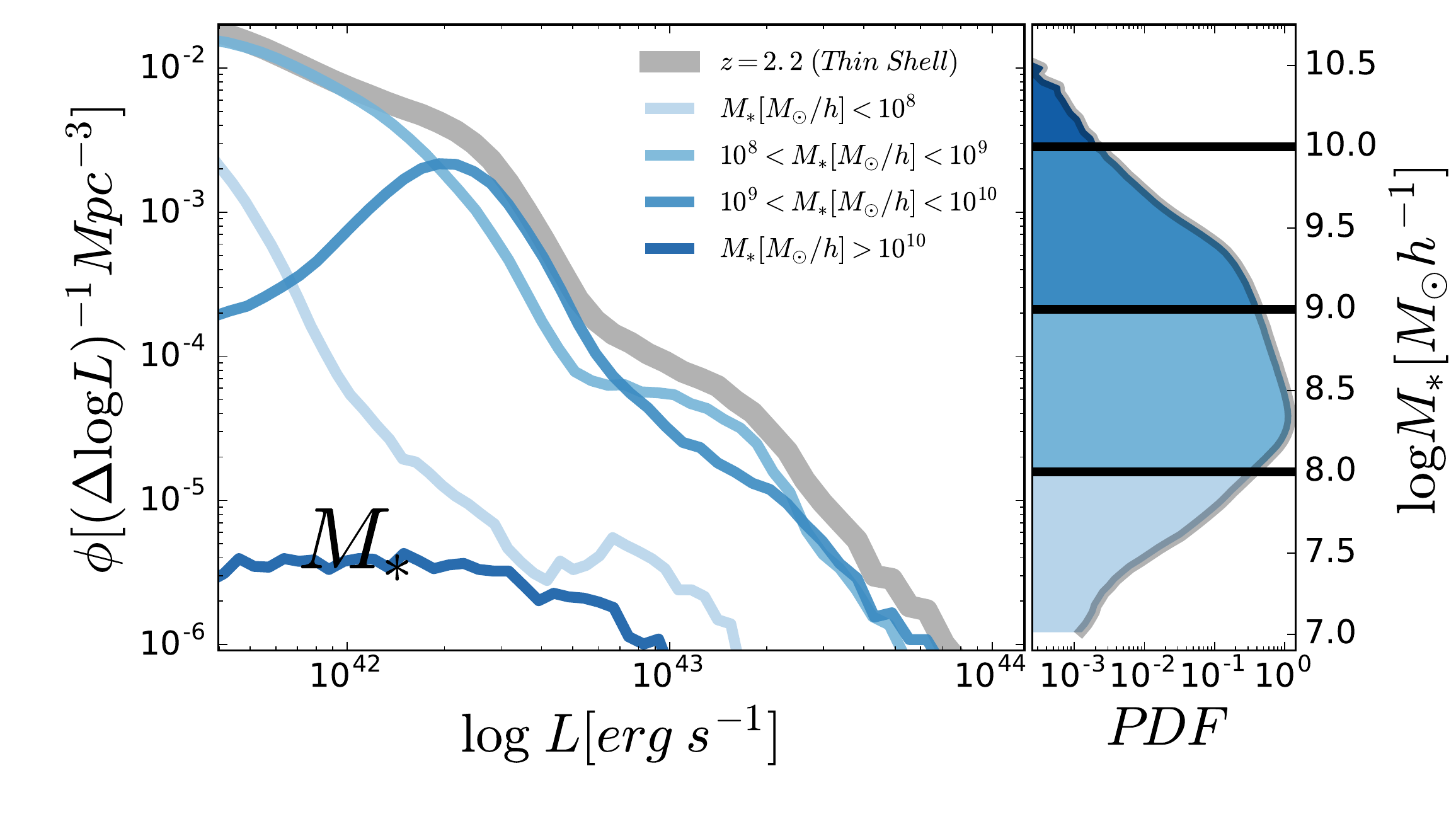}
        \includegraphics[width=.48\linewidth]{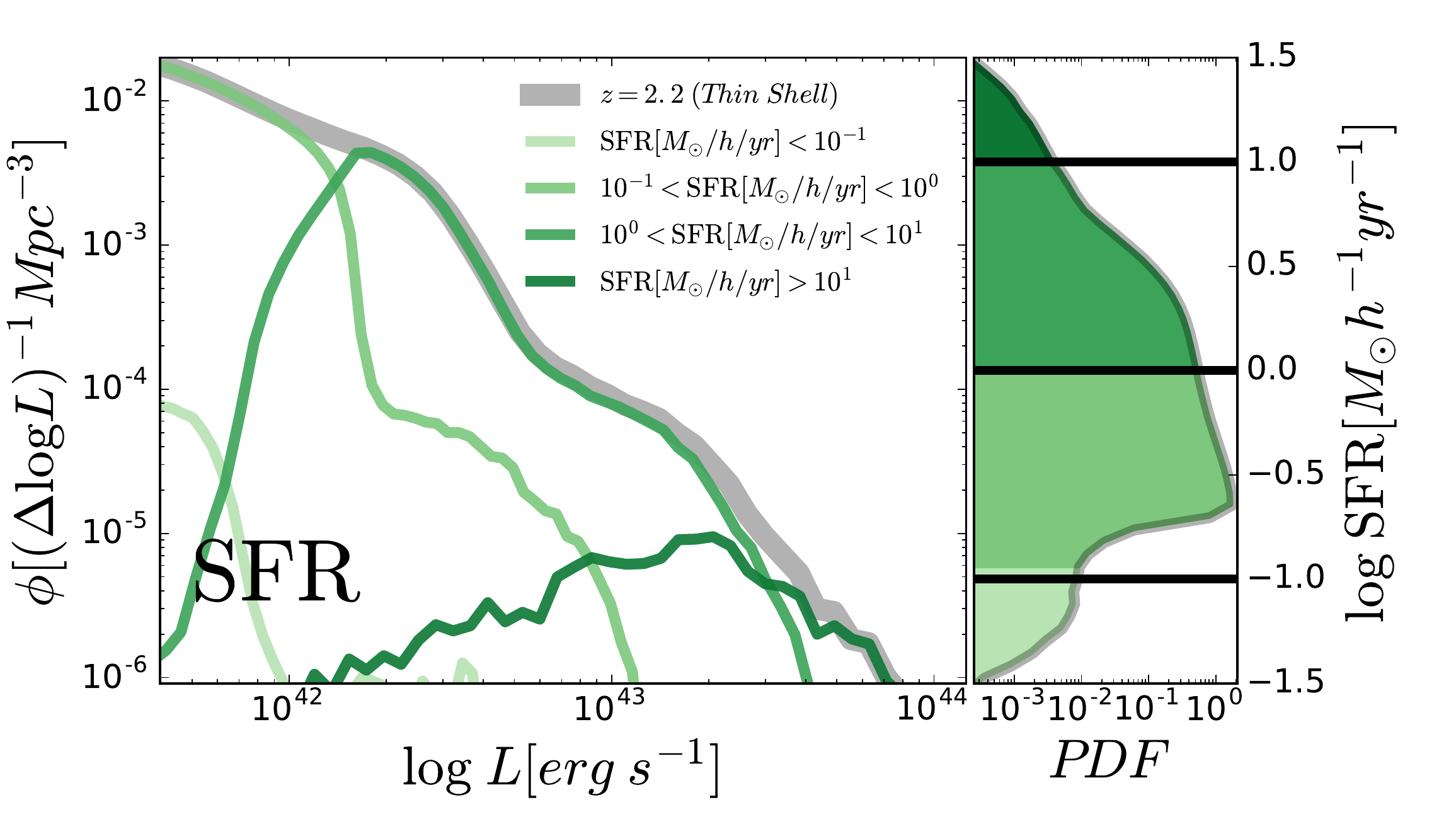}
        \includegraphics[width=.48\linewidth]{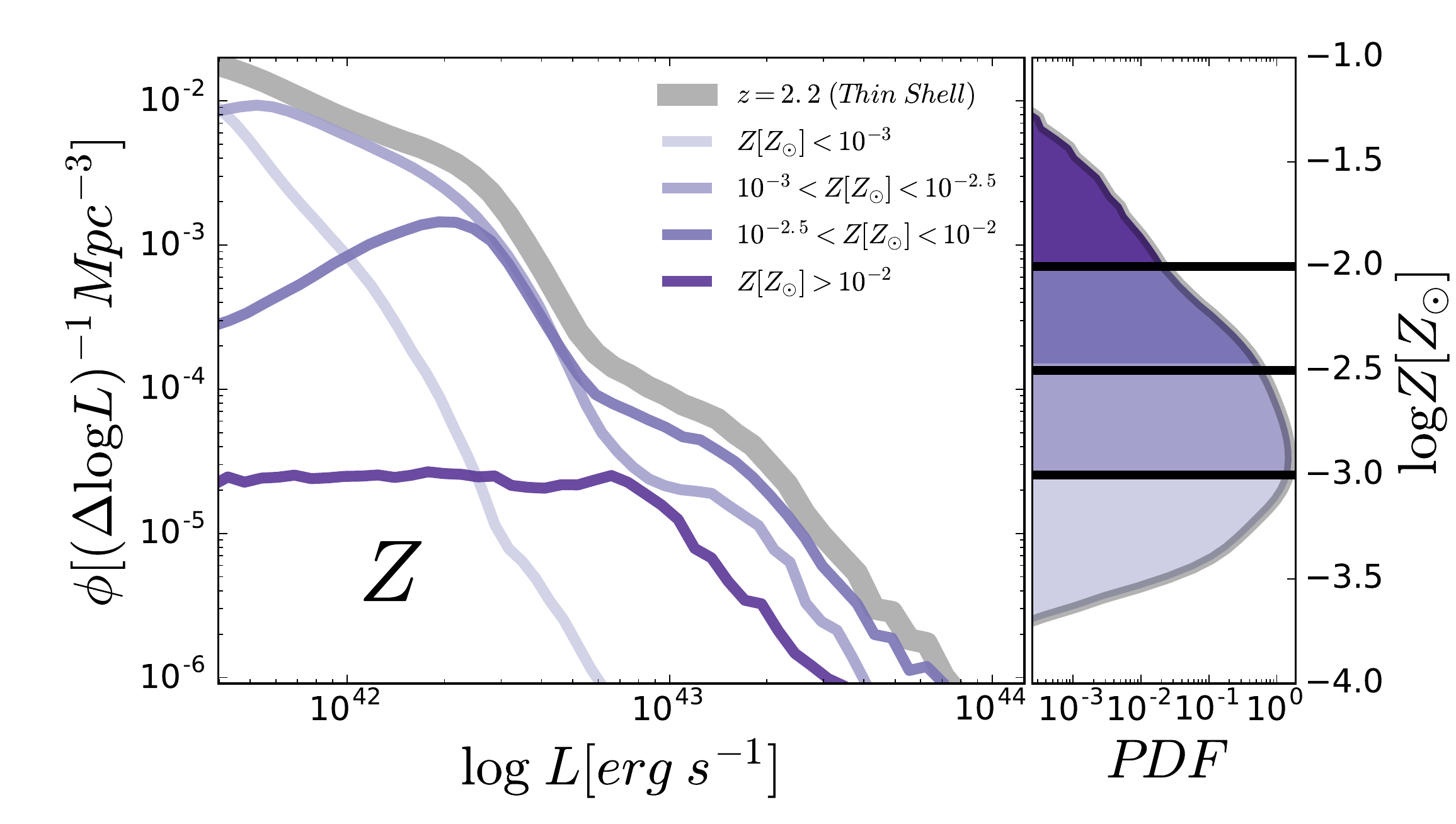}
        \caption{ {\bf Left panels} : Break down of the \lya\ LF or galaxies with \lya\ $\rm EW_0>20$\AA{} in bins of halo mass (upper left), stellar mass (upper right), star formation rate (lower left) and metallicity (lower right). The bins are indicated in the legends. In each quantity the bins are represented in lighter colors for low values and darker as they increase. The total LF is plotted in thick gray line. {\bf Right panels} : The probability distribution function of the different properties. In black we show the bin cuts.}\label{fig:line-profile-comparison-bicone}
\label{fig:LFbreakdown}
\end{figure*}

\indent To illustrate the properties of LAEs, Fig.\ref{fig:LFbreakdown} shows the \lya\ LF obtained with the {\it Thin Shell} geometry at $z=3.0$, split by the contribution of different ranges of halo and stellar mass, star formation rate and gas metallicity. We note that other redshifts and geometries show a similar behavior to what is shown in  Fig.~\ref{fig:LFbreakdown}. Here we are analyzing a subsample composed of every LAE ($\rm EW_{0} > 20$\AA{}) with \lya\ luminosity $> 10 ^{41.5}erg \; s^{-1} $.

\indent When splitting the LF based on the halo mass of LAEs (upper-left panel), we find that the majority of LAEs are hosted by haloes of moderate mass, $M_{\rm halo} \sim 10^{11-12} \munits$ which dominates the bright and moderate luminosities. LAEs with host halo masses below $M_{\rm halo} \lesssim 10^{11} \munits$ dominate the very faint end of the LF, with $\llya \approx 10^{41} \lunits$. Finally, the most massive haloes host galaxies do not contribute significantly to the LF shape. Furthermore, we have checked that there is no clear correlation between  halo mass and  \lya\ luminosity.

\indent In the upper right panel in Fig.~\ref{fig:LFbreakdown}  the LF is split according to the stellar mass of the emitting galaxy. The whole body of the LF is dominated by LAEs with stellar mass about $M_{\rm stellar} \sim 10^{8-10}\munits$. Moreover, galaxies with a very low ($M_{\rm stellar} < 10^{8} \munits$) or a very high ($M_{\rm stellar} > 10^{9} \munits$) stellar mass do not contribute to bright or the faint ends. As in the $M_{\rm h}$ case, we do not find any clear correlation between  stellar mass and  \lya\ luminosity.

\indent The star formation rate, as expected, contributes in a roughly monotonic way to the \lya\ LF. The faint-end of the \lya\ LF is dominated by galaxies with low $\log({\rm SFR}[M_{\odot}/h/yr]) \sim  -0.5$. Additionally, the intermediate luminosities are dominated by moderate ${\rm SFR} \sim 1-10[M_{\odot}/h/yr]$ while the bright end is populated by galaxies with the highest $\rm SFR$ (although with a significant scatter).  Note that this trend only means that the $\rm L_{Ly\alpha}$ of LAEs scales with SFR, but not that every galaxy with high SFR  would result in a LAE. Finally, we note that typically, galaxies with ${\rm SFR} < 0.1[M_{\odot}/h/yr]$ do not contribute to the LF.

\indent The break down of the \lya\ LF in terms of gas metallicity is less intuitive.  Naively one would expect to find an anti-correlation between metallicity and \lya\ luminosity, since  \fesc\ decreases with increasing dust, and thus, metallicity. However, we find the opposite: for LAEs with $\log(Z) < -2$, the low metallicity bins contribute to the lower luminosities and vice versa. This trend is broken for $\log(Z) > -2$ due to the low \fesc\ at this metallicity range. The  galaxies with highest $\rm Z$ do not contribute anymore to the bright end but to low and average luminosities . This leads to the bulk of the \lya\ emitter population being dominated by galaxies with average metallicities, spanning the range $-3 < \log(Z) < -2$. We dig deeper in this relation in \S \ref{ssec:bulk}

\subsection{The bulk properties of LAEs.}\label{ssec:bulk}
\begin{figure*} 
\includegraphics[width=6.93in]{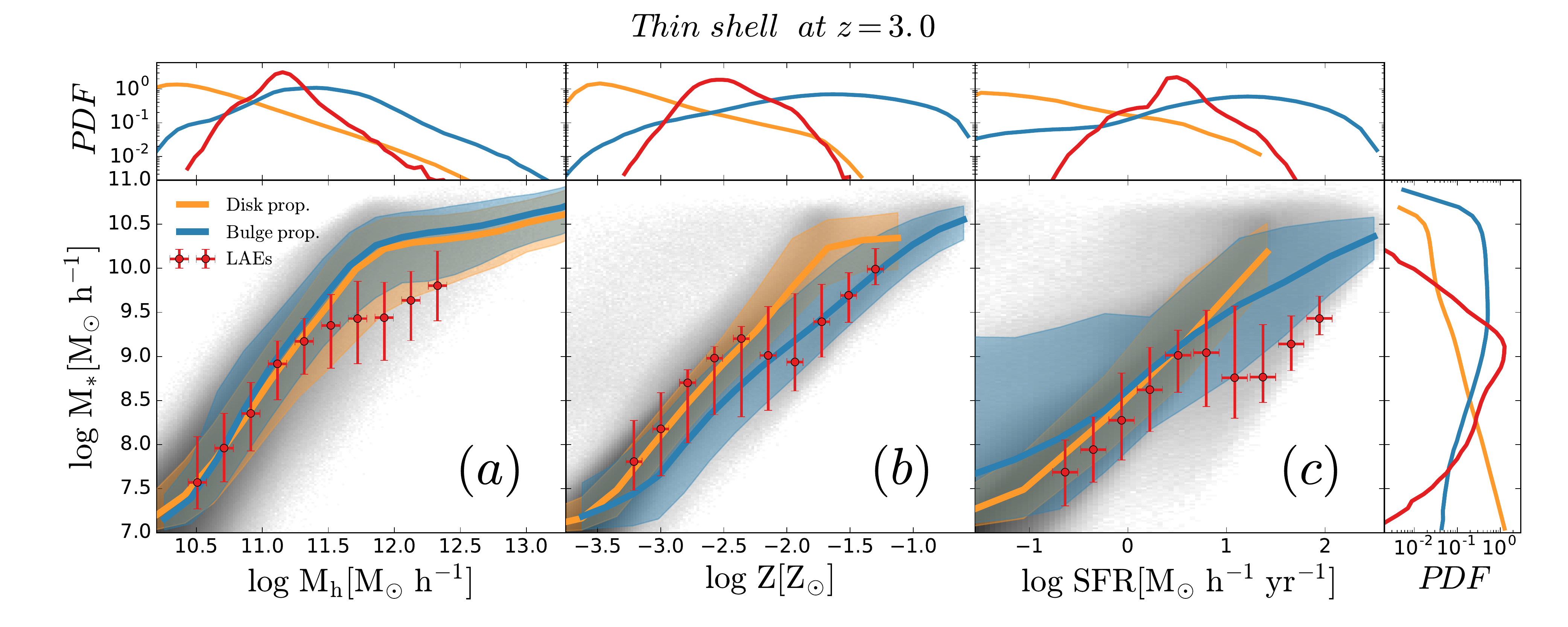} 
\caption{ {\bf a)} The stellar mass - halo mass distribution at $z=3.0.$ The gray shaded region shows the distribution for the full \galform\ sample. The solid yellow and blue lines and correspond to the median of \galform\ central galaxies disk and bulges properties respectively. The shade regions show the 10-90 percentiles. The red dots show the {\it Thin Shell} LAE sample median, 10-90 percentiles (vertical) and the bin size (horizontal). {\bf b)} Same as a) but for the stellar mass - metallicity distribution.  {\bf c)} Same as a) but for the stellar mass - star formation distribution. The top panels show the distributions of the halo mass, star formation and metallicity, respectively, for the full \galform\ (yellow and blue for disk and bulge dominated respectively) and the {\it Thin Shell} model (red). The stellar mass distribution is shown in the right vertical panel.}\label{fig:properties_Ms}
\end{figure*}

\begin{figure*}
\centering
\includegraphics[width=.49\linewidth]{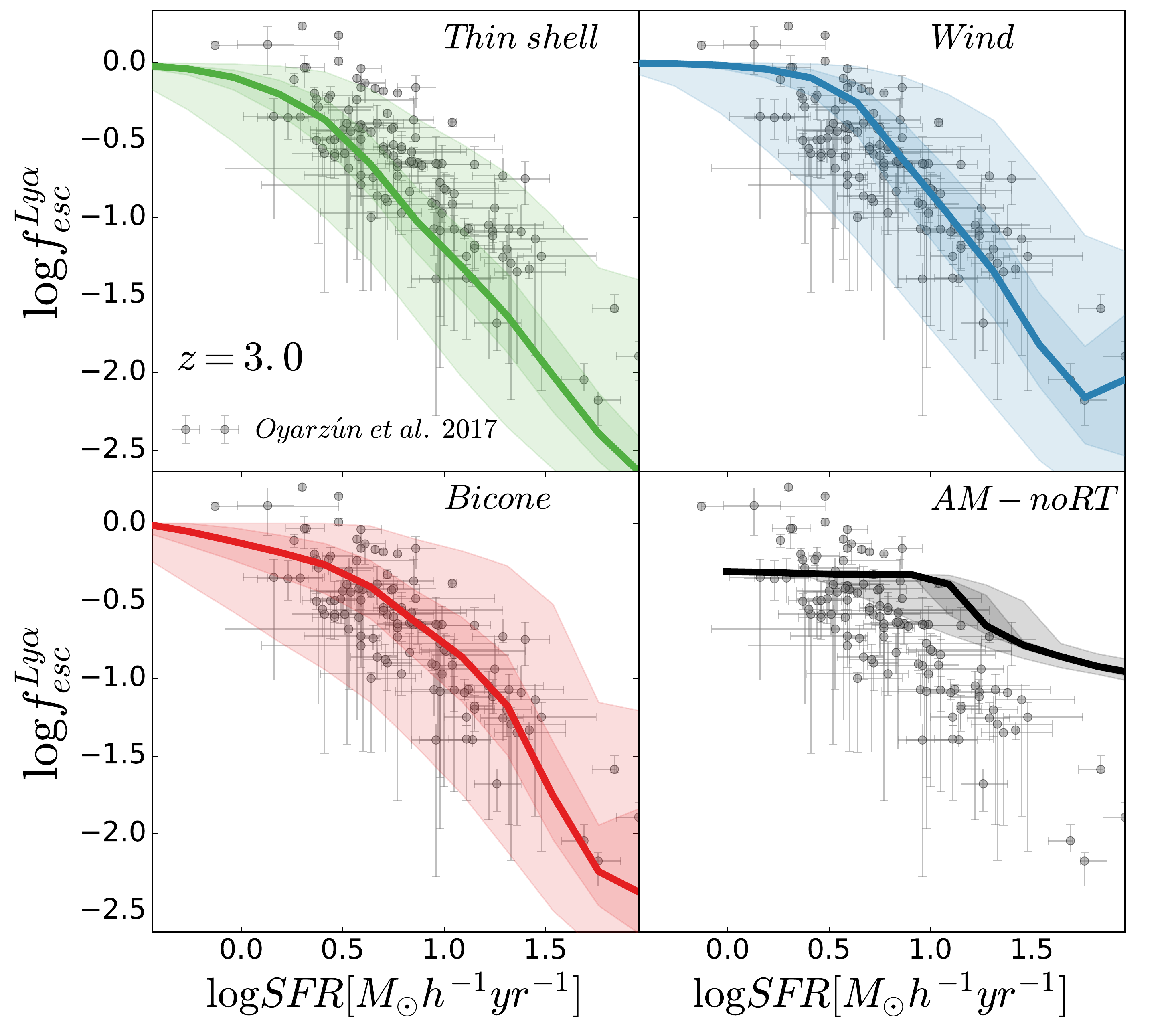}
\includegraphics[width=.49\linewidth]{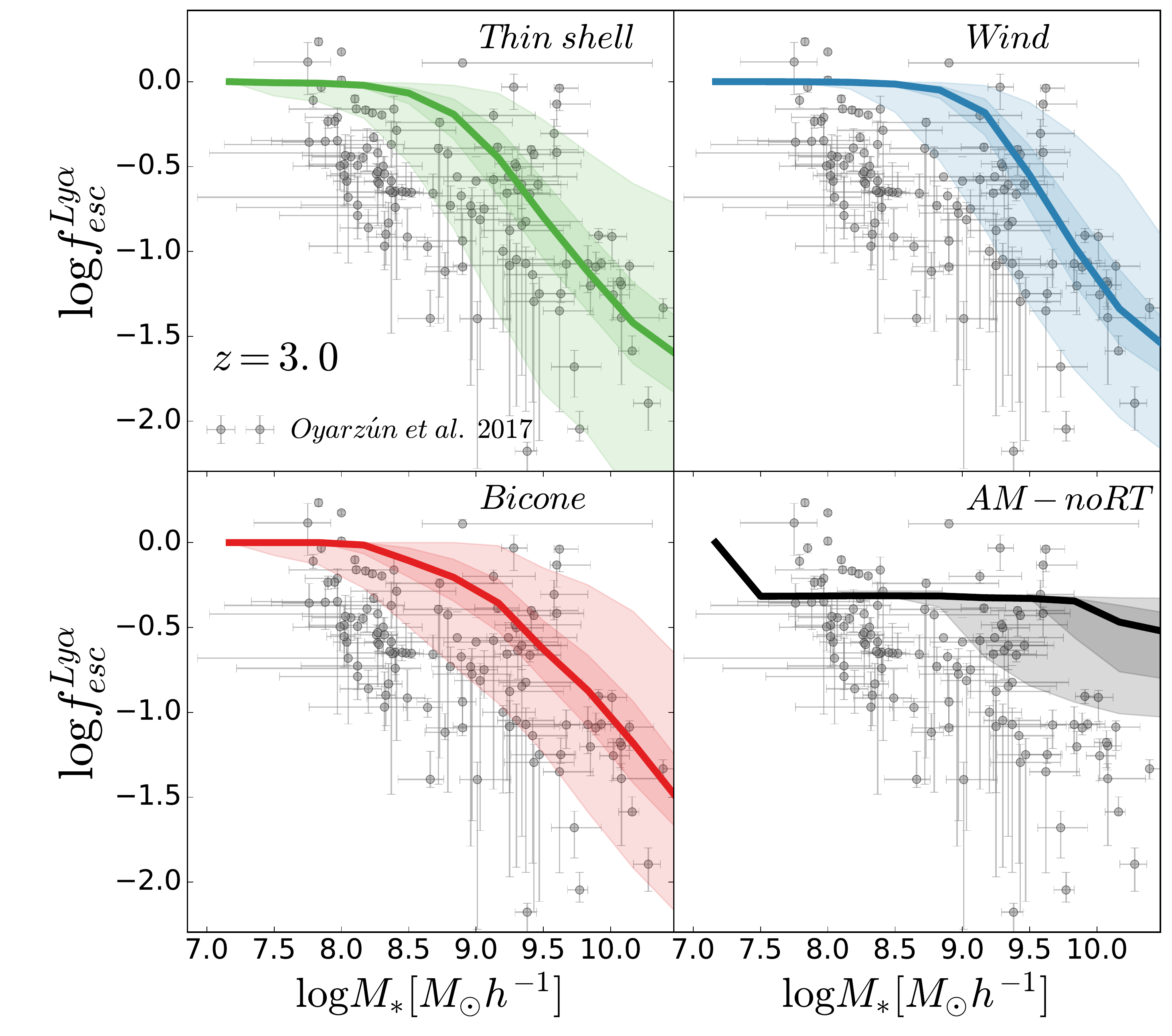}
\caption{The \lya\ \fesc\ as a function of SFR (left panels) and stellar mass (right panels) at $z=3$. 
Gray points are from \citet{oyarzun17}. Each panel displays our model predictions with a different
outflow geometry, as shown in the legend. The bottom-right corner displays the predictions of the model with no radiative transfer. The solid line in each panel is the median of  
\fesc\ predicted by our models. The dark and light coloured shaded regions display the $32-68$ and $5-95$ percentiles of the models predictions, respectively.}
\label{fig:fesc_relations}
\end{figure*}

\indent In this section we analyze the galaxy properties of our simulated LAE, focusing  on the results at redshift $z=3$ and for the {\it Thin Shell} geometry (we checked that different geometries and redshifts give similar results). We restrict our analysis only to central LAEs with a ${\rm 10^{-3}  cMpc^{-3}} h^3 $ number density cut in \lya\ luminosity (we check that different number density cuts produce similar results), and we compare it with the properties of the underlying population of central galaxies, i.e., the full population of galaxies predicted by \galform\ with $M_{\rm stellar} > 10^7 \munits$.  

Figure \ref{fig:properties_Ms} shows some physical properties of the LAEs (red dots) and for the general population of galaxies from \galform\ selected using the same number density cut as the LAEs (yellow for disk properties and blue for bulge properties). Each panel includes the distribution of halo mass $M_{h}$, star formation rate ${\rm SFR}$, metallicity $Z$ and stellar mass $M_{*}$ and the correlation between $M_{*}-M_{h}$, $ M_{*}- {\rm SFR}$ and $M_{*}-Z$. 

The $ M_{h}$ distribution in the LAE sample peaks at intermediate $M_{h}\sim 10^{11}M_{\odot}h^{-1}$ and spans between $10^{10.5}-10^{12}[M_{\odot}h^{-1}]$. LAEs halos trace the massive end of the disk-dominated $\rm M_{h}$ distribution while avoiding the most massive dark matter halos, even if they host the strongest starburst episodes. This is caused by the  ${\rm SFR} - Z$ predicted by \galform\ that associates high metallicites (low \fesc) to high $\rm SFR$.

The metallicity and the $\rm SFR$ of the LAE sample behave in a similar way due to the tight ${\rm SFR} - Z$ relation. The bulk of the LAE sample peaks at intermediate values of $Z$ and $\rm SFR$, avoiding the extremes of the full \galform\ distribution. In particular, the galaxies with the highest SFR are not selected as LAE as the metallicity is also too high, causing a lower $f_{\rm esc}^{Ly\alpha}$. Additionally, the galaxies with extreme low $Z$ are not selected either as their $ \rm SFR $ in too low in these galaxies.

The $ M_{*} - M_{h}$ relations (Fig. \ref{fig:properties_Ms}) for disk and bulge-dominated galaxies behaves in the same way. On the other hand, in the LAE sample this relation is the same as in the underlying galaxy population up to the peak of the $\rm M_{h}$ and $\rm M_{*}$ distributions, where the relation flattens for higher halo masses. In the high halo mass regime, LAEs typically have lower stellar masses than the overall average. This behavior is given by the tight $\rm SFR-Z$ relation causing $f_{\rm esc}^{Ly\alpha}$ to be lower for galaxies with higher $\rm M_{*}$ as they become more dust rich. 

In the LAE sample, the ${\rm SFR} - Z$ relation is consistent with the bulk of the disk-dominated galaxies for $Z<10^{-2.5}Z_{\odot}$. After a transition around $Z \sim 10^{-2.2}$, $Z_{\odot}$ is consistent with starburst galaxies. At metallicities below that transition the LAE $\rm SFR - Z$ relation is slightly above the overall relation.

In the LAE sample the $M_{*} - {\rm SFR}$ relation is below the full \galform\ relation. This implies that for a fixed stellar mass, galaxies with higher SFR are selected, as the intrinsic $L_{Ly\alpha}$ correlates directly with the SFR.

\subsection{The predicted \lya\ \fesc\ against observational estimates}

\begin{figure*} 
\includegraphics[width=7.0in]{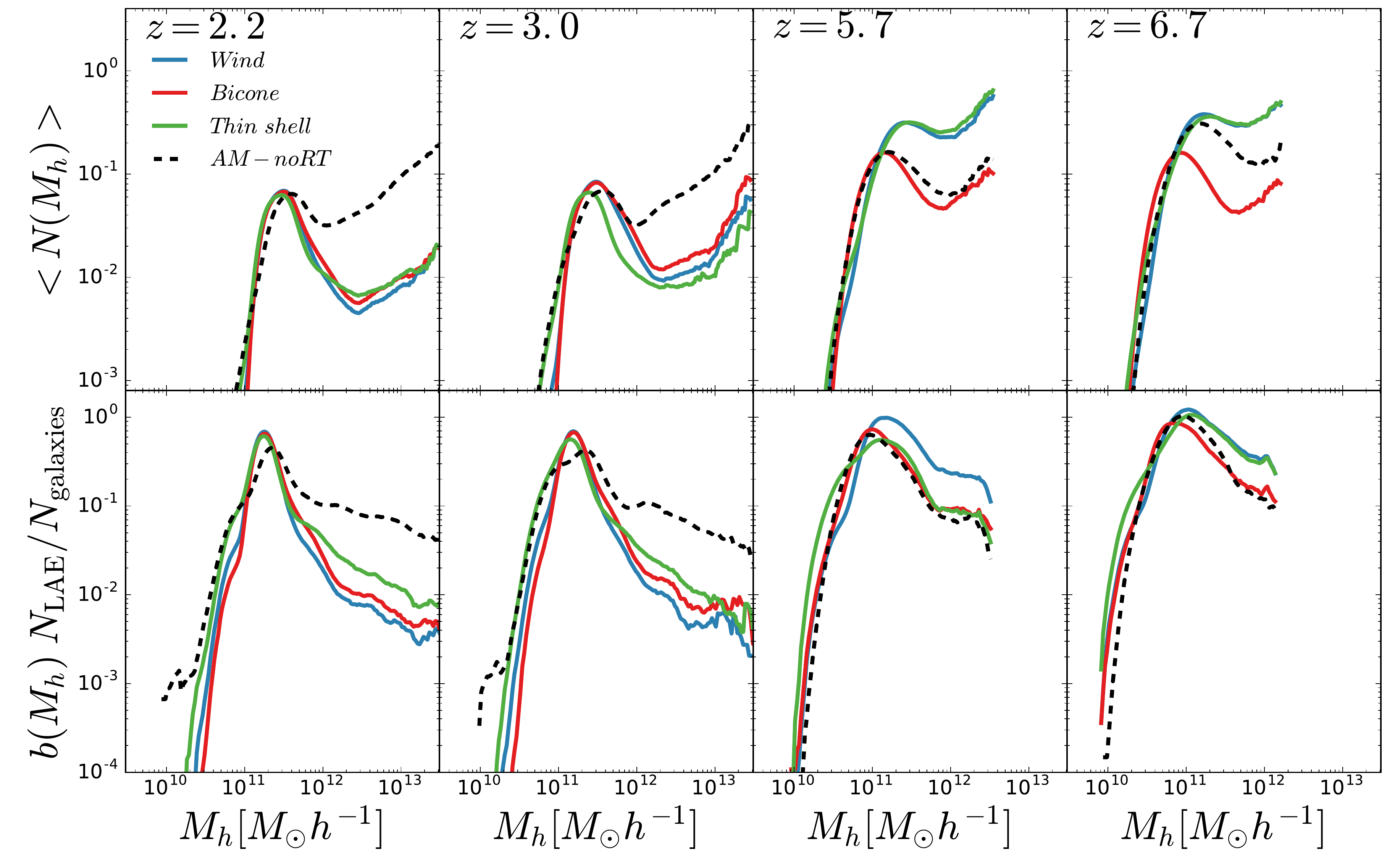} 
\caption{ {\bf Top}: the halo occupation distribution (HOD) at redshift 2.2 , 3.0 , 5.7 and 6.7 from left to right. Model with radiative transfer show as \colorWind , \colorBicone\ and \colorThin\ solid lines for the {\it Wind}, {\it Bicone} and {\it Thin Shell} geometry respectively. The LAE sample \NoRT\ is plotted as dashed black line. {\bf Bottom}: fraction of galaxies that are considered LAE times the bias of the hosting dark matter halo. This quantifies the contribution of the different $\rm M_{h}$ to the overall bias of the population. } \label{fig:halos_across_redshift_fit}
\end{figure*}

In this section we compare our model predictions for the \fesc\ against observational estimates from \citet{oyarzun17} at $z=3$. In order to mimic their sample selection function we select galaxies with $10^{7.6}M_{\odot} < M_{*} < 10^{10.6}M_{\odot}$, and $L_{\rm Ly\alpha} > 10^{41.5} \lunits$.
    
Fig.~\ref{fig:fesc_relations} shows the relation between the \lya\ \fesc\ and the SFR and stellar mass. The \fesc\ computed in \citet{oyarzun17} displays a noticeable anti-correlation between SFR and \fesc\ . In the models including RT galaxies with higher SFR have lower values of \fesc, in remarkable agreement with the observational estimates. The scatter in the observational data of \citet{oyarzun17} is consistent with the spread predicted by our models. This anti-correlation is caused by intrinsic link between SFR and $Z$. Even if the $V_{\rm exp}$ is higher for greater SFR (equation \ref{eq:recipe-thin-V}), dust plays the mayor role in the escape of \lya\ photons and reduces \fesc.  

The stellar mass is also anti-correlated with the \fesc, as shown in the right panel of Fig.~\ref{fig:fesc_relations}. This is due to the known correlation between $M_{*}$  and $Z$. Although our models reproduce the observationally inferred trend, the stellar masses predicted by \galform\ are systematically larger by $\sim 0.5 {\rm dex}$. 
  
Interestingly, the abundance matching model \NoRT\ does not display the same trends found in \citet{oyarzun17}, highlighting the importance of considering radiative transfer effects to predict LAE galaxy properties consistent with observational datasets.

\subsection{The dark matter haloes hosting LAEs }\label{sssec:Halos}

\begin{figure*} 
\includegraphics[width=7.0in]{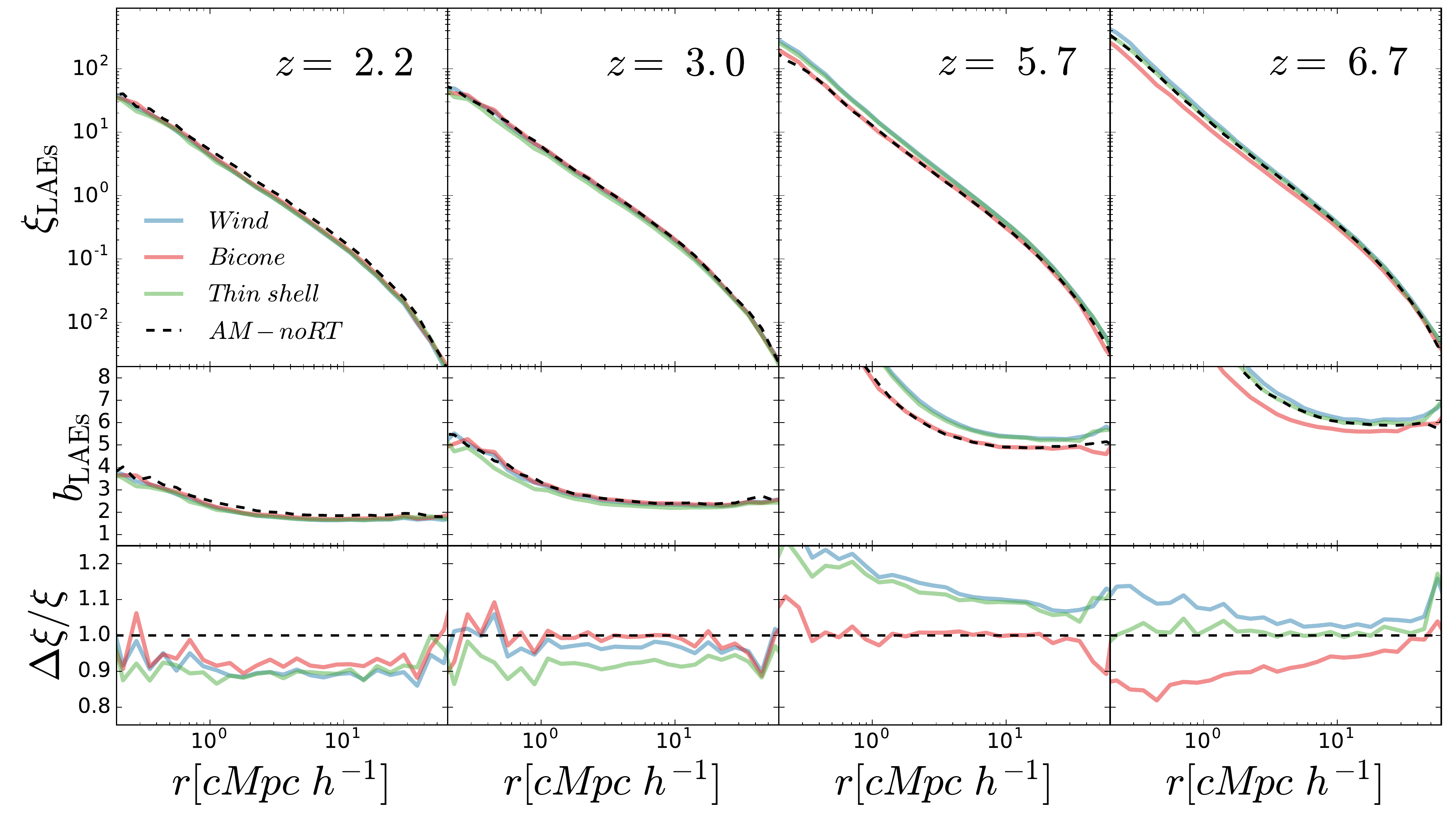} 
\caption{  {\bf Top panels} : Monopole (3D auto-correlation function) for the \NoRT\  sample (black), for the {\it Thin Shell} (\colorThin), galactic wind (\colorWind) and biconical galactic wind (\colorBicone) for redshift 2.2 , 3.0 , 5.7 and 6.7 from left to right. {\bf Middle panels} : The ratio between the different LAE sample and the dark matter correlation function. {\bf Bottom panels} : relative difference between the different samples and the \NoRT\ monopole correlation function.  }\label{fig:monopole_redshift}
\end{figure*}

In the following we study the properties of dark matter halos hosting LAEs. To compare different model predictions, we select the brightest LAEs with a number density cut of $\rm 10^{-3}h^{3}cMpc^{-3}$.  

Fig.~\ref{fig:halos_across_redshift_fit} shows the halo occupation distribution (HOD) at $z=2.2 , 3.0 , 5.7$  and $6.7$. This is constructed by computing the mean number of galaxies within different halo mass bins. 
All models including radiative transfer display a similar HOD at $z=2.2$ and $3.0$. Central galaxies have a peak abundance in haloes of mass $\rm M_{\rm halo} \approx 2\times 10^{11} M_{\odot} h^{-1}$. Satellite galaxies start dominating the abundance of haloes of mass $\rm M_{\rm halo} \gtrsim 10^{12} M_{\odot} h^{-1}$. None of the HODs at these redshifts reach $N(M_h) = 1$. Even at the peak of occupation, less than $10\%$ of haloes host a LAE, regardless of radiative transfer effects.

At $z\geq 5.7$ the HOD of the {\it Bicone} model falls significantly below that from the {\it Thin Shell} and {\it Wind} models. This reflects the differences in the LFs at these high redshifts. As the {\it Bicone} model is not able to reproduce the observed LF the resulting LAE population have quite different properties to the other RT samples.

The model with no radiative transfer systematically places LAEs in higher mass haloes compared to the radiative transfer models at low redshift. The occupation peak for centrals in the \NoRT\  model is  shifted to slightly more massive halos at $z=2.2$ and 3.0. Additionally, at these redshifts, the occupation of dark matter halos with $\rm M_{h} \geq 10^{12} M_{\odot} h^{-1}$ is much greater in the \NoRT\ model than in the models including RT. At redshifts $z=5.7$ and 6.7, the trend is inverted as LAE ({\it Thin Shell} and {\it Wind} geometry) populate halos slightly  more massive than the \NoRT\ model. Also, the occupation of halos with $\rm M_{h} \geq 10^{12} M_{\odot} h^{-1}$ is greater in the RT models.

The bottom panels of Fig.~\ref{fig:halos_across_redshift_fit} show the quantity
\begin{equation}
\label{eq:QQQ_definition}
b(M_h) N_{\rm LAE}(M_h) / N_{\rm galaxies}(M_h),
\end{equation}
where $N_{\rm LAE}(M_h)$ is the number of sources in our LAEs samples in a halo mass bin, $N_{\rm galaxies}(M_h)$ is the number of galaxies in the same $M_h$ bin and the galaxy bias $b(M_h)$ is defined as
\begin{equation}
\label{eq:bias_definition}
\xi_{\rm galaxy} = b^{2} \ \xi_{\rm dark \ matter} ,
\end{equation}
where $\xi_{\rm galaxy}$ and $\xi_{\rm dark \ matter}$ are the two point correlation functions for the galaxies and dark matter. This exhibits the contribution of different mass bins to the overall clustering bias of the LAE population. There is an evolution in the $\rm M_{h}$ that contributes to the bias, being greater at lower redshifts and lower at higher redshift. In particular, the peak values varies from $\rm M_{\rm halo} \approx 2\times 10^{11} M_{\odot} h^{-1}$ at $z=2.2$ to $\rm \approx 6\times 10^{10} M_{\odot} h^{-1}$ at $z=6.7$

At low redshift ($z=2.2$ and 3.0) the greater contribution to the bias come from lower mass halos in the RT models than in the \NoRT\ model. However, this trend is inverted at z=5.7. Additionally, at $z=6.7$ the main contribution to the bias comes from the same halo mass for all the models.\\

\subsection{ The clustering of LAEs. }\label{sssec:Clustreing}
        
In this section we study how \lya\ radiative transfer impacts the clustering of LAEs for each of the outflow geometries implemented. 
The sample used in this section is the same as the one used in \S \ref{sssec:Halos}.
        
In Fig. \ref{fig:monopole_redshift} the top panel shows the spherically-averaged 2-point auto-correlation function (2PCF) in real space at $z= 2.2 , 3.0 , 5.7$ and $6.7$. 
The middle panel shows the bias, defined as in Eq.\ref{eq:QQQ_definition}. Moreover, in order to highlight the differences in the RT samples and the AM-noRT we show in the bottom panel of Fig.\ref{fig:monopole_redshift} the relative difference of the 2PCF of the LAE samples $\xi_{\rm LAE}$ and the AM-noRT, i.e., $\Delta \xi / \xi = ( \xi_{\rm LAE} - \xi_{\rm AM-noRT}) / \xi_{\rm AM-noRT}$, where $\xi_{\rm AM-noRT}$ is the AM-noRT 2PCF. 

Overall, the clustering amplitude increases towards higher redshifts regardless of the LAE model variant. 
In detail, each model predicts a slightly different clustering bias. There is a strong scale-dependence of the clustering bias in all models and at all redshifts for separations below $r \lesssim 15-20 [{\rm Mpc/}h]$.

At $z=2.2$ and $3.0$ the clustering amplitude of the \NoRT\ sample is about 10\% above the one predicted by the RT models. This is a consequence of LAEs being hosted by higher mass 
dark matter halos for this model, as shown in previous sections.  At $z= 5.7$ and $6.7$, the clustering amplitude of the {\it Thin Shell} and {\it Wind} LAE samples are above that of the \NoRT\ and {\it Bicone} models. Interestingly, as shown in the bottom panels of Fig.~\ref{fig:monopole_redshift}, towards redshifts $z>3$ the \NoRT\ sample features a slightly different slope with respect to the RT models. 

In summary, the predicted clustering of LAEs at $z\lesssim 3$ is overall slightly lower when radiative transfer is included, and slightly higher towards $z\gtrsim 3$. 
The relative differences in the amplitude of clustering, with respect to the \NoRT\ model, are of the order of $10\%$. These differences result from the non-trivial relation between the \lya\ luminosity of galaxies and the dark matter halo population hosting these objects.

\subsection{ The clustering in mock catalogs of LAE surveys }\label{sssec:Mocks}

\begin{table*}
\centering
\caption{Mock catalog characteristics including the redshift $z$, the redshift width $\Delta z$, sky coverage (Area), the size along the line of size $\rm L_{\parallel}$, the distance perpendicular to the line of sight $\rm L_{\perp}$, the number of mocks sliced from the simulation box $\rm N_{mocks}$ and the median number of LAEs the mocks $\rm \langle N_{LAE} \rangle$ with the 32 and 68 percentiles. }
\label{tab:mock_properties}
\begin{tabular}{cccccccccccc}
Authors   &  $z$    & $\Delta z$ & $\rm Area$     & $\rm L_{\parallel}$ & $\rm L_{\perp}$ & $\rm N_{mocks}$ & \multicolumn{5}{c}{$\rm \langle N_{LAE} \rangle$} \\ \cline{8-12}
          &         &            & $\rm (deg ^2)$ & $\rm (cMpc)$        & $\rm (cMpc)$    &                 & Survey       & Thin shell      & Wind       & Bicone    & \NoRT       \\ \hline
\cite{Kusakabe2018} & 2.2 & 0.0773 & 0.93 & 104.9 & 93.6 & 448 & 1248 & $1196_{-90}^{+94}$ & $1191_{-80}^{+95}$ & $1189_{-78}^{+105}$ & $1183_{-91}^{+109}$ \\ \\
\cite{Bielby2016} & 3.0 & 0.0633 & 1.07 & 60.0 & 119.1 & 468 & 643 & $639_{-57}^{+48}$ & $639_{-51}^{+53}$ & $637_{-60}^{+52}$ & $631_{-59}^{+66}$ \\ \\
\cite{Ouchi2018a} & 5.7 & 0.0954 & 7.67 & 43.5 & 401.5 & 18 & 734 & $725_{-9}^{+15}$ & $731_{-19}^{+11}$ & $720_{-21}^{+19}$ & $719_{-20}^{+16}$ \\ \\
\cite{Ouchi2018a} & 6.7 & 0.1078 & 21.2 & 41.0 & 696.5 & 19 & 873 & $873_{-30}^{+6}$ & $865_{-19}^{+17}$ & $864_{-20}^{+21}$ & $866_{-6}^{+24}$
\end{tabular}
\end{table*}

\indent In this section we compare our clustering prediction against several measurements of the clustering of LAEs at different redshifts from \citet{Kusakabe2018} at $z=2.2$, \citet{Bielby2016} at $z= 3.0$ and \citet{ouchi10,Ouchi2018a} at $z=5.7$ and $6.7$, respectively. We build LAE mock catalogs mimicking the properties of the different surveys to allow a close comparison with the observational datasets. These surveys use narrow band photometry to detect LAEs over a restricted redshift range. The main difference in the mock catalogs comes from the specific area, flux depth and equivalent width limit ($\rm EW$) of the individual survey.

\begin{figure*} 
\includegraphics[width=7.0in]{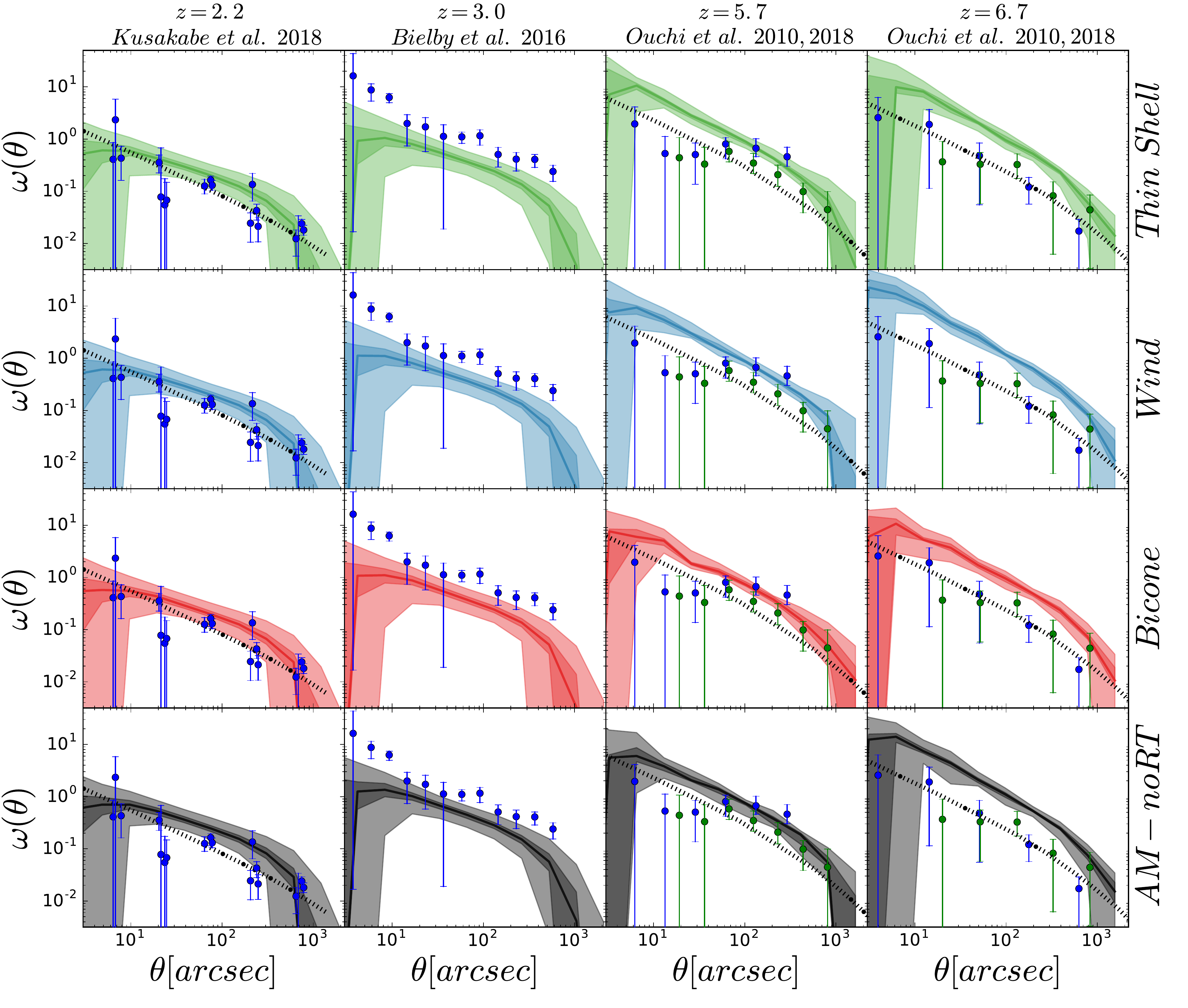} 
\caption{Comparison between different model mocks ({\it Thin Shell}, {\it Wind}, {\it Bicone} and \NoRT\ in rows from top to bottom) and the observed 2-point projected correlation function \citep{Kusakabe2018, Bielby2016, ouchi10, Ouchi2018a} at redshifts 2.2, 3.0, 5.7 and 6.7 in each column from left to right. The observational data is shown by dots and the best fitting power law $\omega(\theta)$ extracted from their original work are plotted as dashed black lines. The solid lines correspond to the median  $\omega(\theta)$ for the mocks and the darker and lighter shades to the 32-68 and 5-95 percentiles respectively.  }
\label{fig:mocks}
\end{figure*} 

To build the mock catalogues, we choose a direction as line of sight (LoS). Assuming a distant observer, a galaxy coordinate is transformed in redshift space using

\begin{equation}\label{eq:redshift_space}
s = x_{\rm LoS} +  \frac{ \rm v_{LoS}}{a(z)H(z)},
\end{equation} 
where $x_{\rm LoS}$ is the galaxy coordinate along the LoS, ${ \rm v_{LoS}}$ is the galaxy peculiar velocity along the LoS and $a(z)$ and $H(z)$ are the scale factor and the Hubble parameter, respectively, at the \lya\ pivot redshift, $z_{\rm pivot}$, of the NB filter. Additionally, we conserve the periodicity of the box along the LoS direction.

Although some surveys have complicated footprints due to multiple pointings, our mocks are constructed as squares comprising an area equal to that of the target survey. Thus, the simulation box is simply split in slices along the LoS. The size perpendicular to the LoS is computed as 

\begin{equation}\label{eq:L_perpendicular}
L_{\perp} = \sqrt{ \rm A_{survey} },
\end{equation} 
where $\rm A_{survey}$ is the survey sky coverage. The thickness (along the LoS) of the slice is computed as

\begin{equation}\label{eq:L_parallel}
L_{\parallel} = D_{\rm co}(z=z_{\rm +}) - D_{\rm co}(z=z_{\rm -}),
\end{equation} 
where $D_{\rm co}(z)$ is the comoving distance at the geometric redshift $z$. Additionally, 

\begin{equation}\label{eq:redshift_filter}
z_{\pm} = {{ \lambda_p \pm 0.5 \ {\rm FWHM}}  \over  {\lambda _{Ly\alpha}}}  - 1,
\end{equation} 
where $\lambda_p$ and ${\rm FWHM}$ are the pivot wavelength and the full width half maximum of the narrow band filter and ${\lambda _{Ly\alpha}}$ is the Lyman $\alpha$ wavelength.


We calculate the limiting luminosity $L_{\rm cut}$ and the minimum rest frame equivalent width $\rm EW_{0,cut}$ for each survey by matching the LAE number density, $n_{LAE}$ of the surveys to the one in the whole simulation box (see Appendix \ref{Ap:B}). Then, our mock catalogs consist of galaxies with luminosity above $L_{cut}$ and $\rm EW_0$ above $\rm EW_{0,\rm cut}$. Table \ref{tab:mock_properties} lists the properties of the mocks, including the parallel and transverse sizes along the LoS, the redshift window $\Delta z = z_{+} - z_{-}$, the number of mocks, $\rm N_{mock}$, sliced from the simulation box and the number of LAE in each survey, and the median with 32-68 percentiles of the number of LAEs in the mocks.

The value of $L_{\parallel}$ for narrow-band surveys is typically very small compared to the box length of the simulation. This allows for a big fragmentation of the simulation box along the LoS. On the other hand, $L_{\perp}$ can vary significantly between surveys. While, at low redshift ($z=2.2,3.0$) $L_{\perp}$ is relatively small and allows a large number of mock surveys, at $z=5.7,6.7$ only one cut is possible due to the large size required for the mock surveys. As a result of this, the number of mocks at $z=2.2,3.0$ (448 and 468 respectively) is much larger than that at $z=5.7,6.7$ (18 and 19 respectively).

Since $n_{LAE}$ in the simulation box is set to match the observed $n_{\rm LAE}$ of each survey (see Appendix \ref{Ap:B}), the observed number of LAE and the median number of LAE in our mocks, $\langle N_{\rm LAE} \rangle$, are compatible within 1 sigma. Additionally, the dispersion of $\langle N_{\rm LAE} \rangle$ is higher (lower) at $z=2.2$ and 3.0 (5.7 and 6.7), since the comoving volume is smaller (larger). Hence, the impact of cosmic variance on clustering measurements is stronger (weaker).

We construct mock catalogs of LAE surveys from \citet{Kusakabe2018} at $z\approx 2.2$, \citet{Bielby2016} at $z \approx 3$, \citet{ouchi10} at $z\approx 5.7$ and \citet{Ouchi2018a} at $z\approx 6.7$. Figure \ref{fig:mocks} shows the comparison between the observed angular 2-point correlation function of these surveys, $\omega_{\rm survey}$, and that computed from the mock catalogues, $\omega_{\rm mock}$.

Overall, $\omega_{\rm mock}$ is very similar among our different model variations, including the \NoRT\ model. The  differences in the clustering due to the different bias of the samples are small in comparison with the scatter due to cosmic variance, making all models indistinguishable from each other.

At redshift 2.2 there is a good agreement between the the mocks and the clustering measurements in \citet{Kusakabe2018}. At $z=3.0$ $\omega_{\rm mock}$ is significantly below the  $\omega_{\rm survey}$. However, the slope of the different samples are very similar to observations. At higher redshifts the  LAE clustering predicted by the mocks is overestimated in our models. In particular, at $z=5.7$, for angular distances $\theta < 50 \; {\rm arcsec}$, $\omega_{\rm mock}$ overestimates the clustering, while at larger $\theta$ the mocks match very well $\omega_{\rm survey}$. Additionally, at redshift 6.7 the $\omega_{\rm mock}$ bias is significantly (about 2-sigma) overestimated in comparison with $\omega_{\rm survey}$. This discrepancy could be caused by multiple reasons. The moderate contamination of interlopers ($\sim 10\%$) in the \citet{ouchi10} sample could decrease the measured clustering amplitude. Also, the observed LAE population at this redshift might contain a significant contribution of objects at the mass resolution limit of our simulation ($M_{\rm halo, min} \approx 3\times 10^9 [\rm M_{\odot}/h]$), thus making our predictions biased towards higher masses and clustering amplitudes.


\section{ Discussion. }\label{}
Here we discuss some of the results found in previous sections. In particular, in subsection \ref{ssec:LAE_differences} we discuss how the different outflow geometries impact the predicted properties of the LAE populations. Then, in subsections \ref{ssec:No_RT_limitations} and \ref{ssec:RT_limitations} we discuss the limitations of our methodology. 

\begin{table}
\centering
\caption{Fraction of shared galaxies between pairs of models at the same redshift.}
\label{tab:shared_galaxies}
\begin{tabular}{cccccc}
\multicolumn{1}{c}{$z$} & \multicolumn{1}{c}{Model}           & \multicolumn{1}{c}{Thin Shell} & \multicolumn{1}{c}{Wind} & \multicolumn{1}{c}{Bicone} & \multicolumn{1}{c}{\NoRT} \\ \hline
\multicolumn{1}{c}{ 2.2 }   & \multicolumn{1}{c}{Thin Shell} & \multicolumn{1}{c}{ 1.000 }           & \multicolumn{1}{c}{ 0.814 }     & \multicolumn{1}{c}{ 0.555 }       & \multicolumn{1}{c}{ 0.229 }      \\
                             & Wind                           &                     0.814             &                     1.000       &                     0.592         &                     0.189        \\
                             & Bicone                         &                     0.555             &                     0.592       &                     1.000         &                     0.197        \\
                             & \NoRT                          &                     0.229             &                     0.189       &                     0.197         &                     1.000        \\ \hline
\multicolumn{1}{c}{ 3.0 }   & \multicolumn{1}{c}{Thin Shell} & \multicolumn{1}{c}{ 1.000 }           & \multicolumn{1}{c}{ 0.805 }     & \multicolumn{1}{c}{ 0.401 }       & \multicolumn{1}{c}{ 0.188 }      \\
                             & Wind                           &                     0.805             &                     1.000       &                     0.427         &                     0.151        \\
                             & Bicone                         &                     0.401             &                     0.427       &                     1.000         &                     0.227        \\
                             & \NoRT                          &                     0.188             &                     0.151       &                     0.227         &                     1.000        \\ \hline
\multicolumn{1}{c}{ 5.7 }   & \multicolumn{1}{c}{Thin Shell} & \multicolumn{1}{c}{ 1.000 }           & \multicolumn{1}{c}{ 0.413 }     & \multicolumn{1}{c}{ 0.798 }       & \multicolumn{1}{c}{ 0.108 }      \\
                             & Wind                           &                     0.413             &                     1.000       &                     0.322         &                     0.063        \\
                             & Bicone                         &                     0.798             &                     0.322       &                     1.000         &                     0.160        \\
                             & \NoRT                          &                     0.108             &                     0.063       &                     0.160         &                     1.000        \\ \hline
\multicolumn{1}{c}{ 6.7 }   & \multicolumn{1}{c}{Thin Shell} & \multicolumn{1}{c}{ 1.000 }           & \multicolumn{1}{c}{ 0.354 }     & \multicolumn{1}{c}{ 0.663 }       & \multicolumn{1}{c}{ 0.104 }      \\
                             & Wind                           &                     0.354             &                     1.000       &                     0.229         &                     0.076        \\
                             & Bicone                         &                     0.663             &                     0.229       &                     1.000         &                     0.259        \\
                             & \NoRT                          &                     0.104             &                     0.076       &                     0.259         &                     1.000        \\ \hline
\end{tabular}
\end{table}

\subsection{Differences between the RT models.}\label{ssec:LAE_differences}

In this work we have used three different gas outflow geometries ({\it Thin Shell}, spherical galactic wind and biconical galactic wind) to model the \lya\ radiative transfer inside galaxies. The galaxy properties predicted for LAEs are very similar. The only significant difference between the predictions of different geometries is on the required distributions of column density and expansion velocity. 

In Table \ref{tab:shared_galaxies} we list the fraction of galaxies shared by pairs of LAE models imposing $\rm EW_{0}>20$\AA{} and a number density cut of $\rm 10^{-3} h^{3} cMpc^{-3}$ in $\rm L_{Ly\alpha}$. We find that the {\it Wind} and {\it Thin Shell} geometries share a high fraction of galaxies ($\sim80\%$) at redshifts 2.2 and 3.0. However, at high redshift these geometries select different galaxies as the shared fraction in relatively low ($\sim40\%$ overlap). This might be due to the fact that there is a necessity of $f_{\rm esc}^{\rm Ly\alpha} \sim 1$ and the recipes to compute $\rm N_{H_{I}}$ and $\rm V_{exp}$ are different. However, quite the opposite relation is seen between the {\it Thin Shell} and {\it Bicone}, as at low redshift they share a relatively low percentage of galaxies ($\sim 45\%$) and this increase at higher redshifts ($\sim 70\%$). 

Finally, when comparing the galaxies in the {\it Wind} and {\it Bicone} geometry  we surprisingly  find a low overlap between them. In particular, the maximum overlap happens at $z=2.2$ ($\sim 55\%$) and it drops down to only  $\sim 20\%$ at $z=6.7$. This shows the impact of the gas geometry on how the RT shapes the LAE selection function; even though the intrinsic galaxy population and the recipes to derive  $\rm N_{H}$ and $\rm V_{exp}$ are the same, the two geometries predicts different populations (although with similar characteristics). 

We conclude that the RT LAE samples, in general, share a big fraction of galaxies ($\geq 50\%$) although the implemented gas geometries are very different.  This is due to the fact that \fesc\ behaves similarly for all of them. In particular, even if the exact dependence is different for each geometry, decreasing $\rm N_{H}$, increasing $\rm V_{exp}$ and decreasing $\tau_{a}$ increase \fesc\, thus the visibility of the object for all of them. This makes the RT LAE samples  very similar, as galaxies with properties that maximize $\rm L_{Ly\alpha}$ and \fesc\ are selected.

\subsection{Limitations of the simple \NoRT\ model.}\label{ssec:No_RT_limitations}

\indent We have also used a very simplistic LAE model where  were radiative transfer effects are not taken into account and $\rm L_{Ly\alpha}$ depends monotonically on the SFR. In Table \ref{tab:shared_galaxies} we also list the overlap between the radiative transfer and \NoRT\ LAE sample. We find that the fraction of  galaxies shared between the \NoRT\ and RT catalogs is low, reaching its maximum value at z=2.2 ($\sim 20\%$) and then decreasing to $\sim 7\%$ at redshift 6.7. 

\indent As shown in Fig. \ref{fig:fesc_relations} the \NoRT\ sample  fails to match not only the observed \fesc $\rm -SFR$ and \fesc $-M_{*}$ relations but also the overall trend where \fesc anti-correlates with these two properties due to the RT (as described above). Additionally, the dark matter halo population, and thus the clustering, is different in comparison with the RT samples. 

\indent This work highlights the importance of taking into account the \lya\ RT inside galaxies when modeling LAEs. In particular, unlike in RT LAE samples, the galaxy properties model \NoRT\ differ from observations, making them less attractive to study galaxy formation and evolution. 

\subsection{Limitations of the RT models.}\label{ssec:RT_limitations}


The IGM also plays a mayor role in the detectability of galaxies based on \lya\ flux \citep{dijkstra06, zheng11, Behrens2017}. The IGM opacity becomes more important at higher redshifts ($\sim 7$) where the universe is denser and colder. However, the IGM might already also have an impact on the LAE selection function at $z=2.2$ as, even if the universe is highly ionized, the cross-section of neutral hydrogen atoms for scattering \lya\ photons is very high. The IGM impact might alleviate some of the tension that we find when we compare LAE models with observations. We will implement the effect of the IGM opacity in future work.

In Fig. \ref{fig:fesc_relations} we found that, although the observed \fesc $\rm -SFR$ relation is perfectly reproduced by our RT models, the \fesc $-\rm M_{*}$ relation is not. Even if the overall trend is similar, we find a significant difference (about 0.5 dex) in the stellar mass. This is probably not caused by our implementation of RT in a semi-analytic model, but by \galform\ itself, as we note that full \galform\ $\rm M_{h}-M_{*}$  relation at redshift 3.0 is overestimated (also about 0.5 dex) in comparison with the observed one \citep{Behroozi_2010}. Another possible source for this discrepancy is the different stellar population synthesis models used by \cite{oyarzun17} and \galform .

Another limitation of the RT models is that they predict very similar galaxy properties for the three different geometries. This degeneracy makes it difficult to determine from observations which geometry is the one driving the \lya\ photons escape. Nonetheless, the three gas geometries used in this work have very different \lya\ line profiles (as shown in figure \ref{fig:fesc_comparison}) which might break the degeneracies and lead to a better understanding of the escape channels of \lya\ radiation. We will implement line profiles in a upcoming work.


\section{ Conclusions and future work. }

\indent Lyman-$\alpha$ emitters are a promising galaxy population to trace the large scale structure of the Universe at high redshifts, $z\gtrsim 2$. One of the main advantages of LAEs is their high luminosity at the \lya\ rest frame wavelength, making them easy to detect. Additionally, due to the Hubble expansion \citep{hubble29}, the \lya\ line is observable in the optical from $z\sim2$ to $\sim7$, allowing  ground-based measurement of these galaxies. However, their selection function is quite complex as it depends upon \lya\ radiative transfer, which is sensitive to local astrophysical conditions. 

\indent We have designed a  theoretical model of LAE based on a Monte Carlo Radiative Transfer code that can be applied to huge cosmological volumes. In particular, we have applied our model the N-body only-dark-matter simulation \pmill\ and the semi-analytical model of galaxy formation and evolution \galform\ \citep{lacey16}. 

\indent Monte Carlo Radiative Transfer codes have demonstrated to be a powerful tool to understand how \lya\ photons escape from galaxies. Unfortunately, the high computational cost prohibits the capability of being directly run over cosmological volumes. In order to avoid this problem we have developed analytical expressions for the \lya\ escape fraction \fesc\ that are quite accurate for a wide range of outflow expansion velocities $V_{\rm exp}$, neutral hydrogen column densities $N_H$ and metallicities $Z$. 

\indent Our methodology computes \fesc\ for each galaxy as a function of $Z$, $V_{\rm exp}$ and $N_{\rm H_{\rm I}}$, which characterise the gas outflows from which \lya\ photons escape. We compute these quantities using galaxy properties such as the size, SFR or halo mass. Free parameters to compute these quantities are chosen to fit the observed luminosity function over a wide range of redshifts. After calibration we find that every geometry reproduces well the observed LAEs LF at low redshift while only the {\it Thin Shell} and {\it Wind} manage to match them at high redshift. We conclude that our {\it Bicone} geometry (as described in this work), at high redshift, is less favoured with respect to the others.

\indent We have analysed the relative abundance of \lya\ emitters by breaking down their LF in terms of several properties. Halo or stellar masses are not significantly correlated with \lya\ luminosities. The LF is actually mostly dominated by relatively low mass galaxies. However, when the LF is split in SFR bins we find a clear positive correlation with \lya\ luminosity. Finally, when the LF is divided into metallicity bins we find a scattered correlation for $\log(Z)<-2$. Moreover, the contribution of high metallicities ($\log(Z)>-2$) to the bright end of the LF is small.

\indent We also compared the properties of a \lya\ selected sample to the bulk of the galaxy population at high redshifts. 
We find that LAEs lie in relatively low mass halos. Additionally, the galaxies with the strongest starburst episodes are not selected as LAE since these galaxies typically have higher metallicities, and thus their \fesc\ is low.

\indent To validate our predicted \fesc, We have compared our LAE samples to the observational data from \cite{oyarzun17}. We find a remarkable good agreement between our predictions and the observationally measured \fesc\ - SFR relation. The LAE samples including RT reproduce successfully this anti-correlation and the scatter found between these quantities. 
However, the predicted \fesc\ - $M_{*}$ plane is offset by $\sim 0.5 \; {\rm dex}$ in $M_{*}$ with respect to the data from \citet{oyarzun17}. This difference can be due to the different assumptions about the stellar population synthesis models used by \cite{oyarzun17} and \galform, the impact of a different IMF in \galform, or simply that \galform\ predicts significantly more massive star-forming galaxies at these higher redshifts with respect to observational estimates. Finally, we find that our LAE \NoRT\ sample based on assuming a monotonic relation between SFR and $L_{Ly \alpha}$ is not able to reproduce any of the observed trends. This highlights the crucial role of RT in shaping the LAE selection function.

\indent We have also studied the dark matter halo population hosting LAEs in our models. We find differences between the samples including RT and the sample without RT. At low redshift, in comparison with the \NoRT , the RT models predicts lower mass dark matter halos host LAE. This trend reverses at high redshift, as LAEs lie in more massive halos in the RT samples. We also find that the satellite fraction is low at all redshifts ($\sim 2\%$) and similar for all of the model variants.

\indent The difference in the DM halo populations is directly translated into clustering discrepancies between the \NoRT\ and RT samples. At low redshift, as a consequence of LAEs modeled with RT lying in lower mass DM halos, we find that they have a lower galaxy bias than the \NoRT\ sample. This trend is reversed at high redshifts, when RT LAEs lie in more massive dark matter halos. Thus, we find that the RT models have a steeper galaxy bias evolution than the model excluding RT.

\indent Finally, we have compared our model clustering predictions with observations finding some tension. While at redshifts 2.2 and 5.7 the observed clustering is well reproduced, at redshifts 3.0 and 6.7 the galaxy bias is poorly constrained. As studied in previous works \citep{zheng11} the IGM transmission could have an impact on \lya\ selected samples that might alleviate this tension. 

\indent We have demonstrated the importance of RT in shaping the selection function of LAEs for galaxy properties as metallicity, SFR or DM halo properties. On one hand, the peculiar observational trends found can not be reproduce with a simple monotonic relation between SFR and $L_{Ly\alpha}$. On the other hand, the inclusion of RT changes in a very particular way the clustering of \lya\ selected samples. All this make extremely important to construct models with \lya\ RT in order to understand the galaxy properties, formation and evolution of LAEs. Moreover, future surveys tracing the large scale structure of the Universe through LAEs will require a deep understanding of the channels through which \lya\ photons escape in order to  obtain unbiased cosmological constrains.

\indent In future work we plan to implement the transmission of \lya\ photons through the IGM, which is especially important at high redshifts. In order to do so we will develop analytic expression for the \lya\ line profile and a model to compute the IGM transmission in large cosmological volumes. These tools will enable us to explore how the IGM shapes the LAE galaxy properties and clustering.

\section*{Acknowledgements}

The authors acknowledge the useful discussions with Zheng Zheng, Mark Dijkstra, Anne Verhamme in addition to the whole CEFCA team. We also thank Grecco Oyarzun for making their observational data available to us. The authors also acknowledge the support of the Spanish Ministerio de Economiaa y Competividad project No. AYA2015-66211-C2-P-2. We acknowledge also STFC Consolidated Grants ST/L00075X/1 and ST/P000451/1 at Durham University. This work used the DiRAC Data Centric system at Durham University, operated by the Institute for Com- putational Cosmology on behalf of the STFC DiRAC HPC Fa- cility (www.dirac.ac.uk). This equipment was funded by BIS Na- tional E-infrastructure capital grant ST/K00042X/1, STFC capital grants ST/H008519/1 and ST/K00087X/1, STFC DiRAC Opera- tions grant ST/K003267/1 and Durham University. DiRAC is part of the National E-Infrastructure.





\bibliographystyle{mnras}
\bibliography{ref} 

\begin{thebibliography}{}
\makeatletter
\relax
\def\mn@urlcharsother{\let\do\@makeother \do\$\do\&\do\#\do\^\do\_\do\%\do\~}
\def\mn@doi{\begingroup\mn@urlcharsother \@ifnextchar [ {\mn@doi@}
  {\mn@doi@[]}}
\def\mn@doi@[#1]#2{\def\@tempa{#1}\ifx\@tempa\@empty \href
  {http://dx.doi.org/#2} {doi:#2}\else \href {http://dx.doi.org/#2} {#1}\fi
  \endgroup}
\def\mn@eprint#1#2{\mn@eprint@#1:#2::\@nil}
\def\mn@eprint@arXiv#1{\href {http://arxiv.org/abs/#1} {{\tt arXiv:#1}}}
\def\mn@eprint@dblp#1{\href {http://dblp.uni-trier.de/rec/bibtex/#1.xml}
  {dblp:#1}}
\def\mn@eprint@#1:#2:#3:#4\@nil{\def\@tempa {#1}\def\@tempb {#2}\def\@tempc
  {#3}\ifx \@tempc \@empty \let \@tempc \@tempb \let \@tempb \@tempa \fi \ifx
  \@tempb \@empty \def\@tempb {arXiv}\fi \@ifundefined
  {mn@eprint@\@tempb}{\@tempb:\@tempc}{\expandafter \expandafter \csname
  mn@eprint@\@tempb\endcsname \expandafter{\@tempc}}}

\bibitem[\protect\citeauthoryear{{Ahn}}{{Ahn}}{2003}]{ahn03}
{Ahn} S.,  2003, Journal of Korean Astronomical Society, \href
  {http://adsabs.harvard.edu/abs/2003JKAS...36..145A} {36, 145}

\bibitem[\protect\citeauthoryear{{Ahn}}{{Ahn}}{2004}]{ahn04}
{Ahn} S.,  2004, \mn@doi [\apjl] {10.1086/381750}, \href
  {http://adsabs.harvard.edu/abs/2004ApJ...601L..25A} {601, L25}

\bibitem[\protect\citeauthoryear{{Ahn}, {Lee}  \& {Lee}}{{Ahn}
  et~al.}{2000}]{ahn00}
{Ahn} S.-H.,  {Lee} H.-W.,   {Lee} H.~M.,  2000, Journal of Korean Astronomical
  Society, \href {http://adsabs.harvard.edu/abs/2000JKAS...33...29A} {33, 29}

\bibitem[\protect\citeauthoryear{{Barnes} \& {Haehnelt}}{{Barnes} \&
  {Haehnelt}}{2010}]{barnes10}
{Barnes} L.~A.,  {Haehnelt} M.~G.,  2010, \mn@doi [\mnras]
  {10.1111/j.1365-2966.2009.16172.x}, \href
  {http://adsabs.harvard.edu/abs/2010MNRAS.403..870B} {403, 870}

\bibitem[\protect\citeauthoryear{{Baugh}}{{Baugh}}{2006}]{baugh06}
{Baugh} C.~M.,  2006, \mn@doi [Reports on Progress in Physics]
  {10.1088/0034-4885/69/12/R02}, \href
  {http://adsabs.harvard.edu/abs/2006RPPh...69.3101B} {69, 3101}

\bibitem[\protect\citeauthoryear{{Baugh}, {Lacey}, {Frenk}, {Granato}, {Silva},
  {Bressan}, {Benson}  \& {Cole}}{{Baugh} et~al.}{2005}]{baugh05}
{Baugh} C.~M.,  {Lacey} C.~G.,  {Frenk} C.~S.,  {Granato} G.~L.,  {Silva} L.,
  {Bressan} A.,  {Benson} A.~J.,   {Cole} S.,  2005, \mn@doi [\mnras]
  {10.1111/j.1365-2966.2004.08553.x}, \href
  {http://adsabs.harvard.edu/abs/2005MNRAS.356.1191B} {356, 1191}

\bibitem[\protect\citeauthoryear{{Behrens}, {Byrohl}, {Saito}  \&
  {Niemeyer}}{{Behrens} et~al.}{2017}]{Behrens2017}
{Behrens} C.,  {Byrohl} C.,  {Saito} S.,   {Niemeyer} J.~C.,  2017, preprint,
  \href {http://adsabs.harvard.edu/abs/2017arXiv171006171B} {} (\mn@eprint
  {arXiv} {1710.06171})

\bibitem[\protect\citeauthoryear{{Behroozi}, {Conroy}  \&
  {Wechsler}}{{Behroozi} et~al.}{2010}]{Behroozi_2010}
{Behroozi} P.~S.,  {Conroy} C.,   {Wechsler} R.~H.,  2010, \mn@doi [\apj]
  {10.1088/0004-637X/717/1/379}, \href
  {http://adsabs.harvard.edu/abs/2010ApJ...717..379B} {717, 379}

\bibitem[\protect\citeauthoryear{{Benitez} et~al.,}{{Benitez}
  et~al.}{2014}]{J-PAS}
{Benitez} N.,  et~al., 2014, preprint, \href
  {http://adsabs.harvard.edu/abs/2014arXiv1403.5237B} {} (\mn@eprint {arXiv}
  {1403.5237})

\bibitem[\protect\citeauthoryear{{Bielby} et~al.,}{{Bielby}
  et~al.}{2016}]{Bielby2016}
{Bielby} R.~M.,  et~al., 2016, \mn@doi [\mnras] {10.1093/mnras/stv2914}, \href
  {http://adsabs.harvard.edu/abs/2016MNRAS.456.4061B} {456, 4061}

\bibitem[\protect\citeauthoryear{{Bower}, {Benson}, {Malbon}, {Helly}, {Frenk},
  {Baugh}, {Cole}  \& {Lacey}}{{Bower} et~al.}{2006}]{bower06}
{Bower} R.~G.,  {Benson} A.~J.,  {Malbon} R.,  {Helly} J.~C.,  {Frenk} C.~S.,
  {Baugh} C.~M.,  {Cole} S.,   {Lacey} C.~G.,  2006, \mn@doi [\mnras]
  {10.1111/j.1365-2966.2006.10519.x}, \href
  {http://adsabs.harvard.edu/abs/2006MNRAS.370..645B} {370, 645}

\bibitem[\protect\citeauthoryear{{Cassata} et~al.,}{{Cassata}
  et~al.}{2011}]{Cassata_2011}
{Cassata} P.,  et~al., 2011, \mn@doi [\aap] {10.1051/0004-6361/201014410},
  \href {http://adsabs.harvard.edu/abs/2011A%26A...525A.143C} {525, A143}

\bibitem[\protect\citeauthoryear{{Chisholm}, {Orlitov{\'a}}, {Schaerer},
  {Verhamme}, {Worseck}, {Izotov}, {Thuan}  \& {Guseva}}{{Chisholm}
  et~al.}{2017}]{Chisholm2017}
{Chisholm} J.,  {Orlitov{\'a}} I.,  {Schaerer} D.,  {Verhamme} A.,  {Worseck}
  G.,  {Izotov} Y.~I.,  {Thuan} T.~X.,   {Guseva} N.~G.,  2017, \mn@doi [\aap]
  {10.1051/0004-6361/201730610}, \href
  {http://adsabs.harvard.edu/abs/2017A%26A...605A..67C} {605, A67}

\bibitem[\protect\citeauthoryear{{Cole}, {Lacey}, {Baugh}  \& {Frenk}}{{Cole}
  et~al.}{2000}]{cole00}
{Cole} S.,  {Lacey} C.~G.,  {Baugh} C.~M.,   {Frenk} C.~S.,  2000, \mn@doi
  [\mnras] {10.1046/j.1365-8711.2000.03879.x}, \href
  {http://adsabs.harvard.edu/abs/2000MNRAS.319..168C} {319, 168}

\bibitem[\protect\citeauthoryear{{Dayal}, {Maselli}  \& {Ferrara}}{{Dayal}
  et~al.}{2011}]{dayal11}
{Dayal} P.,  {Maselli} A.,   {Ferrara} A.,  2011, \mn@doi [\mnras]
  {10.1111/j.1365-2966.2010.17482.x}, \href
  {http://adsabs.harvard.edu/abs/2011MNRAS.410..830D} {410, 830}

\bibitem[\protect\citeauthoryear{{Dijkstra}}{{Dijkstra}}{2017}]{dijkstra17}
{Dijkstra} M.,  2017, preprint, \href
  {http://adsabs.harvard.edu/abs/2017arXiv170403416D} {} (\mn@eprint {arXiv}
  {1704.03416})

\bibitem[\protect\citeauthoryear{{Dijkstra}, {Haiman}  \& {Spaans}}{{Dijkstra}
  et~al.}{2006}]{dijkstra06}
{Dijkstra} M.,  {Haiman} Z.,   {Spaans} M.,  2006, \mn@doi [\apj]
  {10.1086/506243}, \href {http://adsabs.harvard.edu/abs/2006ApJ...649...14D}
  {649, 14}

\bibitem[\protect\citeauthoryear{{Dijkstra}, {Mesinger}  \&
  {Wyithe}}{{Dijkstra} et~al.}{2011}]{dijkstra11}
{Dijkstra} M.,  {Mesinger} A.,   {Wyithe} J.~S.~B.,  2011, \mn@doi [\mnras]
  {10.1111/j.1365-2966.2011.18530.x}, \href
  {http://adsabs.harvard.edu/abs/2011MNRAS.414.2139D} {414, 2139}

\bibitem[\protect\citeauthoryear{{Foreman-Mackey}, {Hogg}, {Lang}  \&
  {Goodman}}{{Foreman-Mackey} et~al.}{2013a}]{emcee}
{Foreman-Mackey} D.,  {Hogg} D.~W.,  {Lang} D.,   {Goodman} J.,  2013a, \mn@doi
  [\pasp] {10.1086/670067}, \href
  {http://adsabs.harvard.edu/abs/2013PASP..125..306F} {125, 306}

\bibitem[\protect\citeauthoryear{{Foreman-Mackey}, {Hogg}, {Lang}  \&
  {Goodman}}{{Foreman-Mackey} et~al.}{2013b}]{Foreman_Mackey_2013}
{Foreman-Mackey} D.,  {Hogg} D.~W.,  {Lang} D.,   {Goodman} J.,  2013b, \mn@doi
  [\pasp] {10.1086/670067}, \href
  {http://adsabs.harvard.edu/abs/2013PASP..125..306F} {125, 306}

\bibitem[\protect\citeauthoryear{{Garel}, {Blaizot}, {Guiderdoni}, {Schaerer},
  {Verhamme}  \& {Hayes}}{{Garel} et~al.}{2012}]{garel12}
{Garel} T.,  {Blaizot} J.,  {Guiderdoni} B.,  {Schaerer} D.,  {Verhamme} A.,
  {Hayes} M.,  2012, \mn@doi [\mnras] {10.1111/j.1365-2966.2012.20607.x}, \href
  {http://adsabs.harvard.edu/abs/2012MNRAS.422..310G} {422, 310}

\bibitem[\protect\citeauthoryear{{Gawiser} et~al.,}{{Gawiser}
  et~al.}{2007}]{gawiser07}
{Gawiser} E.,  et~al., 2007, \mn@doi [\apj] {10.1086/522955}, \href
  {http://adsabs.harvard.edu/abs/2007ApJ...671..278G} {671, 278}

\bibitem[\protect\citeauthoryear{{Granato}, {Lacey}, {Silva}, {Bressan},
  {Baugh}, {Cole}  \& {Frenk}}{{Granato} et~al.}{2000}]{granato00}
{Granato} G.~L.,  {Lacey} C.~G.,  {Silva} L.,  {Bressan} A.,  {Baugh} C.~M.,
  {Cole} S.,   {Frenk} C.~S.,  2000, \mn@doi [\apj] {10.1086/317032}, \href
  {http://adsabs.harvard.edu/abs/2000ApJ...542..710G} {542, 710}

\bibitem[\protect\citeauthoryear{{Gronke}, {Dijkstra}, {McCourt}  \&
  {Oh}}{{Gronke} et~al.}{2016}]{Gronke_2016}
{Gronke} M.,  {Dijkstra} M.,  {McCourt} M.,   {Oh} S.~P.,  2016, \mn@doi
  [\apjl] {10.3847/2041-8213/833/2/L26}, \href
  {http://adsabs.harvard.edu/abs/2016ApJ...833L..26G} {833, L26}

\bibitem[\protect\citeauthoryear{{Guaita} et~al.,}{{Guaita}
  et~al.}{2010}]{guaita10}
{Guaita} L.,  et~al., 2010, \mn@doi [\apj] {10.1088/0004-637X/714/1/255}, \href
  {http://adsabs.harvard.edu/abs/2010ApJ...714..255G} {714, 255}

\bibitem[\protect\citeauthoryear{{Guaita} et~al.,}{{Guaita}
  et~al.}{2017}]{Guaita2017}
{Guaita} L.,  et~al., 2017, \mn@doi [\aap] {10.1051/0004-6361/201730603}, \href
  {http://adsabs.harvard.edu/abs/2017A%26A...606A..19G} {606, A19}

\bibitem[\protect\citeauthoryear{{Harrington}}{{Harrington}}{1973}]{harrington73}
{Harrington} J.~P.,  1973, \mnras, \href
  {http://adsabs.harvard.edu/abs/1973MNRAS.162...43H} {162, 43}

\bibitem[\protect\citeauthoryear{{Hill} et~al.,}{{Hill}
  et~al.}{2008}]{Hill2008}
{Hill} G.~J.,  et~al., 2008, in {Kodama} T.,  {Yamada} T.,   {Aoki} K.,  eds,
  Astronomical Society of the Pacific Conference Series Vol. 399, Panoramic
  Views of Galaxy Formation and Evolution. p.~115 (\mn@eprint {arXiv}
  {0806.0183})

\bibitem[\protect\citeauthoryear{{Hu}, {Cowie}  \& {McMahon}}{{Hu}
  et~al.}{1998}]{hu98}
{Hu} E.~M.,  {Cowie} L.~L.,   {McMahon} R.~G.,  1998, \mn@doi [\apjl]
  {10.1086/311506}, \href {http://adsabs.harvard.edu/abs/1998ApJ...502L..99H}
  {502, L99+}

\bibitem[\protect\citeauthoryear{{Hubble}}{{Hubble}}{1929}]{hubble29}
{Hubble} E.,  1929, \mn@doi [Proceedings of the National Academy of Science]
  {10.1073/pnas.15.3.168}, \href
  {http://adsabs.harvard.edu/abs/1929PNAS...15..168H} {15, 168}

\bibitem[\protect\citeauthoryear{{Inoue} et~al.,}{{Inoue}
  et~al.}{2018}]{Inoue2018}
{Inoue} A.~K.,  et~al., 2018, preprint, \href
  {http://adsabs.harvard.edu/abs/2018arXiv180100067I} {} (\mn@eprint {arXiv}
  {1801.00067})

\bibitem[\protect\citeauthoryear{{Kashikawa} et~al.,}{{Kashikawa}
  et~al.}{2006}]{kashikawa06}
{Kashikawa} N.,  et~al., 2006, \mn@doi [\apj] {10.1086/504966}, \href
  {http://adsabs.harvard.edu/abs/2006ApJ...648....7K} {648, 7}

\bibitem[\protect\citeauthoryear{{Kobayashi}, {Totani}  \&
  {Nagashima}}{{Kobayashi} et~al.}{2007}]{kobayashi07}
{Kobayashi} M.~A.~R.,  {Totani} T.,   {Nagashima} M.,  2007, \mn@doi [\apj]
  {10.1086/522200}, \href {http://adsabs.harvard.edu/abs/2007ApJ...670..919K}
  {670, 919}

\bibitem[\protect\citeauthoryear{{Konno}, {Ouchi}, {Nakajima}, {Duval},
  {Kusakabe}, {Ono}  \& {Shimasaku}}{{Konno} et~al.}{2016}]{Konno2016}
{Konno} A.,  {Ouchi} M.,  {Nakajima} K.,  {Duval} F.,  {Kusakabe} H.,  {Ono}
  Y.,   {Shimasaku} K.,  2016, \mn@doi [\apj] {10.3847/0004-637X/823/1/20},
  \href {http://adsabs.harvard.edu/abs/2016ApJ...823...20K} {823, 20}

\bibitem[\protect\citeauthoryear{{Konno} et~al.,}{{Konno}
  et~al.}{2018}]{Konno_2018}
{Konno} A.,  et~al., 2018, \mn@doi [\pasj] {10.1093/pasj/psx131}, \href
  {http://adsabs.harvard.edu/abs/2018PASJ...70S..16K} {70, S16}

\bibitem[\protect\citeauthoryear{{Kulas}, {Shapley}, {Kollmeier}, {Zheng},
  {Steidel}  \& {Hainline}}{{Kulas} et~al.}{2011}]{kulas11}
{Kulas} K.~R.,  {Shapley} A.~E.,  {Kollmeier} J.~A.,  {Zheng} Z.,  {Steidel}
  C.~C.,   {Hainline} K.~N.,  2011, preprint, \href
  {http://adsabs.harvard.edu/abs/2011arXiv1107.4367K} {11074367} (\mn@eprint
  {arXiv} {1107.4367})

\bibitem[\protect\citeauthoryear{{Kusakabe} et~al.,}{{Kusakabe}
  et~al.}{2018}]{Kusakabe2018}
{Kusakabe} H.,  et~al., 2018, \mn@doi [\pasj] {10.1093/pasj/psx148}, \href
  {http://adsabs.harvard.edu/abs/2018PASJ..tmp...11K} {}

\bibitem[\protect\citeauthoryear{{Lacey} et~al.,}{{Lacey}
  et~al.}{2016}]{lacey16}
{Lacey} C.~G.,  et~al., 2016, \mn@doi [\mnras] {10.1093/mnras/stw1888}, \href
  {http://adsabs.harvard.edu/abs/2016MNRAS.462.3854L} {462, 3854}

\bibitem[\protect\citeauthoryear{{Laursen} \& {Sommer-Larsen}}{{Laursen} \&
  {Sommer-Larsen}}{2007}]{laursen07}
{Laursen} P.,  {Sommer-Larsen} J.,  2007, \mn@doi [\apjl] {10.1086/513191},
  \href {http://adsabs.harvard.edu/abs/2007ApJ...657L..69L} {657, L69}

\bibitem[\protect\citeauthoryear{{Laursen}, {Razoumov}  \&
  {Sommer-Larsen}}{{Laursen} et~al.}{2009}]{laursen09a}
{Laursen} P.,  {Razoumov} A.~O.,   {Sommer-Larsen} J.,  2009, \mn@doi [\apj]
  {10.1088/0004-637X/696/1/853}, \href
  {http://adsabs.harvard.edu/abs/2009ApJ...696..853L} {696, 853}

\bibitem[\protect\citeauthoryear{{Laursen}, {Sommer-Larsen}  \&
  {Razoumov}}{{Laursen} et~al.}{2011}]{Laursen2011}
{Laursen} P.,  {Sommer-Larsen} J.,   {Razoumov} A.~O.,  2011, \mn@doi [\apj]
  {10.1088/0004-637X/728/1/52}, \href
  {http://adsabs.harvard.edu/abs/2011ApJ...728...52L} {728, 52}

\bibitem[\protect\citeauthoryear{{Le Delliou}, {Lacey}, {Baugh}, {Guiderdoni},
  {Bacon}, {Courtois}, {Sousbie}  \& {Morris}}{{Le Delliou}
  et~al.}{2005}]{ledelliou05}
{Le Delliou} M.,  {Lacey} C.,  {Baugh} C.~M.,  {Guiderdoni} B.,  {Bacon} R.,
  {Courtois} H.,  {Sousbie} T.,   {Morris} S.~L.,  2005, \mn@doi [\mnras]
  {10.1111/j.1745-3933.2005.00007.x}, \href
  {http://adsabs.harvard.edu/abs/2005MNRAS.357L..11L} {357, L11}

\bibitem[\protect\citeauthoryear{{Le Delliou}, {Lacey}, {Baugh}  \&
  {Morris}}{{Le Delliou} et~al.}{2006}]{ledelliou06}
{Le Delliou} M.,  {Lacey} C.~G.,  {Baugh} C.~M.,   {Morris} S.~L.,  2006,
  \mn@doi [\mnras] {10.1111/j.1365-2966.2005.09797.x}, \href
  {http://adsabs.harvard.edu/abs/2006MNRAS.365..712L} {365, 712}

\bibitem[\protect\citeauthoryear{{Malhotra} \& {Rhoads}}{{Malhotra} \&
  {Rhoads}}{2002}]{malhotra02}
{Malhotra} S.,  {Rhoads} J.~E.,  2002, \mn@doi [\apjl] {10.1086/338980}, \href
  {http://adsabs.harvard.edu/abs/2002ApJ...565L..71M} {565, L71}

\bibitem[\protect\citeauthoryear{{Nagamine}, {Ouchi}, {Springel}  \&
  {Hernquist}}{{Nagamine} et~al.}{2010}]{nagamine10}
{Nagamine} K.,  {Ouchi} M.,  {Springel} V.,   {Hernquist} L.,  2010, \pasj,
  \href {http://adsabs.harvard.edu/abs/2010PASJ...62.1455N} {62, 1455}

\bibitem[\protect\citeauthoryear{{Nelson} et~al.,}{{Nelson}
  et~al.}{2015}]{Nelson2015}
{Nelson} D.,  et~al., 2015, \mn@doi [Astronomy and Computing]
  {10.1016/j.ascom.2015.09.003}, \href
  {http://adsabs.harvard.edu/abs/2015A%26C....13...12N} {13, 12}

\bibitem[\protect\citeauthoryear{{Neufeld}}{{Neufeld}}{1990}]{neufeld90}
{Neufeld} D.~A.,  1990, \mn@doi [\apj] {10.1086/168375}, \href
  {http://adsabs.harvard.edu/abs/1990ApJ...350..216N} {350, 216}

\bibitem[\protect\citeauthoryear{{Orsi} \& {Angulo}}{{Orsi} \&
  {Angulo}}{2018}]{orsi18}
{Orsi} {\'A}.~A.,  {Angulo} R.~E.,  2018, \mn@doi [\mnras]
  {10.1093/mnras/stx3349}, \href
  {http://adsabs.harvard.edu/abs/2018MNRAS.475.2530O} {475, 2530}

\bibitem[\protect\citeauthoryear{{Orsi}, {Lacey}, {Baugh}  \& {Infante}}{{Orsi}
  et~al.}{2008}]{orsi08}
{Orsi} A.,  {Lacey} C.~G.,  {Baugh} C.~M.,   {Infante} L.,  2008, \mn@doi
  [\mnras] {10.1111/j.1365-2966.2008.14010.x}, \href
  {http://adsabs.harvard.edu/abs/2008MNRAS.391.1589O} {391, 1589}

\bibitem[\protect\citeauthoryear{{Orsi}, {Lacey}  \& {Baugh}}{{Orsi}
  et~al.}{2012}]{orsi12}
{Orsi} A.,  {Lacey} C.~G.,   {Baugh} C.~M.,  2012, \mn@doi [\mnras]
  {10.1111/j.1365-2966.2012.21396.x}, \href
  {http://adsabs.harvard.edu/abs/2012MNRAS.425...87O} {425, 87}

\bibitem[\protect\citeauthoryear{{Osterbrock}}{{Osterbrock}}{1989}]{osterbrock89}
{Osterbrock} D.~E.,  1989, {Astrophysics of gaseous nebulae and active galactic
  nuclei}

\bibitem[\protect\citeauthoryear{{Ouchi} et~al.,}{{Ouchi}
  et~al.}{2008}]{ouchi08}
{Ouchi} M.,  et~al., 2008, \mn@doi [\apjs] {10.1086/527673}, \href
  {http://adsabs.harvard.edu/abs/2008ApJS..176..301O} {176, 301}

\bibitem[\protect\citeauthoryear{{Ouchi} et~al.,}{{Ouchi}
  et~al.}{2010}]{ouchi10}
{Ouchi} M.,  et~al., 2010, \mn@doi [\apj] {10.1088/0004-637X/723/1/869}, \href
  {http://adsabs.harvard.edu/abs/2010ApJ...723..869O} {723, 869}

\bibitem[\protect\citeauthoryear{{Ouchi} et~al.,}{{Ouchi}
  et~al.}{2018}]{Ouchi2018a}
{Ouchi} M.,  et~al., 2018, \mn@doi [\pasj] {10.1093/pasj/psx074}, \href
  {http://adsabs.harvard.edu/abs/2018PASJ...70S..13O} {70, S13}

\bibitem[\protect\citeauthoryear{{Oyarz{\'u}n}, {Blanc}, {Gonz{\'a}lez},
  {Mateo}  \& {Bailey}}{{Oyarz{\'u}n} et~al.}{2017}]{oyarzun17}
{Oyarz{\'u}n} G.~A.,  {Blanc} G.~A.,  {Gonz{\'a}lez} V.,  {Mateo} M.,
  {Bailey} III J.~I.,  2017, \mn@doi [\apj] {10.3847/1538-4357/aa7552}, \href
  {http://adsabs.harvard.edu/abs/2017ApJ...843..133O} {843, 133}

\bibitem[\protect\citeauthoryear{{Planck Collaboration} et~al.,}{{Planck
  Collaboration} et~al.}{2016}]{Planck_2016}
{Planck Collaboration} et~al., 2016, \mn@doi [\aap]
  {10.1051/0004-6361/201525830}, \href
  {http://adsabs.harvard.edu/abs/2016A%26A...594A..13P} {594, A13}

\bibitem[\protect\citeauthoryear{{Rhoads}, {Malhotra}, {Dey}, {Stern},
  {Spinrad}  \& {Jannuzi}}{{Rhoads} et~al.}{2000}]{rhoads00}
{Rhoads} J.~E.,  {Malhotra} S.,  {Dey} A.,  {Stern} D.,  {Spinrad} H.,
  {Jannuzi} B.~T.,  2000, \mn@doi [\apjl] {10.1086/317874}, \href
  {http://adsabs.harvard.edu/abs/2000ApJ...545L..85R} {545, L85}

\bibitem[\protect\citeauthoryear{{Santos}, {Ellis}, {Kneib}, {Richard}  \&
  {Kuijken}}{{Santos} et~al.}{2004}]{santos04}
{Santos} M.~R.,  {Ellis} R.~S.,  {Kneib} J.,  {Richard} J.,   {Kuijken} K.,
  2004, \mn@doi [\apj] {10.1086/383080}, \href
  {http://adsabs.harvard.edu/abs/2004ApJ...606..683S} {606, 683}

\bibitem[\protect\citeauthoryear{{Schaerer} \& {Verhamme}}{{Schaerer} \&
  {Verhamme}}{2008}]{schaerer08}
{Schaerer} D.,  {Verhamme} A.,  2008, \mn@doi [\aap]
  {10.1051/0004-6361:20078913}, \href
  {http://adsabs.harvard.edu/abs/2008A%26A...480..369S} {480, 369}

\bibitem[\protect\citeauthoryear{{Shapley}, {Steidel}, {Pettini}  \&
  {Adelberger}}{{Shapley} et~al.}{2003}]{shapley03}
{Shapley} A.~E.,  {Steidel} C.~C.,  {Pettini} M.,   {Adelberger} K.~L.,  2003,
  \mn@doi [\apj] {10.1086/373922}, \href
  {http://adsabs.harvard.edu/abs/2003ApJ...588...65S} {588, 65}

\bibitem[\protect\citeauthoryear{{Sobral} et~al.,}{{Sobral}
  et~al.}{2017}]{Sobral2017}
{Sobral} D.,  et~al., 2017, \mn@doi [\mnras] {10.1093/mnras/stw3090}, \href
  {http://adsabs.harvard.edu/abs/2017MNRAS.466.1242S} {466, 1242}

\bibitem[\protect\citeauthoryear{{Steidel}, {Giavalisco}, {Pettini},
  {Dickinson}  \& {Adelberger}}{{Steidel} et~al.}{1996}]{steidel96}
{Steidel} C.~C.,  {Giavalisco} M.,  {Pettini} M.,  {Dickinson} M.,
  {Adelberger} K.~L.,  1996, \mn@doi [\apjl] {10.1086/310029}, \href
  {http://adsabs.harvard.edu/abs/1996ApJ...462L..17S} {462, L17+}

\bibitem[\protect\citeauthoryear{{Steidel}, {Erb}, {Shapley}, {Pettini},
  {Reddy}, {Bogosavljevi{\'c}}, {Rudie}  \& {Rakic}}{{Steidel}
  et~al.}{2010}]{steidel10}
{Steidel} C.~C.,  {Erb} D.~K.,  {Shapley} A.~E.,  {Pettini} M.,  {Reddy} N.,
  {Bogosavljevi{\'c}} M.,  {Rudie} G.~C.,   {Rakic} O.,  2010, \mn@doi [\apj]
  {10.1088/0004-637X/717/1/289}, \href
  {http://adsabs.harvard.edu/abs/2010ApJ...717..289S} {717, 289}

\bibitem[\protect\citeauthoryear{{Steidel}, {Bogosavljevi{\'c}}, {Shapley},
  {Kollmeier}, {Reddy}, {Erb}  \& {Pettini}}{{Steidel}
  et~al.}{2011}]{steidel11}
{Steidel} C.~C.,  {Bogosavljevi{\'c}} M.,  {Shapley} A.~E.,  {Kollmeier} J.~A.,
   {Reddy} N.~A.,  {Erb} D.~K.,   {Pettini} M.,  2011, \mn@doi [\apj]
  {10.1088/0004-637X/736/2/160}, \href
  {http://adsabs.harvard.edu/abs/2011ApJ...736..160S} {736, 160}

\bibitem[\protect\citeauthoryear{{Taniguchi} et~al.,}{{Taniguchi}
  et~al.}{2005}]{taniguchi05}
{Taniguchi} Y.,  et~al., 2005, \pasj, \href
  {http://adsabs.harvard.edu/abs/2005PASJ...57..165T} {57, 165}

\bibitem[\protect\citeauthoryear{{Verhamme}, {Schaerer}  \&
  {Maselli}}{{Verhamme} et~al.}{2006}]{verhamme06}
{Verhamme} A.,  {Schaerer} D.,   {Maselli} A.,  2006, \mn@doi [\aap]
  {10.1051/0004-6361:20065554}, \href
  {http://adsabs.harvard.edu/abs/2006A%26A...460..397V} {460, 397}

\bibitem[\protect\citeauthoryear{{Wang} et~al.,}{{Wang} et~al.}{2018}]{wang18}
{Wang} Y.,  et~al., 2018, preprint, \href
  {http://adsabs.harvard.edu/abs/2018arXiv180201539W} {} (\mn@eprint {arXiv}
  {1802.01539})

\bibitem[\protect\citeauthoryear{{Zheng} \& {Miralda-Escud{\'e}}}{{Zheng} \&
  {Miralda-Escud{\'e}}}{2002}]{zheng02}
{Zheng} Z.,  {Miralda-Escud{\'e}} J.,  2002, \mn@doi [\apj] {10.1086/342400},
  \href {http://adsabs.harvard.edu/abs/2002ApJ...578...33Z} {578, 33}

\bibitem[\protect\citeauthoryear{{Zheng}, {Cen}, {Trac}  \&
  {Miralda-Escud{\'e}}}{{Zheng} et~al.}{2010}]{zheng10}
{Zheng} Z.,  {Cen} R.,  {Trac} H.,   {Miralda-Escud{\'e}} J.,  2010, \mn@doi
  [\apj] {10.1088/0004-637X/716/1/574}, \href
  {http://adsabs.harvard.edu/abs/2010ApJ...716..574Z} {716, 574}

\bibitem[\protect\citeauthoryear{{Zheng}, {Cen}, {Trac}  \&
  {Miralda-Escud{\'e}}}{{Zheng} et~al.}{2011}]{zheng11}
{Zheng} Z.,  {Cen} R.,  {Trac} H.,   {Miralda-Escud{\'e}} J.,  2011, \mn@doi
  [\apj] {10.1088/0004-637X/726/1/38}, \href
  {http://adsabs.harvard.edu/abs/2011ApJ...726...38Z} {726, 38}

\makeatother
\end{thebibliography}




\appendix
\section{Validating the \fesc\ fitting formulae}\label{Ap:A}

\begin{figure*} 
\includegraphics[width=6.93in]{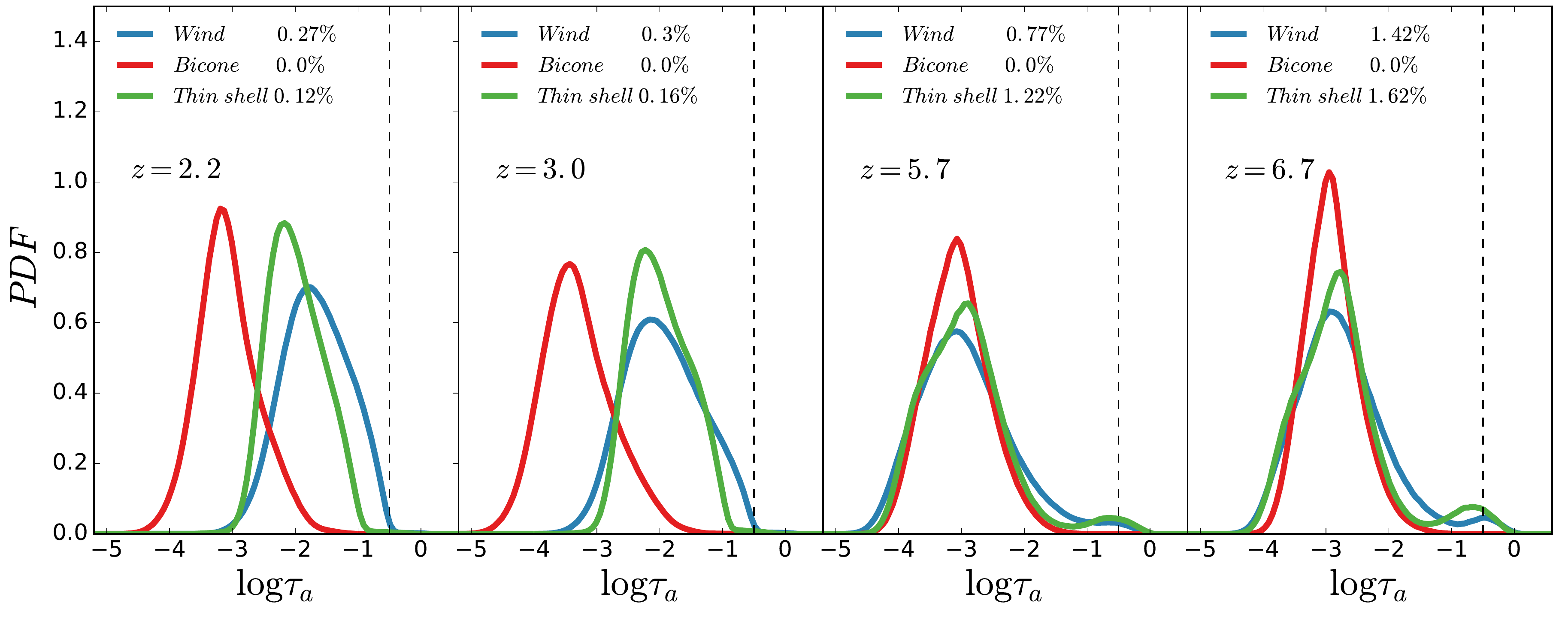} 
\caption{ Distribution  of the dust optical depth  for the RT LAE samples for $z=2.2$, 3.0, 5.7 and 6.7 from left to right. Solid lines represent the {\it Thin Shell} (\colorThin), galactic {\it Wind} (\colorWind) and biconical galactic wind (\colorBicone) models. In black dash lines we show the $\tau_a$ value below which the typical discrepancies between our $f^{Ly\alpha}_{esc}$ model and the MC RT code are $<10\%$ ($\log \tau _a = -0.5$). In each panel we also indicate the  percentage of LAEs with $\log \tau _a > -0.5$   }\label{fig:galaxy_ta}
\end{figure*}   

\begin{figure*}
    \centering
    \textbf{Thin Shell \fesc\ radiative transfer code - model comparison}\par\medskip
    \includegraphics[width=6.0in]{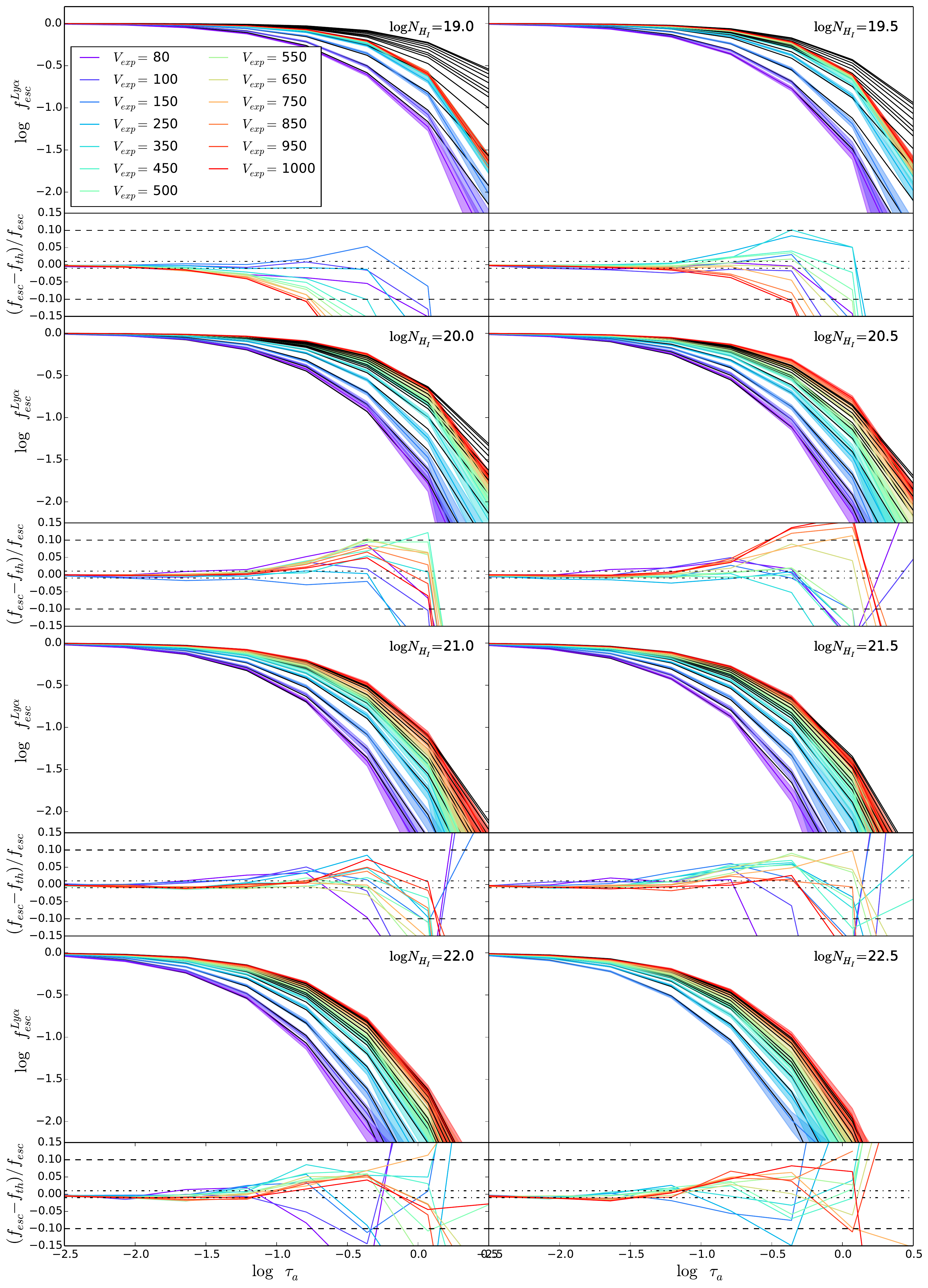}
    \caption{Comparison between the output of the radiative transfer code and our model for the $f^{Ly\alpha}_{esc}$ in the {\it Thin Shell} geometry. Each panel is divided in top ( the values of the escape fraction) and button ( relative difference between our model and the radiative transfer code). In top panels  the output from the radiative transfer code in plotted in colored lines ( color coded by the velocity of the system) with their errors ( shades with the same color) and our model prediction in black. In button panels the relative difference between our model and our code are plotted in colored lines and the $\pm 1\%$ and $\pm 10\%$ are represented by black dashed-dotted and dashed lines respectively. Note that he color code is the same in every panel. }\label{fig:fesc-comparison-thin}
\end{figure*}

\begin{figure*}
    \centering
    \textbf{Galactic wind \fesc\ radiative transfer code - model comparison}\par\medskip
    \includegraphics[width=6.5in]{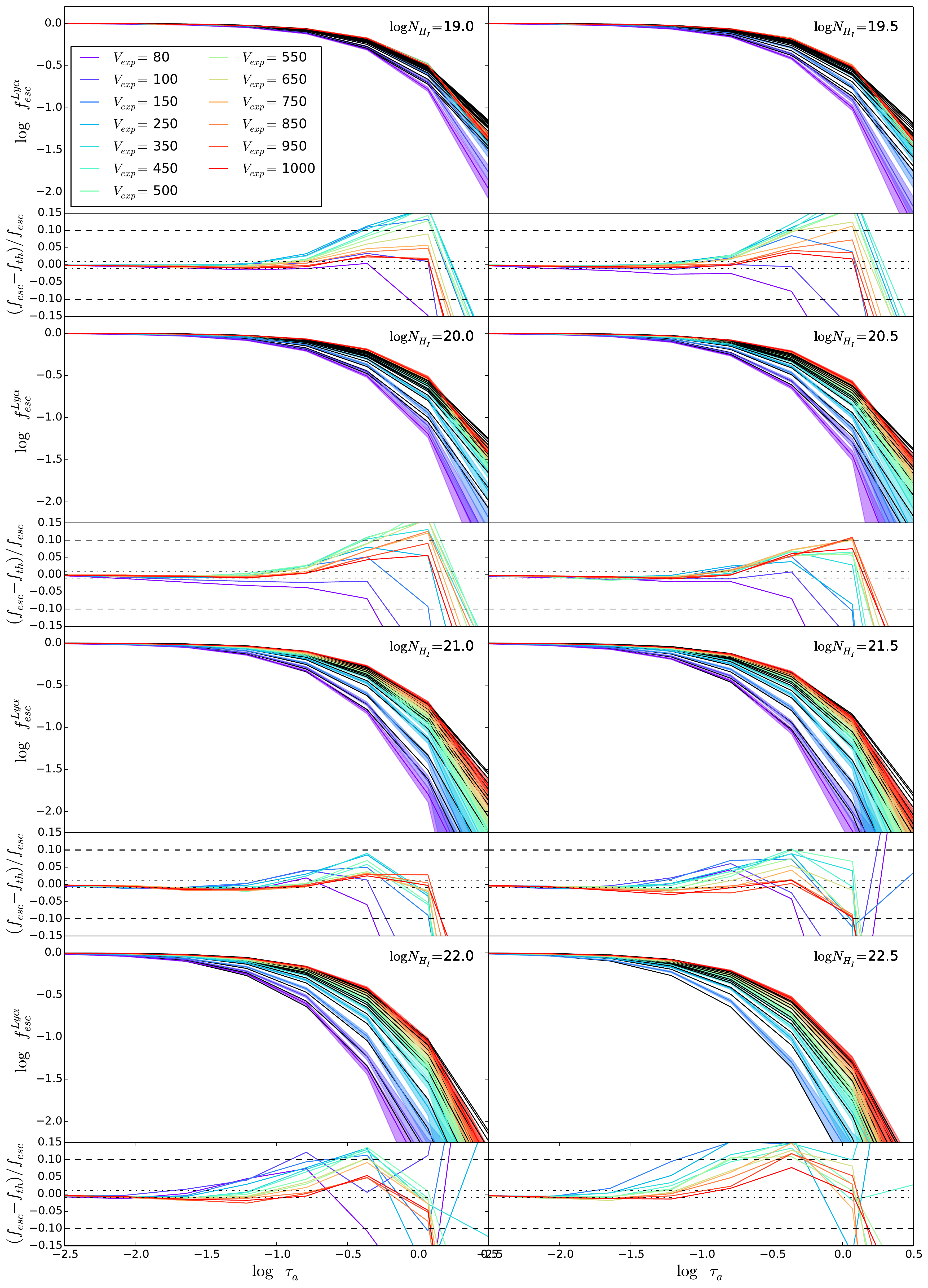}
    \caption{Same as figure \ref{fig:fesc-comparison-wind} but for the galactic wind. }\label{fig:fesc-comparison-wind}
\end{figure*}

\begin{figure*}
    \centering
    \textbf{Biconical galactic wind \fesc\ radiative transfer code - model comparison}\par\medskip
    \includegraphics[width=6.5in]{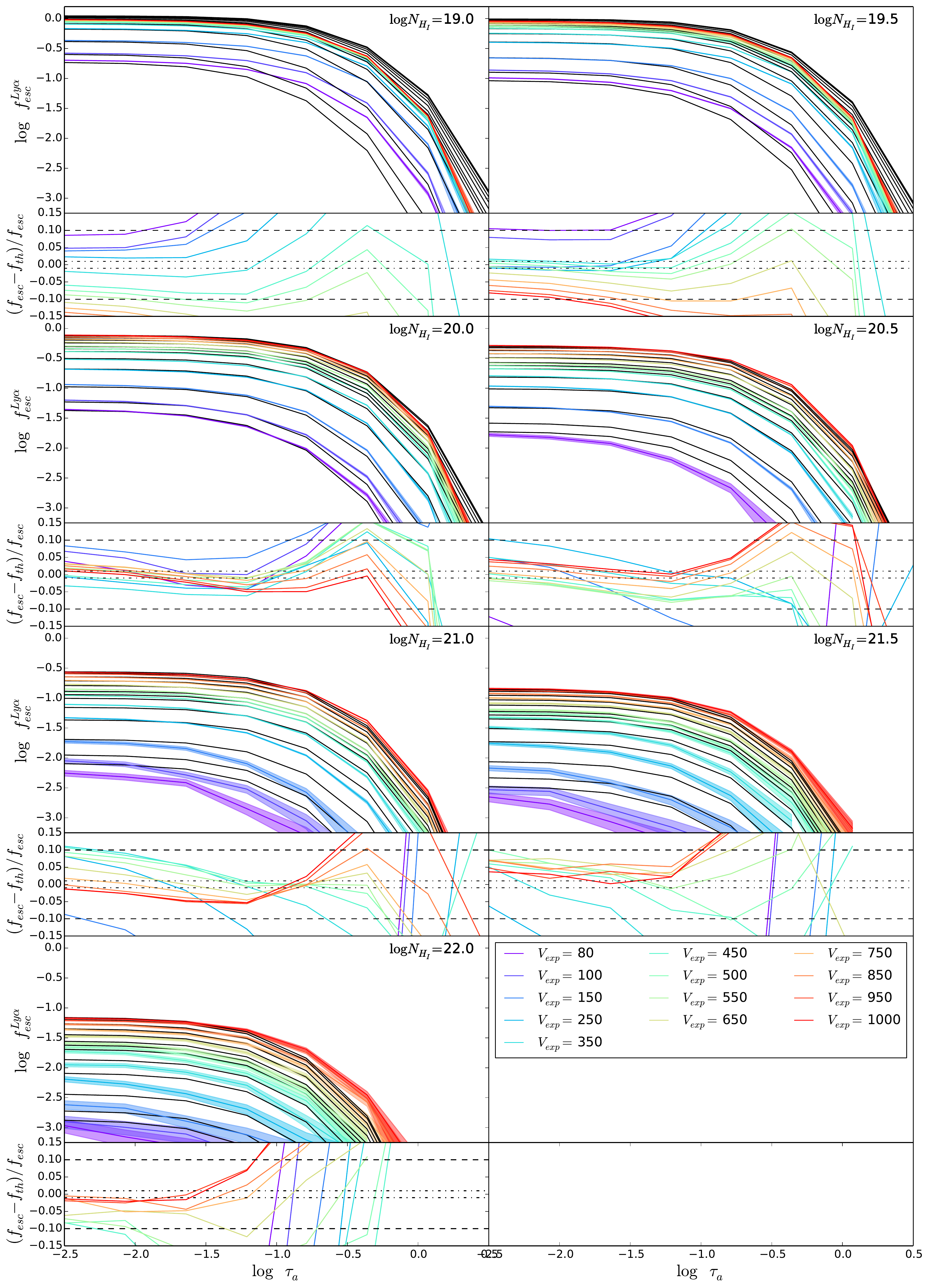}
    \caption{Same as figure \ref{fig:fesc-comparison-bicone} but for the biconical galactic wind.}\label{fig:fesc-comparison-bicone}
\end{figure*}

\indent A direct comparison between the Monte Carlo radiative transfer (MCRT) code output and or \fesc\ model for the {\it Thin Shell} geometry is shown in Fig. \ref{fig:fesc-comparison-thin}, for the galactic wind in Fig. \ref{fig:fesc-comparison-wind} and for the {\it Bicone} in Fig. \ref{fig:fesc-comparison-bicone}. These figures are divided in 8 panels sub divided in another two: a) escape fraction v.s. dust optical depth and b) the relative difference between our model and the radiative transfer code output. In panels a) the output of the radiative transfer code for a fixed $N_H$ is plotted color coded in solid lines with their respective errors ( same colored shade region ) computed using eq. \ref{eq:fescError} and our \fesc\ model is plotted in black solid lines. In type b) panels we show the relative difference between our model and the MCRT code with the same color code than above. 

In general,  the performance of our model decrease with $\tau_a$, this is because, as discussed above, decreasing \fesc\ increases the errors. This disagreement, in some cases, leads to an overestimation of \fesc\ when its true value is $\lesssim 0.01$. For our work, these low values are very rare and so do not affect our results.    Overall, we find that the typical discrepancies are below 10\% and 1\% for $\log \tau_a <-0.5$ and $<-1.0$ respectively.

Our model for the {\it Thin Shell} \fesc\ is able to reproduce the whole velocity range of our grid for $\rm N_{H}<10^{19.5} cm^{-2}$, reaching a $99\%$ accuracy in most cases.

Our \fesc\ model using the {\it Wind} geometry is also able to reproduce the output of our radiative transfer code through most of the parameter, only failing at very high $\rm N_H$ and low $V_{exp}$ combinations, where \fesc $<0.1$. As in the {\it Thin Shell} geometry, our model behaves better for $\rm N_{H}>10^{19.5} cm^{-2}$. In particular, in most of out grid,  the disagreement is lower than $10\%$ and for low $\tau_a<-1$ the typical agreement is $1\%$. 

The {\it Bicone} geometry is more complex than the other geometries, and its \fesc\ model has the worst performance of all. However, for most of the grid the model is within $10\%$ errors. As explained in \S 2,  the maximum \fesc\ depends on the properties of the outflowing gas, causing that only systems with very low optical depth 
(low $\rm N_{H}$ and/or high $\rm V_{exp}$) manage to reach $f_{esc}^{Ly\alpha} = 1$. This also causes that in very optically thick systems \fesc\ reaches 0.001 (even if there is no dust). We decided not to include $\rm N_{H} = 10^{22.5} cm^{-2}$ in our model because the maximum value of \fesc\ at $\rm V_{exp}=1000 \; km/s $ would be about 0.01 and, as discussed above, it is unnecessary to reproduce such low values.

In Fig. \ref{fig:galaxy_ta} we show the distribution of dust optical depth for our RT LAE samples (selected as in \S \ref{sec:outprop}) at redshifts 2.2 , 3.0 , 5.7 and 6.7 from left to right. At redshifts 2.2 and 3.0 the {\it Wind} and {\it Thin Shell} $\tau_a$ distributions are very similar in width and center ($\log \tau_a \sim -2$) while the {\it Bicone} model predict $\log \tau_a \sim -3$. Since the {\it Bicone} \fesc\ exhibits an upper limit $<1$, it requires low column densities 
(see Fig. \ref{fig:vexp_and_nh}) and $Z$, thus low $\tau_a$ values. At high redshifts (5.7 and 6.7) the dust optical depth distributions for the three geometries are very similar and peak at $\log \tau \sim -3$. In addition to the bulk of the distribution, the {\it Thin Shell} and {\it Wind} geometries also present a small bump around $\log \tau \sim -0.5$. 

In the legend of each panel of Fig. \ref{fig:galaxy_ta} we indicate the percentage of galaxies with $\log \tau_a >-0.5$, where the typical discrepancies in our \fesc\ model reach 10\%. This fraction $<2\%$ for all the configurations studied in this work. Additionally, as shown in Fig. \ref{fig:vexp_and_nh} most ($>95\%$) galaxies are inside the $\rm V_{exp} - N_H$ explored region. We conclude that the amount of galaxies with discrepancies $>10\%$ is negligible.

\section{Choosing an EW and luminosity cut for the mock catalogues}\label{Ap:B}

\begin{table*}
\centering
\caption{Properties of the different mock catalogs and surveys.}
\label{tab:mock_EW_L}
\begin{tabular}{ccccccclccccc}
Authors      &   $z$         &  \multicolumn{5}{c}{$\rm EW_{0,cut} [ $\AA{}$ ]$}      & & \multicolumn{5}{c}{$\rm L_{Ly\alpha , cut } \left[ erg \ s^{-1} \right] $}       \\ \cline{3-7} \cline{9-13}
             &               &   Survey   &   Thin Shell &  Wind     &  Bicone   &  AM    & &  Survey   & Thin Shell  &  Wind     &  Bicone   &  AM                         \\ \cline{1-13}
\\
\cite{Kusakabe2018} &  2.2 &  $20.0$ & $19.91$ & $20.3$ & $18.79$ & $19.52$ & & $1.62\ 10 ^ {42}$ & $1.54\ 10 ^ {42}$ & $1.92\ 10 ^ {42}$ & $1.38\ 10 ^ {42}$ & $1.6\ 10 ^ {42}$ \\
\cite{Bielby2016} &  3.0 &  $65.0$ & $42.0$ & $48.37$ & $46.05$ & $20.42$ & & $1.62\ 10 ^ {42}$ & $1.33\ 10 ^ {42}$ & $1.57\ 10 ^ {42}$ & $1.47\ 10 ^ {42}$ & $1.48\ 10 ^ {42}$ \\
\cite{Ouchi2018a} &  5.7 &  $20.0$ & $20.06$ & $20.06$ & $18.77$ & $21.45$ & & $6.3\ 10 ^ {42}$ & $7.39\ 10 ^ {42}$ & $6.89\ 10 ^ {42}$ & $2.61\ 10 ^ {42}$ & $6.78\ 10 ^ {42}$ \\
\cite{Ouchi2018a} &  6.7 &  $20.0$ & $20.06$ & $20.06$ & $15.88$ & $21.45$ & & $7.9\ 10 ^ {42}$ & $7.7\ 10 ^ {42}$ & $6.11\ 10 ^ {42}$ & $1.86\ 10 ^ {42}$ & $8.37\ 10 ^ {42}$
\end{tabular}
\end{table*}

In order to compare our clustering predictions with observations we construct mock catalogs that mimic the properties of several surveys at different redshifts. In general, there are several options for building mock catalogs to measure clustering. 

The first one, for example, is to use the same selection criteria (flux depth, equivalent width cut, etc) than the observed samples. This first option is useful if all the properties used in the selection criteria are well reproduced by the models. 

The LAE surveys studied in this work are limited by $\rm L_{Ly \alpha} > L_{Ly \alpha,cut}$ and $\rm EW_{0} > EW_{0,cut}$. In general $L_{Ly \alpha,cut}$ and $\rm EW_{0,cut}$ are different for every survey. These values are listed in Table \ref{tab:mock_EW_L}. 

Our models are designed so they reproduce the abundance and luminosity distribution LAEs as we force them to fit, as good as possible, the observed LF at different redshifts. In detail, we combine different observations of the \lya\ LF at the same redshift in order to calibrate our models. Because of this, the surveys that we use to study the clustering and calibrate our models, in general, use different selection criteria or the source sample is different. This could lead to discrepancies in the predicted number density of sources by our models imposing the clustering studies restrictions and the observed abundance of sources in these ones. 

In particular, at $z=2.2$ the survey constraining the clustering \citep{Kusakabe2018} is, at least, partially included in one of the surveys used to calibrate the LF \citep{Konno2016}. Additionally, $\rm EW_{0,cut}$ is the same for all the surveys used to fit the LF \citep{Cassata_2011,Konno2016,Sobral2017} and \cite{Kusakabe2018}. 

However, at $z=3.0 $ the selection criteria of the surveys used to fit the LF \citep{Cassata_2011,ouchi08} has $\rm EW_{0,cut}=20$\AA{} and \cite{Bielby2016} (clustering measurements) has $\rm EW_{0,cut}=65$\AA{}. 

The best scenario happens at redshifts 5.7 and 6.7, where the surveys used to calibrate our models \citep{ouchi08,Konno2016} are practically the same in sky coverage and selection criteria than the ones used to constrain the clustering \citep{ouchi10,Ouchi2018a}.

The second method to construct mock catalogs consists in matching the observed number density of sources. This can be achieve by relaxing the selection criteria. To minimize the possible secondary effects in the clustering due to changes in the selection criteria, we choose the combination that minimizes
\begin{equation}\label{eq:LLya_EW_cuts}
\rm
Q = ( log L_{Ly \alpha,n} - \log L_{Ly \alpha,s} )^2 + ( \log EW_{0,n} - \log EW_{0,s} )^2 ,
\end{equation} 
where $\rm L_{Ly \alpha,s}$ and $\rm EW_{0,s}$ are the $\rm L_{Ly \alpha,cut}$ and $\rm EW_{0,cut}$ imposed by each survey and $\rm L_{Ly \alpha,n}$ and $\rm EW_{0,n}$ define the iso-$\rm n_{LAE}$ curve with the LAE observed abundance. In Table \ref{tab:mock_EW_L} we list $\rm L_{Ly \alpha,s}$ and $\rm EW_{0,s}$ for the different surveys and the used values of $\rm L_{Ly \alpha,cut}$ and $\rm EW_{0,cut}$ to construct the mock catalogs.

\begin{figure*} 
\includegraphics[width=6.5in]{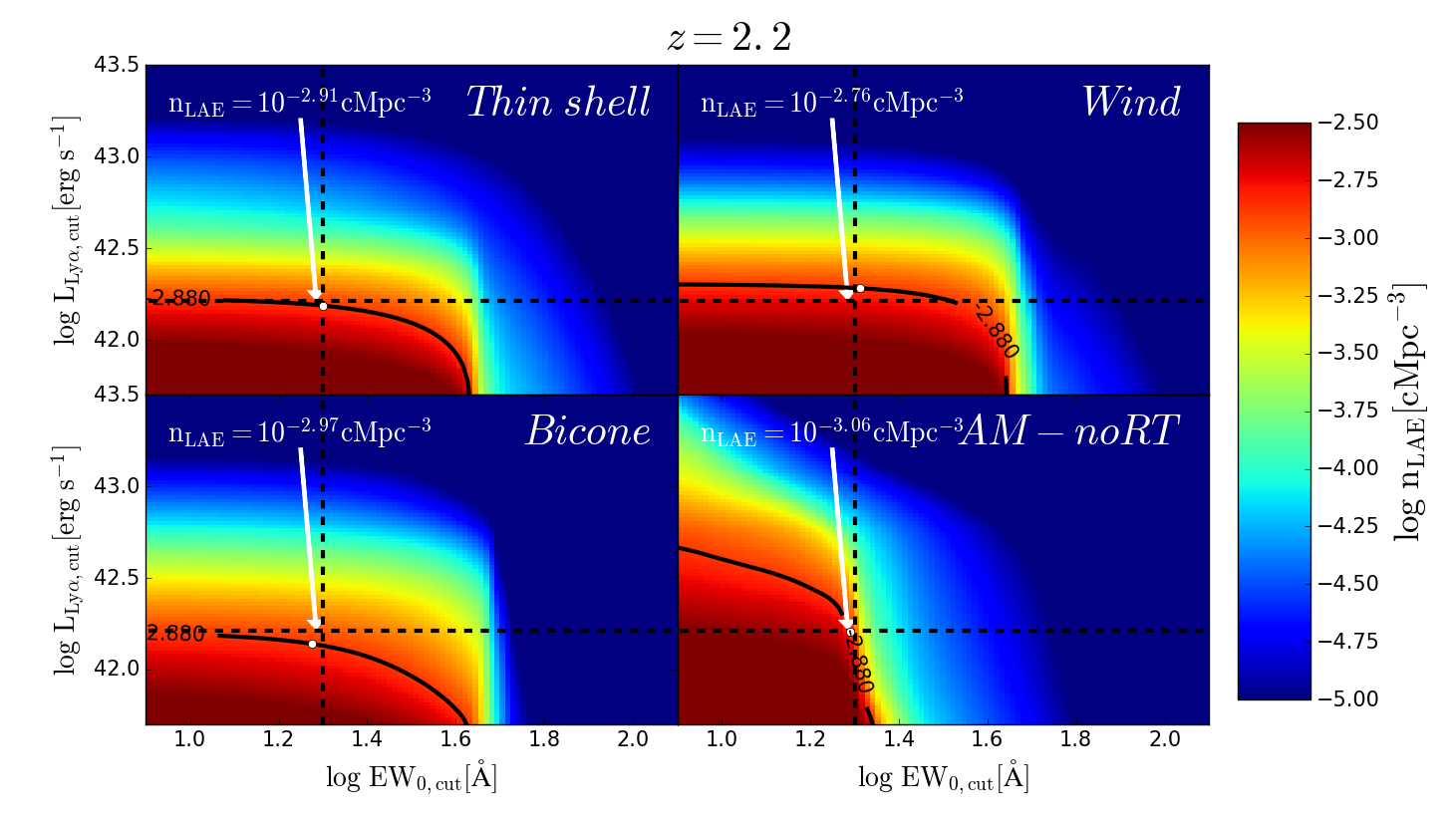} 
\caption{ Number density of LAEs $\rm n_{LAE}$ with \lya\ luminosity $\rm L_{Ly \alpha} > L_{Ly \alpha , cut}$ and \lya\ rest frame equivalent width  $\rm EW_0 > EW_{0,cut}$  at redshift 2.2 for the {\it Thin Shell} (top left), {\it Wind} (top right), {\it Bicone} (bottom left) and \NoRT\ (bottom right) model. In horizontal and vertical dashed black line we show the cut in $\rm L_{Ly \alpha}$ and $\rm EW_0$  respectively, in the survey at this redshift  \citep{Kusakabe2018}. The place where these lines intersect sets the predicted $\rm n_{LAE}$ by our models which value is indicated in the same panel. The solid back line is the iso-number density curve of the observed $\rm n_{LAE}$. The white dot indicates the position in the iso-number density curve that minimize the distance between our model prediction and the observed $\rm n_{LAE}$.}

\label{fig:compo_z2}. 
\end{figure*}

\begin{figure*} 
\includegraphics[width=6.93in]{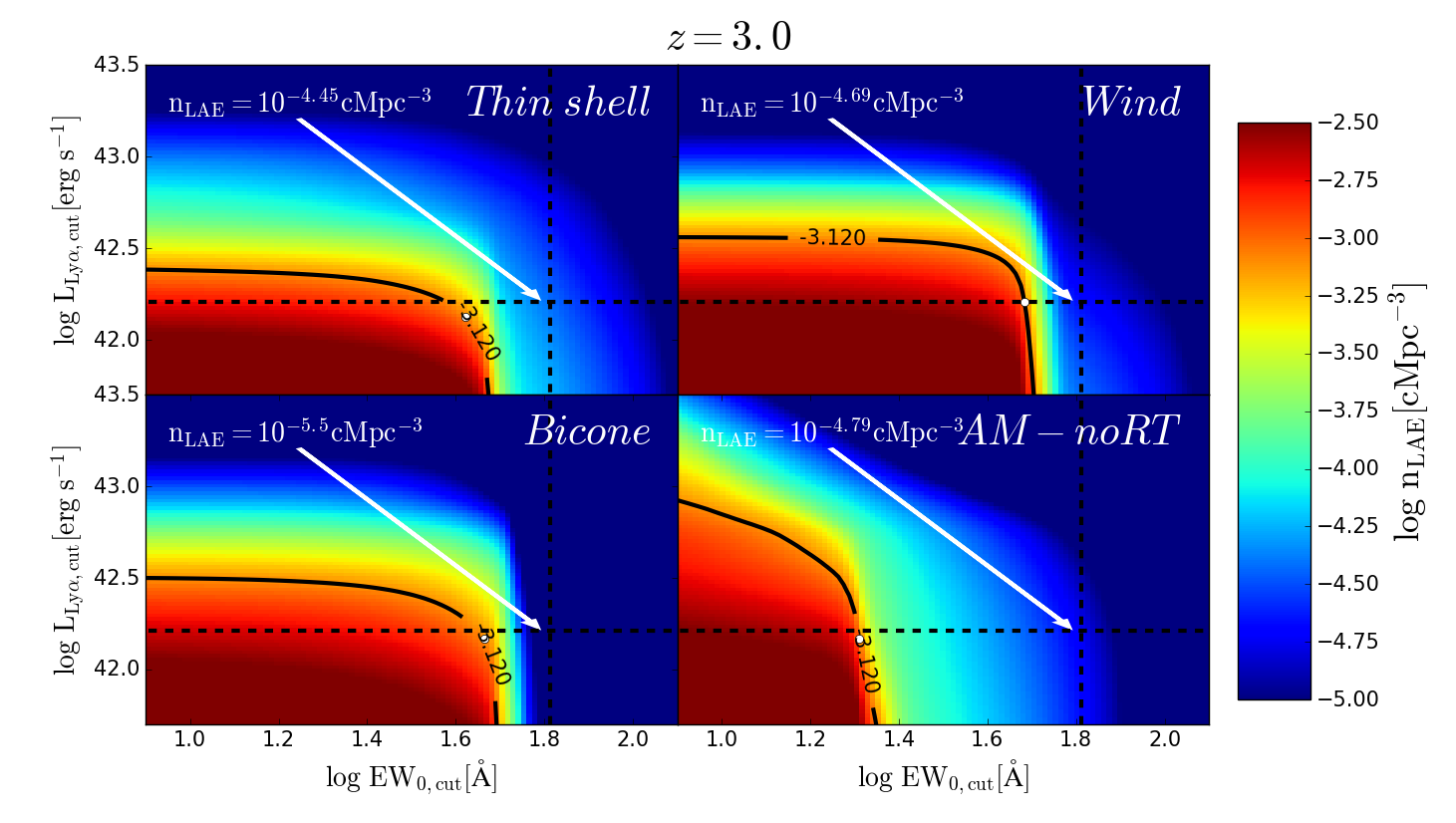} 
\caption{Same as Fig. \ref{fig:compo_z2} but at redshift 3.0  \citep{Bielby2016} }
\label{fig:compo_z3}
\end{figure*}

\begin{figure*} 
\includegraphics[width=6.93in]{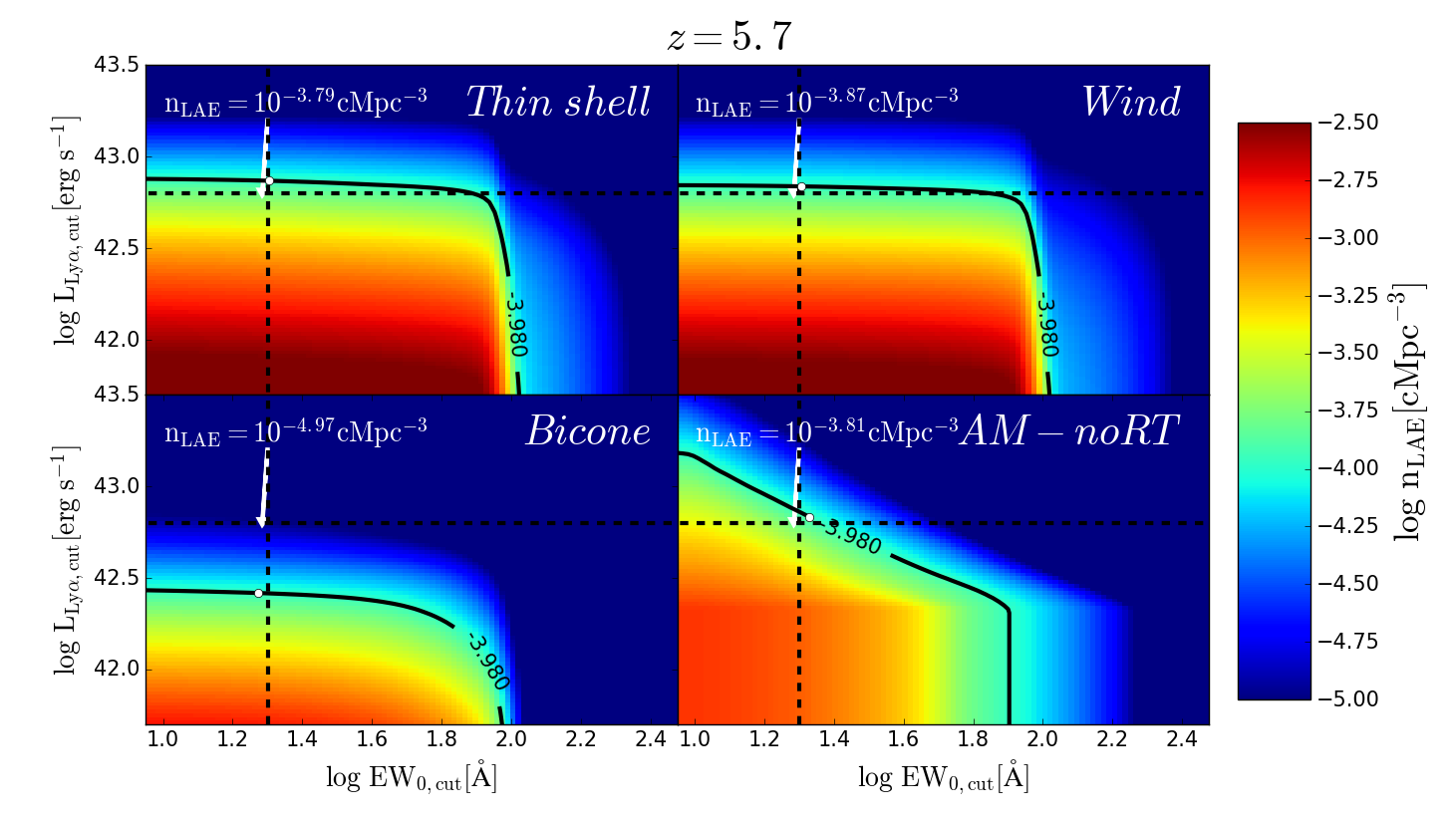} 
\caption{Same as Fig. \ref{fig:compo_z2} but at redshift 5.7  \citep{Ouchi2018a}   }
\label{fig:compo_z5}
\end{figure*}

\begin{figure*} 
\includegraphics[width=6.93in]{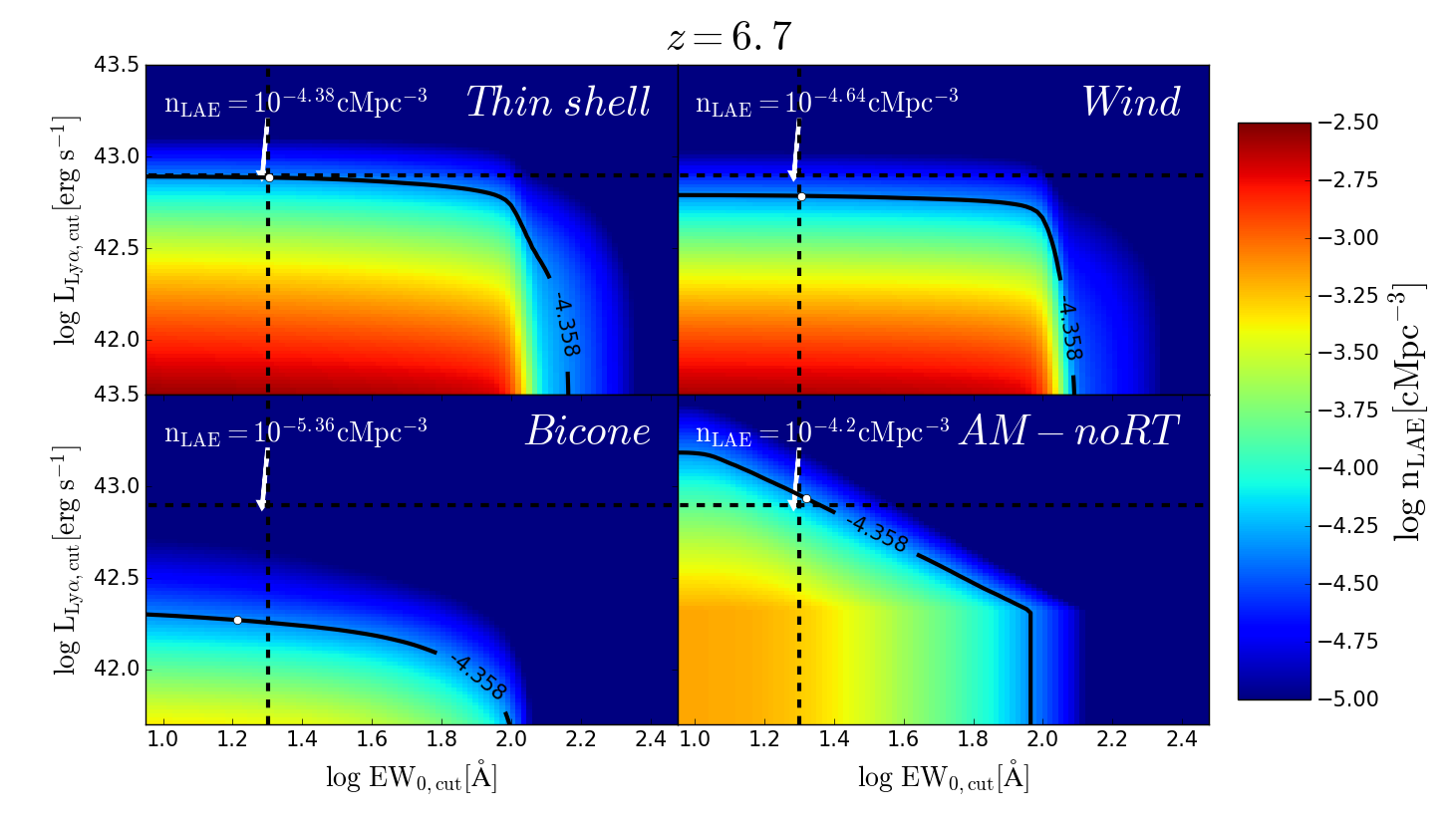} 
\caption{ Same as Fig. \ref{fig:compo_z2} but  at redshift 6.7  \citep{Ouchi2018a}  }
\label{fig:compo_z6}
\end{figure*}

In Figs. \ref{fig:compo_z2}, \ref{fig:compo_z3}, \ref{fig:compo_z5} and \ref{fig:compo_z6} we show the predicted $\rm n_{LAE}$ by our different models for several $\rm L_{Ly \alpha,cut}$-$\rm EW_{0,cut}$ combinations at $z=2.2$, 3.0, 5.7 and 6.7 respectively. In these figures we also show $\rm L_{Ly \alpha,cut}$ and $\rm EW_{0,cut}$ of each of the surveys used for clustering in black dashed lines. The intersection between these shows the location of the clustering surveys selection criteria. Additionally, it is shown the individual value of $\rm n_{LAE}$ predicted by our models imposing the observational cuts (indicated with the white arrow). We also show the curve with constant $\rm n_{LAE}$ matching the observed abundance (solid black line). Finally, the $\rm L_{Ly \alpha,cut}$-$\rm EW_{0,cut}$ combination that minimize Eq. \ref{eq:LLya_EW_cuts} is shown as a white dot. 

At redshift 2.2 the predicted (using the survey selection criteria) and observed $\rm n_{LAE}$ match quite well. Thus, $\rm L_{Ly \alpha,cut}$ and $\rm EW_{0,cut}$ are very similar to $\rm L_{Ly \alpha,s}$ and $\rm EW_{0,s}$. However, the opposite case is found at $z=3.0$, where predicted $\rm n_{LAE}$ is heavily underestimated in comparison with observations. This is mainly due to the mismatch between the predicted $\rm EW_0$ distribution and the observed one. This might be due to the difference in selection criteria used the authors of the works for constraining the LF and the work building the clustering sample. While $\rm L_{Ly \alpha,cut}$ is relatively similar to $\rm L_{Ly \alpha,s}$, in order to recover the observed $\rm n_{LAE}$, in all models, the value of $\rm EW_{0,cut}$ is significantly lower than $\rm EW_{0,s}$. 

The scenarios at redshift 5.7 and 6.7 are quite similar. At both redshifts the predicted number density, using the survey selection criteria, and observed $\rm n_{LAE}$ match quite well for the {\it Thin Shell}, {\it Wind} and \NoRT\ samples. However, in the {\it Bicone} model $\rm L_{Ly \alpha,cut}$ and $\rm L_{Ly \alpha,s}$ are very different. In particular, the {\it Bicone} model requires a low   $\rm L_{Ly \alpha,cut}$ in order to balance underestimation of abundance (see Fig. \ref{fig:LF_across_redshift_fit}).


\bsp	
\label{lastpage}
\end{document}